\documentclass[aps,prb,showpacs,amsmath,twocolumn,reprint,amssymb,nofootinbib,superscriptaddress]{revtex4-2}
\usepackage{graphicx}
\graphicspath{{}}
\usepackage{dcolumn}
 \AtBeginDocument{\usepackage{booktabs}}
\usepackage{siunitx}
\usepackage[]{hyperref}
\usepackage{xcolor}
\usepackage{physics}
\usepackage{amsmath}
\usepackage{placeins}
\newcommand{\ionm}[2]{${}^{#1}$#2${}^+$}
\newcommand{\ion}[1]{#1${}^+$}
\newcommand{\Yb}{\ionm{172}{Yb}}
\newcommand{\YbNM}{\ion{Yb}}
\newcommand{\statels}[3]{\mbox{${}^{#1}$#2${}_{#3}$}}
\newcommand{\CoolTrans}{\statels{2}{S}{\frac{1}{2}}$\,\leftrightarrow\,$\statels{2}{P}{\frac{1}{2}}}

\begin{document}

\preprint{}

\title[]{Finite temperature spectrum at the symmetry-breaking linear-zigzag transition}

\author{Jan Kiethe}
\affiliation{Physikalisch-Technische Bundesanstalt, Bundesallee 100, 38116 Braunschweig, Germany}
\author{Lars Timm}
\affiliation{Institut f\"ur Theoretische Physik, Leibniz Universit\"at Hannover, Appelstr. 2, 30167 Hannover, Germany}
\author{Haggai Landa}
\affiliation{Institut de Physique Th\'eorique, Universit\'e Paris-Saclay, CEA, CNRS, 91191 Gif-sur-Yvette, France}
\affiliation{IBM Quantum, IBM Research Haifa, Haifa University Campus, Mount Carmel, Haifa 31905, Israel}
\author{Dimitri Kalincev}
\affiliation{Physikalisch-Technische Bundesanstalt, Bundesallee 100, 38116 Braunschweig, Germany}
\author{Giovanna Morigi}
\affiliation{Theoretische Physik, Saarland University, Campus E26, 66123 Saarbr\"ucken, Germany}
\author{Tanja E. Mehlst\"aubler}\email[Corresponding author:]{Tanja.Mehlstaeubler@ptb.de}
\affiliation{Physikalisch-Technische Bundesanstalt, Bundesallee 100, 38116 Braunschweig, Germany}
\affiliation{Institut f\"ur Quantenoptik, Leibniz Universit\"at Hannover, Welfengarten 1, 30167 Hannover, Germany}

\date{\today}

\begin{abstract}
We investigate the normal mode spectrum of a trapped ion chain at the symmetry-breaking linear to zigzag transition and at finite temperatures. For this purpose we modulate the amplitude of the Doppler cooling laser in order to excite and measure mode oscillations. The expected mode softening at the critical point, a signature of the second-order transition, is not observed. Numerical simulations show that this is mainly due to the finite temperature of the chain. Inspection of the trajectories suggest that the thermal shifts of the normal-mode spectrum can be understood by the ions collectively jumping between the two ground state configurations of the symmetry broken phase. We develop an effective analytical model, which allows us to reproduce the low-frequency spectrum as a function of the temperature and close to the transition point. In this model the frequency shift of the soft mode is due to the anharmonic coupling with the high frequency modes of the spectrum, acting as an averaged effective thermal environment. Our study could prove important for implementing ground-state laser cooling close to the critical point.
\end{abstract}

\maketitle

\section{Introduction}
Ion Coulomb crystals are an unusual form of condensed matter, where crystalline order emerges from the interplay between Coulomb repulsion and the external trapping potential, while the temperature is controlled by means of lasers \cite{Dubin1999}. These properties make them versatile and controllable systems \cite{Dubin1999,Leibfried2003}, which are among the most prominent platforms for quantum computation 	
\cite{Cirac1995,Leibfried2003a,SchmidtKaler2003,Schindler2013,Monroe2014,Harty2014,Monz2016,Wright2019} and for the simulation of the equilibrium and out-of-equilibrium dynamics of many-body systems \cite{Bylinskii2015,Kiethe2017,Matjeschk2012,Toyoda2013,Martinez2016,Zhang2017,Gaerttner2017,Brox2017,Gorman2018,Zhang2018,Kokail2019,Tamura2020,Zhang2017,Jurcevic2017}.

Amongst others, the Kibble-Zurek mechanism \cite{Kibble1976,Zurek1985,DelCampo2010} and creation of topological defects have been demonstrated \cite{Ulm2013,Pyka2013,Mielenz2013,Ejtemaee2013}.
Two widely discussed transitions are the linear to zigzag \cite{Dubin1993,Schiffer1993,Fishman2008,Piacente2010} and the pinning to 
sliding (Aubry) transition \cite{Garcia-Mata2007,Benasi2011,Mandelli2013}. These were 
shown to be second-order phase transitions \cite{Fishman2008,Aubry1983},
that exhibit a soft mode with vanishing frequency at a critical point. The system however is critical solely at zero temperature. Therefore, the observation of critical behavior requires one to characterize and understand finite temperature effects at the transition, such as the size of the crossover region due to temperature and how thermal excitations modify the normal mode spectrum.

The frequency spectrum at zero temperature is well described by the harmonic crystal approximation. Deviations
to this analytical solution and in particular finite frequencies close to the critical point have been observed for the soft mode of the Aubry-type transition in trapped ion chains, at a temperature of around 1mK \cite{Kiethe2017}. Here, we focus on the experimentally more accessible linear to zigzag transition and investigate the coupling of the soft mode to the thermal phonon environment.
We develop a theoretical model that allows one to reproduce the presented spectroscopic measurements by means of a harmonic chain, whose normal-mode spectrum at low frequencies results from the temperature dependent coupling with vibrational modes at high frequencies. In this sense, the high-frequency modes can be considered a thermal phonon environment. We discuss this result in connection to earlier works \cite{Gong2010,Li2019}, that described finite temperature effects in terms of an effective shift of the transition point. Our findings deepen the understanding of the complex dynamics of ion Coulomb crystals. They could prove important, for instance, for laser cooling the linear ion chain to the ground state in the vicinity of the transition.

This paper is organized as follows: In Section \ref{sec:LinZZ} we briefly review the linear to zigzag transition. In Section \ref{sec:experiments} we present our experimental methods and results of vibrational mode measurements, using resonant light force modulation. Subsequently, in Section \ref{sec:numerics} we compare our findings to molecular dynamics simulations. In Section \ref{sec:analytic}, we discuss a simplified analytical model which allows one to gain insight into the temperature dependence of the spectroscopic measurements. In Section \ref{sec:conclusions} the conclusions are drawn. The appendices provide supplementary material to the studies presented in Sec. \ref{sec:numerics} and Sec. \ref{sec:analytic}.

\section{\label{sec:LinZZ}The linear--zigzag transition}

We consider $N$ ions with charge $e$ and mass $m$, which are confined by a linear Paul trap. The trap potential is described in ponderomotive approximation by three trapping frequencies $\omega_z$, $\omega_x$ and $\omega_y$. The total potential energy $V$ is the sum of the trap confinement and of the unscreened Coulomb interaction between the ions:
\begin{multline}\label{eq:totalPotentialEnergy}
V=\sum_{i=1}^{N}\frac{m}{2}\left(\omega_x^2 x_i^2+\omega_y^2 y_i^2 + \omega_z^2 z_i^2\right) \\
+\frac{e^2}{4\pi\varepsilon_0}\sum_{i=1}^{N}\sum_{j<i}\left|\vb{r}_i-\vb{r}_j\right|^{-1},
\end{multline}
where $\vb{r}_i=(x_i,y_i,z_i)^T$ denotes the position of the ion $i$ ($i=1,\ldots,N$) and $\varepsilon_0$ is the vacuum permittivity.  For later convenience, we introduce the vector $\vb{u}=(x_1,x_2,\dots,x_N,$ $y_1,y_2,\dots,y_N,$ $z_1,z_2,\dots,z_N)^T$, which gives the configuration of the crystal. 

At sufficiently low temperature, the ions localize at the equilibrium positions $\vb{u}(0)$ of the potential $V$, for which the equations $\partial V/\partial u_j=0$ holds. In this configuration the dynamics of the chain is characterized by the matrix $K'$, with elements
\begin{equation}\label{eq:HessianMatrix}
K'_{ij}=\left.\frac{\partial^2 V}{\partial u_i u_j}\right|_{\vb{u}\left(0\right)}\,.
\end{equation}
For stable equilibrium $K'$ has finite and positive eigenvalues. In the rest of this paper we choose $\omega_z<\omega_x<\omega_y$, focusing particularly on the aspect ratio  $\alpha=\omega_x/\omega_z$ for which the ions can either form an one-dimensional crystal along the $z$ axis, the linear chain, or form a two-dimensional crystal in the form of a zigzag configuration on the $x-z$ plane with two degenerate ground states \cite{Birkl1992,Piacente2004}.

\begin{figure*}
	\includegraphics[width=\textwidth]{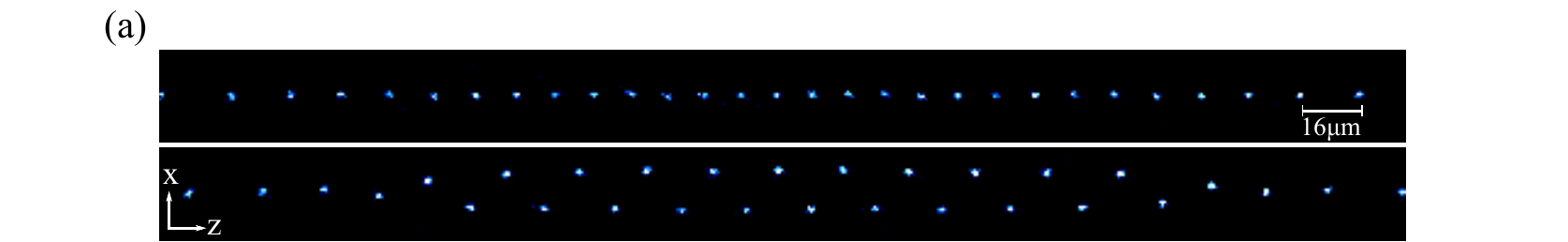}\\
	\hspace{0.25cm}
	\includegraphics[width=7.477cm]{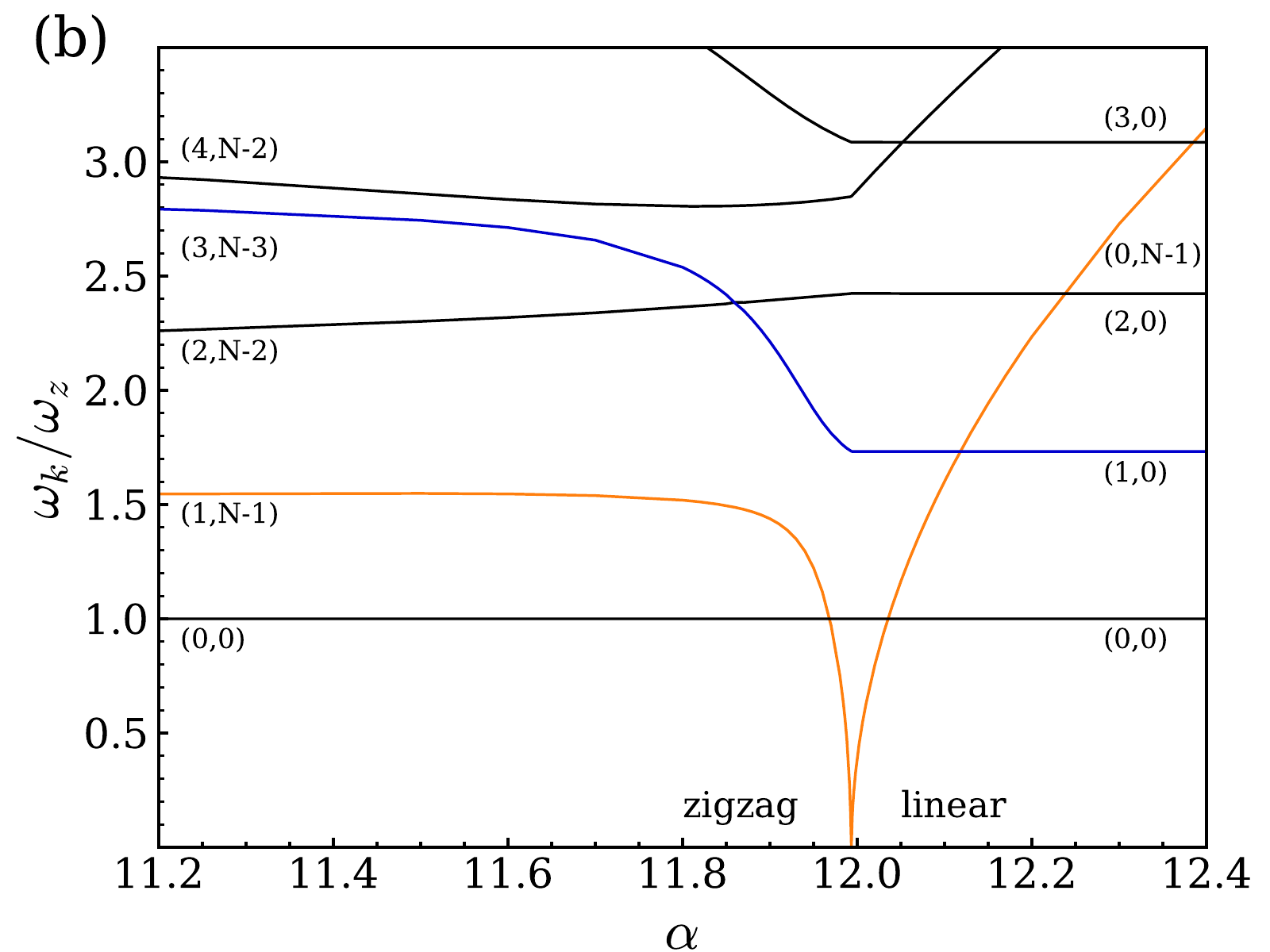}	\includegraphics[height=5.608cm]{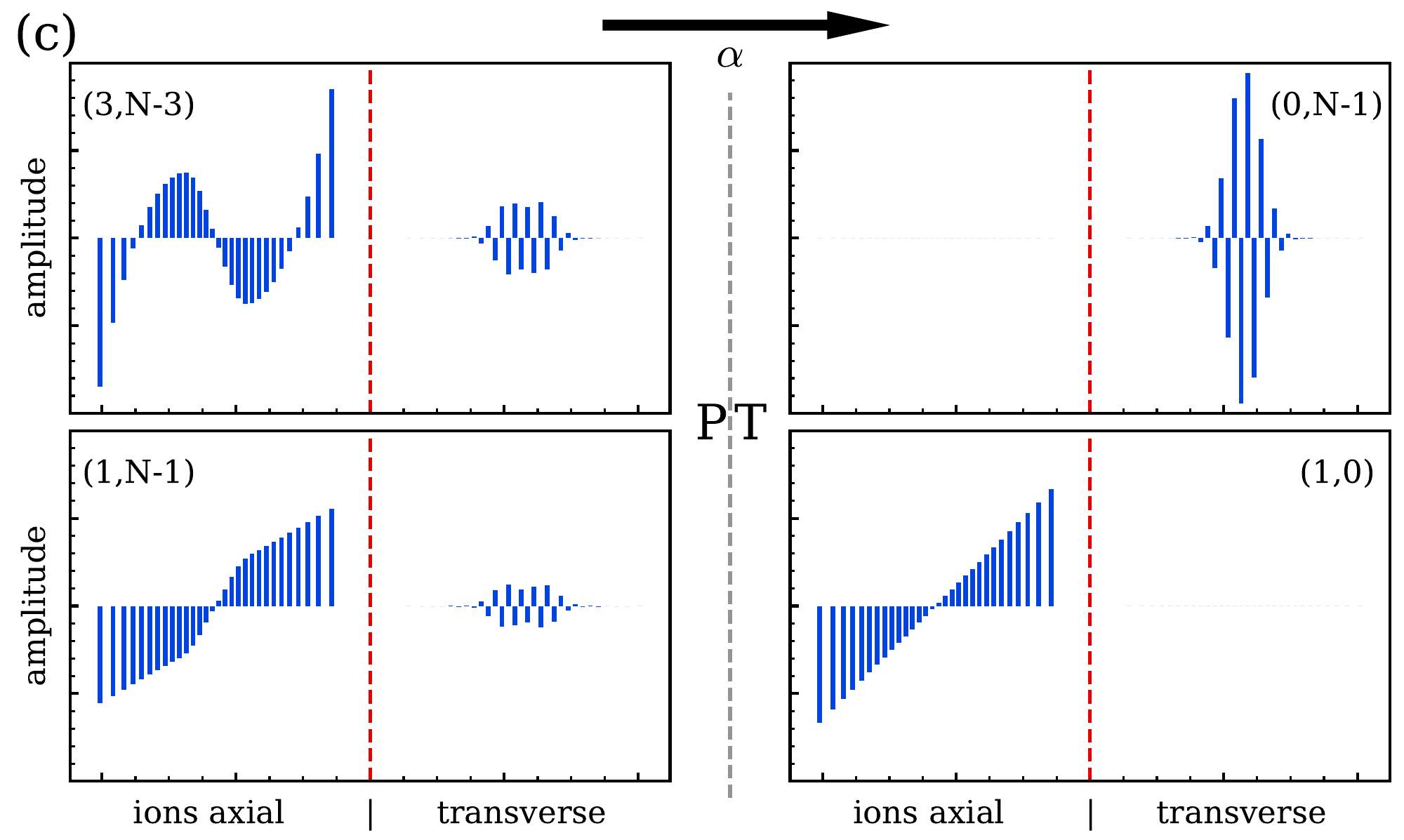}
	\caption{\label{fig:Modevec} (a) Experimental pictures of the linear chain (top) and of the zigzag configuration (bottom). The crystals are formed by 30 laser-cooled \YbNM ions in a linear Paul trap. The images were taken under an angle of \SI{45}{\degree} to the crystal plane and focused on the bottom row of the zigzag crystal. The right most ion for the linear chain is out of view.
	(b) Normal modes as a function of the aspect ratio $\alpha$ and in the vicinity of the linear to zigzag transition at $\alpha_c\approx 12.0$ for $N=30$. The modes are evaluated in the theoretical limit $T=0$. For $\alpha<\alpha_c$ the ions form a zigzag structure, for $\alpha>\alpha_c$ a linear chain. The vanishing of the zigzag mode frequency at $\alpha_c$ signals the phase transition (PT) point. The notation $(n,p)$ indicates the number of axial and transverse nodal points of the corresponding mode vector.
	(c) Normal mode vector for the breathing and the zigzag mode below [above] the phase transition with nodal points (3,N-3) [(1,0)] and (1,N-1) [(0,N-1)], respectively. PT indicates the phase transition.
	The third direction $y$ is not shown, as it has 0 amplitudes for all ions in these modes. Here, the axial trapping frequency is in the range of $\omega_z\approx2\pi\times$\SIrange[range-phrase=--,range-units = brackets]{20}{35}{\kilo\hertz}.
	The transverse trapping frequency is chosen such that $\alpha$ varies in the interval $[11.0,12.5]$.}
\end{figure*}

Figure \ref{fig:Modevec}(a) displays an experimental photo of a linear and of a zigzag chain of 30 ions. The two structures are separated by the critical value of the aspect ratio $\alpha_c\approx 12$. A numerical estimate of the scaling of the transition point with the number of ions gives $\alpha_c(N)\sim 0.556N^{0.915}$, see Refs. \cite{Steane1997,Schiffer1993,Enzer2000}, which gives an approximate location of the transition point \cite{Dubin1993}. The shift of the transition point due to quantum fluctuations have been determined in Refs. \cite{Silvi2014,Podolsky2014}.

The linear to zigzag instability is a continuous phase transition in the thermodynamic limit, corresponding to letting $N\to\infty$ and to rescaling the trap frequencies with $N$ in order to keep the critical aspect ratio $\alpha_c$, constant \cite{Morigi2004,Fishman2008}.
It is associated with breaking of reflection symmetry about the $z$ axis (for $\omega_x=\omega_y$ the broken symmetry is rotational and the transition is characterized by a Goldstone mo\-de) \cite{Fishman2008}.
As for ferromagnetism in one dimension, these properties are strictly valid in the limit of $T=0$, while at finite temperature the transition becomes a "crossover". Let us now make our statement more precise.
In our case, where laser cooling of the chain can be modeled by an effective thermal reservoir \cite{Stenholm1986}, one can use a canonical ensemble to model the properties at steady state. A phase transition, like the linear-zigzag structural instability, is then identified in the thermodynamic limit by discontinuities in the derivatives of the free energy.
The linear-zigzag instability can be mapped to the Ising model for ferromagnetism, where the phase transition is present only at zero temperature and is a quantum phase transition \cite{Fishman2008,Shimshoni2011,Sachdev2000}. At finite but low temperatures, when $k_BT$ is smaller than the gap between the ground and the first excited state of the quantum model, the properties are universal \cite{Shimshoni2011}.
At higher temperatures, such as the ones we consider in this paper, the transition becomes non-universal and abrupt changes and power law scaling characteristics of a phase transition are replaced by a smooth behavior which we here denote by "crossover" (and shall not be confused with the crossover due to finite-size effects) \cite{Sachdev2000}.

In order to understand the effect of temperatures on the vibrational spectrum across the linear to zigzag transition and in a finite chain, we first discuss the normal mode spectrum at $T=0$.
The normal mode spectrum is determined by assuming that the ion displacements due to thermal noise are small in comparison to the equilibrium ion distances. The normal mode frequencies are related to the eigenvalues $\lambda_j$ of the matrix $K'$ by the relation $\omega_j=\sqrt{\lambda_j/m}$ ($j=1,\ldots, 3N$).
The corresponding mode vectors are given by the columns of the dynamical matrix $\lambda_{ij}$, that diagonalizes $K'$ and the mode amplitudes are denoted as $\Theta_j$. 
We use the notation ($n,p$) to identify the mode vectors by the number of nodal points (phase flips between ions) along the axial $(n)$ and transverse direction $(p)$\footnote{This is not an unique identification. In fact, due to the finite size some modes have the same number of nodal points.
A unique identification is achieved, for instance, by specifying also the mode frequency.}. For example the three lowest axial modes in the linear chain are denoted by (0,0), (1,0) and (2,0), while the lowest three transverse modes are (0,$N$-1), (0,$N$-2), (0,$N$-3). The lowest normal mode frequencies for $N=30$ are displayed in Fig. \ref{fig:Modevec}(b) as a function of the aspect ratio $\alpha$ across the linear to zigzag transition.

At the transition point the frequency of one normal mode vanishes. In the linear chain this mode is the zigzag mode and has a purely transverse oscillation with (0,$N$-1) nodal points,  see Fig. \ref{fig:Modevec}(c). In the thermodynamic limit the zigzag mode of the linear chain is the soft mode of the phase transition \cite{Fishman2008}.
It is interesting to analyze the property of the eigenmode at lowest frequency as a function of the aspect ratio $\alpha$. While in the linear chain $(\alpha>\alpha_c)$ is corresponds to the zigzag mode, in the symmetry broken (zigzag) phase at $\alpha<\alpha_c$, the eigenmode at lowest frequency gains an axial nodal point and becomes the new breathing mode (1,$N$-1) of the zigzag configuration. The axial breathing mode (1,0) of the linear chain, instead, gains 2 axial nodes, as well as a transverse zigzag pattern to become the (3,$N$-3) mode.

In the following sections, we will denote the linear chain by 1D phase and the zigzag crystal by 2D phase. We remark that the term zigzag mode refers to the mode with (0,$N$-1) nodal points in the 1D phase and (1,$N$-1) nodal points in the symmetry-broken, 2D phase. Moreover, the breathing mode is the mode with (1,0) nodal points in the 1D phase and (3,$N$-3) nodal points in the 2D phase.

In the rest of this paper we analyze how the normal mode spectroscopy at the structural transition is modified at finite temperatures.

\section{\label{sec:experiments}Measurement of vibrational modes}
In this Section we describe our experimental method for measuring vibrational mode frequencies, that makes use of a single laser beam with frequency near resonant to the Doppler cooling transition. We then present and discuss our measurements of the lowest axial modes near the linear to zigzag transition. This method was originally introduced in \cite{Kiethe2017}.

\subsection{Setup}

\begin{figure*}[htb]
	\centering
	\includegraphics[width=0.8\textwidth]{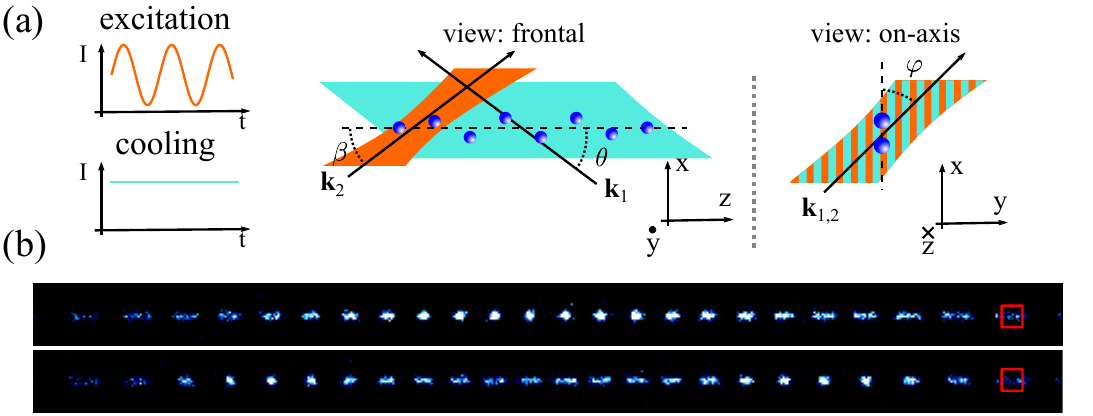}
	\caption{\label{fig:Exp_Setup} Resonant light force modulation. (a) Schematic laser beam set-up. The ion crystal is almost uniformly illuminated by a cooling beam $\vb{k}_1$  (cyan). The laser beam illuminates the crystals axial extent $(z)$ under an angle of $\theta\approx\SI{25}{\degree}$ and the transverse extent $(x)$ under and angle of $\varphi\approx\SI{45}{\degree}$. An excitation beam $\vb{k}_2$ (orange) is focused down to only on a fraction of the crystal. It hits the crystal $z$ axis under an angle $\beta\approx\SI{25}{\degree}$ and the $x$ axis under the same angle $\varphi$ as the cooling beam. The excitation beam is amplitude modulated, resulting in a sinusoidal force. The beam sizes are not to scale. (b) Example crystal photos with excited normal modes. Images taken with an EMCCD camera when the modulation frequency of the excitation laser is resonant with a normal mode. Exposure Time: \SI{100}{\milli\second}; $P_{m}\approx\SI{20}{\micro\watt}$. Top: Breathing mode / (1,0). Bottom: (2,0). Red squares represent possible regions of interest to record fluorescence of single ions.}
\end{figure*}

In order to have a well defined ordered structure, we trap $N=30$ \ionm{172}{Yb} ions in a linear Paul trap with high control of electrical fields \cite{Pyka2014,Keller2019}. The axial trapping frequency is in the range of $\omega_z\approx2\pi\times$\SIrange[range-phrase=--,range-units = brackets]{20}{35}{\kilo\hertz}. The transverse trapping frequency is in the range of $\omega_x\approx2\pi\times$\SIrange[range-phrase=--,range-units = brackets]{220}{440}{\kilo\hertz}, depending on the chosen trapping ratio $\alpha$.
All ions are illuminated by a linearly polarized laser with central wavelength of \SI{369.5}{\nano\metre} addressing the \CoolTrans transition in \YbNM and cooling the ions close to the Doppler cooling temperature of $T_D=\SI{0.5}{\milli\kelvin}$. As shown in Fig. \ref{fig:Exp_Setup}, the laser beam forms an angle of about $\theta=\SI{25}{\degree}$ with the axial direction of the ion crystal and an angle of about $\varphi=\SI{45}{\degree}$ with the transverse direction.
We denote its wave vector as $\vb{k}_1$, which we use in the subsequent text to identify the laser beam itself.
The beam has elliptic shape with waists of approximately \SI{2.6}{\milli\metre} in the horizontal and \SI{80}{\micro\metre} in the vertical direction, resulting in an almost uniform illumination of a \SI{400}{\micro\metre} times \SI{20}{\micro\metre} ion crystal in the z-x plane.
Typical laser powers in the subsequent measurements are ${P_1=\SI{1}{\milli\watt}}$, corresponding to a saturation of  $s_1\approx1.75$ at the beam center (saturation power of $\vb{k}_1$ is $P_{1,s}\approx\SI{570}{\micro\watt})$. 

We also employ a second laser beam at the same wavelength and with the same angles to the crystal, that is focused to a beam waist of about \SI{80}{\micro\metre} in both vertical and horizontal direction, addressing a smaller region of the crystal.
Its wave vector is denoted by $\vb{k}_2$. The beam is amplitude modulated in order to excite the crystal's normal modes.
The amplitude modulation is added by applying a sine wave with frequency $\omega_e$ to the RF amplitude of an acousto-optic modulator used as a fast shutter.
The modulation of the power is given by
\begin{equation*}
	P_2=\frac{P_{m}}{2} \left(1+\cos(\omega_e t)\right)\,,
\end{equation*}
where $P_{m}$ is the maximum power in the beam. The saturation of $\vb{k}_2$ at the beam center is then
\begin{equation*}
	s_2=s_m(1+\cos(\omega_e t)),
\end{equation*}
where 
\begin{equation*}
	s_m=\frac{P_m/2}{P_{2,s}}
\end{equation*}
with saturation power $P_{2,s}\approx\SI{38}{\micro\watt}$ of $\vb{k}_2$.
The total saturation of an ion at the center of the beams is than $s= s_1 + s_m + s_m \cos(\omega_e t)$.
The ions fluorescence is imaged via a lens system of $N/A=0.2$ and recorded by an electron-multiplying (EM)CCD camera, which can resolve individual ions.

\subsection{\label{sec:experiments:met}Method}

We excite the crystal's collective motion with the help of the amplitude modulated cooling laser.
Both cooling lasers $\vb{k}_1$ and $\vb{k}_2$ are continuously incident on the ions during the measurement and both exert a constant light force on the crystal, that shifts the minimum of the trap potential.
The amplitude modulated laser adds an oscillating force $F_m$ with excitation frequency $\omega_e$.
This oscillating force is roughly linear, if $s_m$ is smaller than $s_1$.
For multiple ions, the saturation $s_2$ and $s_1$ will depend on the ion positions with respect to the laser beam center. Specifically, the saturation power $P_{s}(\vb{r}_i)$ will depend on the position of the $i$th ion.

In principle, all normal modes can be excited by means of this technique. In the measurements we present below the waist of laser $\vb{k}_2$ was focused to only \SI{80}{\micro\metre}. It illuminates several ions at the same time, as illustrated in Fig. \ref{fig:Exp_Setup}(a). This prevented the excitation of modes with a higher number of nodal points, due to the small overlap of their mode vector with the laser intensity profile, such as the zigzag mode in the 1D phase, which has $N$-1 nodal points. We note that the excitation of an arbitrarily chosen modes can be realized by implementing single ion addressing.

On resonance, the amplitude of the driven mode increases linearly with $F_m$ for small oscillations around the equilibrium positions. In combination with a constant linear damping $\gamma$ due to laser cooling of $\vb{k}_1$, a steady state with a constant, frequency-dependent mode amplitude $\Theta_j(\omega_e)$ can be reached after several oscillations.
In order to detect an excitation, we record the ions fluorescence with an EMCCD over an exposure time of typically \SI{100}{\milli\second}.
This is long compared to the normal mode oscillation periods, which are on the order of the center of mass oscillation period of about \SI{40}{\micro\second}. Therefore, 
light from all possible ion positions during the oscillations is recorded, leading to an 
apparent increase of the ions' size at the resonance $\omega_e\approx\omega_j$.
The imaged spatial extent of each ion $i$ is proportional to the amplitude of the driven 
normal mode and the ions vector element of the corresponding mode vector $\lambda_{ij}$. 
The resonance frequency is found by identifying the frequency at the maximum amplitude of the ions oscillation.
An experimental photo of the excited (1,0) mode and 
(2,0) mode in the 1D phase is shown in Fig. \ref{fig:Exp_Setup}(b), for which we 
used $P_m\approx \SI{20}{\micro\watt}$ ($s_m\approx0.53$ for an ion at the beam center).
Similar to the ion amplitude, the velocity of the ions increases on resonance, leading to a drop in fluorescence due to the Doppler shift. For a single ion this decrease in fluorescence can be measured with a PMT and enables one to identify the motional resonance.
The described method is similar to what has been employed in dusty plasmas to measure acoustic waves \cite{Liu2003,Piacente2004a}.

\subsection{\label{sec:experiments:res}Experimental Results}

\begin{figure}[tbh]
	\centering
	\includegraphics[width=0.47\textwidth]{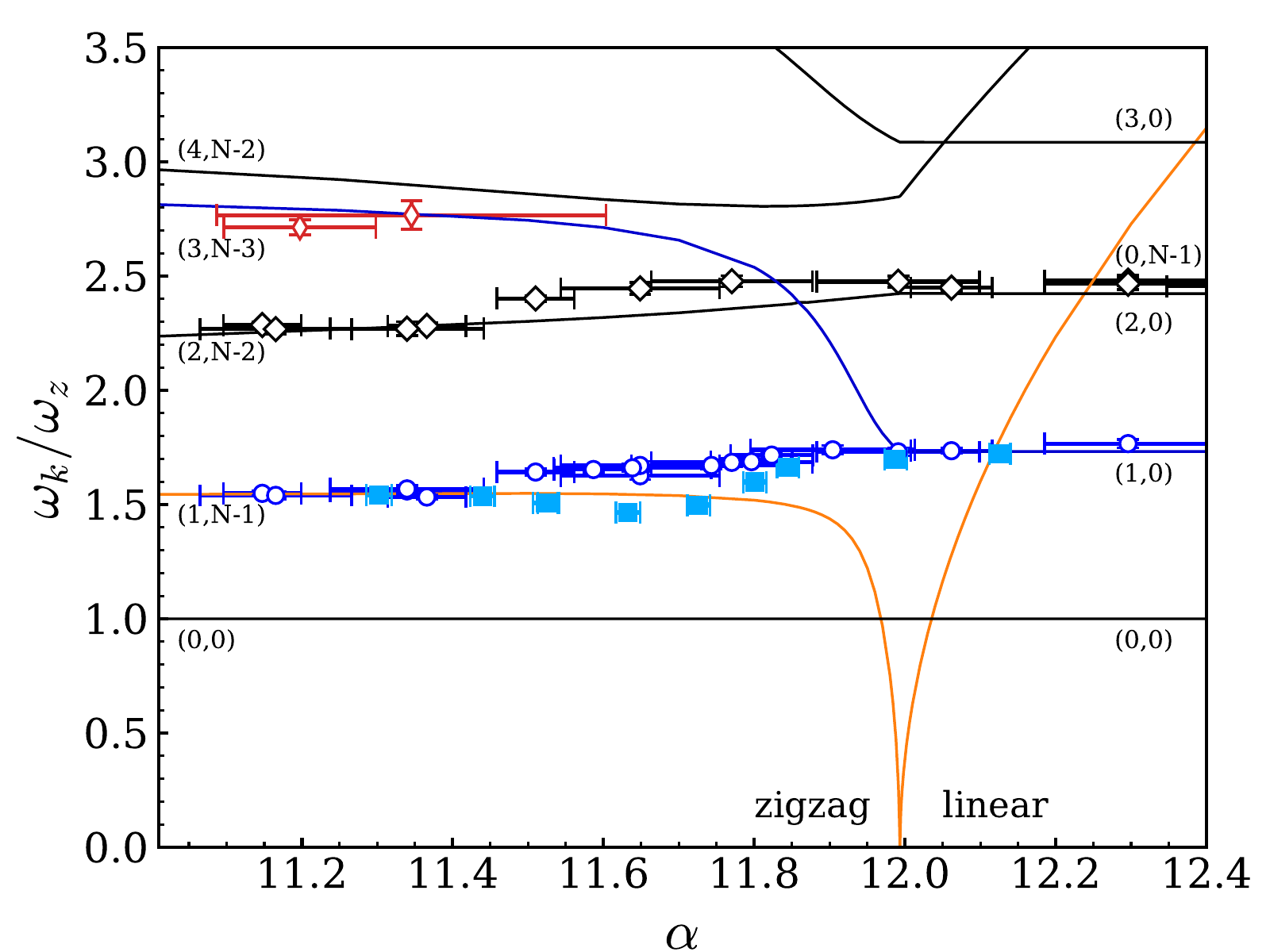}
\caption{\label{fig:exp_analytical}Low frequency spectrum as a function of the aspect ratio $\alpha$. The solid lines are the modes of the harmonic crystal at $T=0$, the symbols refer to the experimental measurements. Experimental data from Series (A) as empty symbols (blue circles, black diamonds, red thin diamonds) with $P^{(A)}_m=\SI{20}{\micro\watt}$ and Series (B) as filled symbols (light blue squares) with $P^{(B)}_m=\SI{6}{\micro\watt}$. For (A): The error bars represent estimated uncertainties in measured frequencies. For (B): the error bars represent fit uncertainties. The ions are laser cooled close to the Doppler limit. From comparison to simulations in section \ref{sec:numerics:res}, we find $T\approx\SI{3.5}{\milli\kelvin}$.}
\end{figure}

We measured several mode frequencies near the linear to zigzag transition in two experiment series, with two different modulation powers: $P_m^{(A)}= \SI{20}{\micro\watt}$ (A) and $P_m^{(B)}=\SI{6}{\micro\watt}$ (B).

For measurement run (A), we determined the center frequency $\omega_{j}$ of any resonance by scanning the excitation frequency $\omega_e$ manually and searching for the maximum amplitude of the ions for the excited mode. The uncertainties were estimated by finding a region, in which the amplitude of the excitation was still maximal.
The width of this region was taken as the error of the measurement, with typical values of about \SIrange{100}{300}{\hertz} for each resonance. The power of the excitation laser was set to $P_m^{(A)}= \SI{20}{\micro\watt}$, chosen such that a resonance of the three to four lowest modes could be observed.

At first, the trapping ratio $\alpha$ is determined by measuring the axial and transverse center-of-mass mode frequencies, i.e. the trapping frequencies. This is followed by searching for the low lying modes with 1,2, and 3 axial nodal points.
In Fig. \ref{fig:exp_analytical} we show the measured vibrational mode frequencies, in comparison to the normal mode frequencies expected from the second-order approximation.
Away from the phase transition the experimental results agree with the theoretical predictions. However, close to the phase transition, the frequency of the zigzag mode does not vanish. The measured frequency of mode (1,$N$-1) (blue empty circles) increases when $\alpha$ approaches $\alpha_c$ until it reaches the expected frequency of the breathing mode of the 1D phase. The purely radial zigzag mode in the 1D phase could not be excited by this measurement, due to the missing overlap between the laser beam profile and the normal mode vector, see section \ref{sec:experiments:met}. While the (2,0) mode frequency (black empty diamonds) was observed over the complete phase transition, close to transition the breathing mode in the 2D phase (red empty thin diamonds) was not detected.

In measurement series (B), a single ions fluorescence was recorded with a region of interest (ROI) on the EMCCD, while sweeping the excitation frequency. Near resonance, a decrease in fluorescence in the ROI is observed, because the excited ion moves partially out of the ROI during exposure and it gains a Doppler shift due to its increased velocity.
We fit the fluorescence drop to a Lorentzian line shape in order to determine the resonance frequencies of the axial center of mass and the normal mode with one axial nodal line.

To obtain a finer resolution in series (B) the maximum power of the amplitude modulated laser was about $P_m^{(B)}=\SI{6}{\micro\watt}$, sufficient to excite the center of mass and breathing mode. A smaller amplitude of the forced oscillation reduces sampling of higher order terms of the Coulomb potential, which leads to asymmetric line shapes and line broadening. In Fig. \ref{fig:exp_analytical}, we show the results of these measurements as cyan squares. The results agree qualitatively with the measurements from series (A). We verify an increased vibrational mode frequency of the zigzag mode in the 2D phase close to $\alpha_c$. In the range of $\alpha=11.5$ to $11.85$ a quantitative difference of the measured frequencies is observed, up to a difference of approximately $0.2\omega_z$.

The quantitative difference between (A) and (B) is due to the smaller power of the amplitude modulated laser used for measurement run (B). A larger driving force increases the mode amplitudes in the steady state and therefore enhances non-linear frequency shifts due to the Coulomb interaction. Increased power leads to a higher observed frequency at maximum excitation. Additionally, the lineshape becomes increasingly asymmetric with increased power. We refer the interested reader to Appendix \ref{sec:app:power_dep}, where we discuss he influence of $P_m$ on the frequency of the (1,$N$-1) mode, see Fig. \ref{fig:app:powersweeps}.

\section{\label{sec:numerics}Molecular dynamics simulations}
As seen in the last section, close to the phase transition the measured excitation frequencies significantly deviate from the vibrational spectrum in the harmonic approximation. We identify two possible sources for this deviation, either the damping due to laser cooling or the interaction with higher order terms in the expansion of the Coulomb potential could be responsible. As the damping, $\gamma=\SI{8.75e3}{\per\second}$, is orders of magnitude smaller than the lowest axial frequencies, approx. $\SI{1.6e5}{\per\second}$, its influence is negligible. Therefore, the higher order terms are the most likely cause for the observed deviations. In the experiment there are two excitation sources, that lead to increased amplitudes: the thermal noise from laser cooling and the sinusoidal driving force. In order to gain deeper insight into the impact of the mode populations on the measurable frequencies, we carry out molecular dynamics simulations of the crystal under a stochastic force.

\begin{figure}
	\centering
	\includegraphics[width=0.47\textwidth]{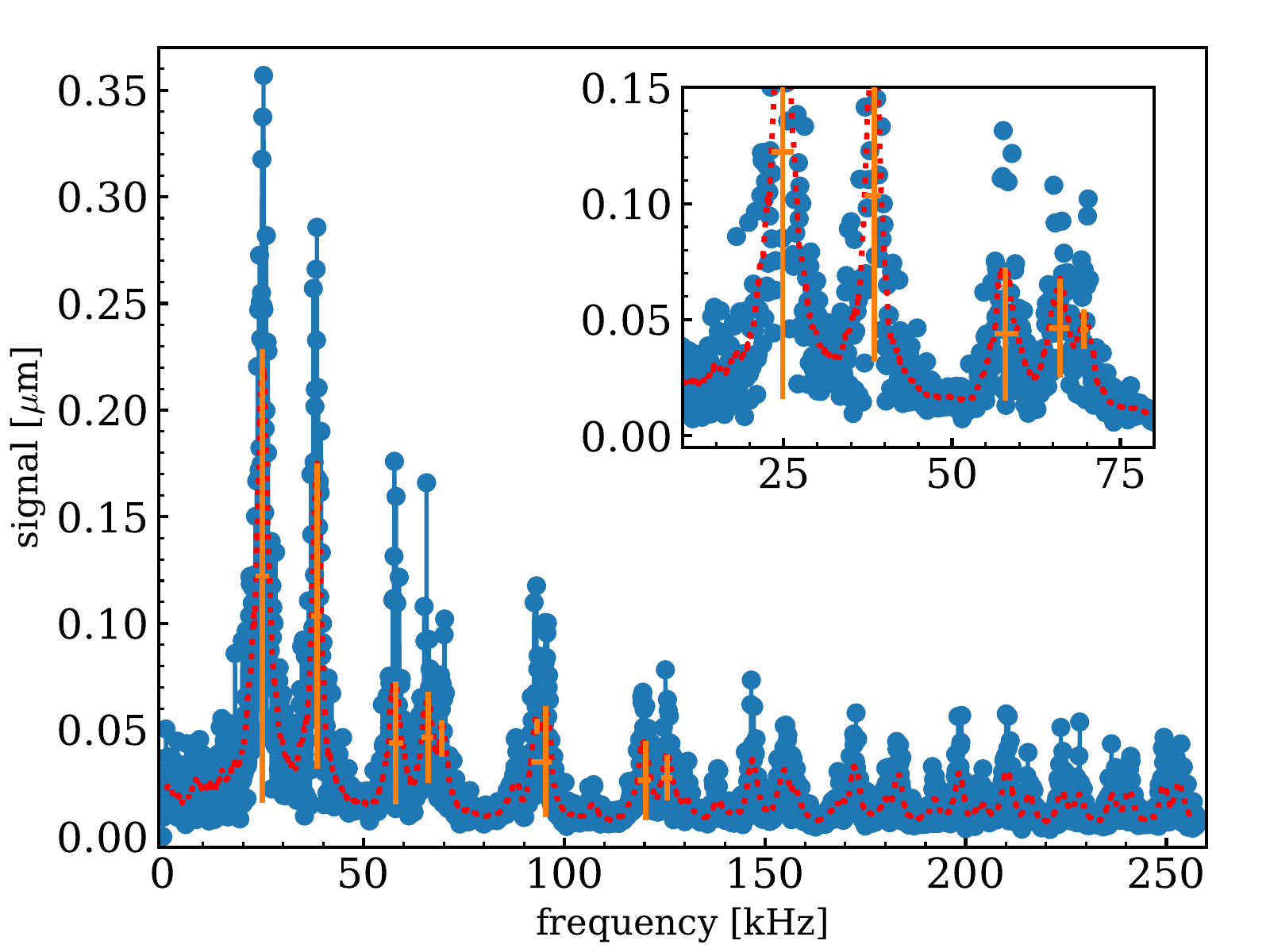}
	\caption{\label{fig:example_fft}Example of a FFT signal $S_z$, Eq. \eqref{eq:S}. The signal is extracted from the trajectory of the $z$ coordinates of 5 molecular dynamics simulations at $\alpha=11.7$ and $T=\SI{0.1}{\milli\kelvin}$. The blue dots show the FFT, the dotted red line is the running mean of the signal over 30 points, the orange lines indicate automatically estimated center and widths of the peaks. The inset shows a zoom up of the first 5 peaks. See text for further details on the simulation.}
\end{figure}

We simulate the dynamics of the ion crystal by numerically solving the classical equations of motion of the ions in the presence of damping and of the Langevin force describing thermal noise \cite{Pyka2013}. This approach is complementary to the Fokker-Planck equation for Doppler cooling of an ion crystal \cite{Morigi2001}. The equation of motion for the $i$th degree of freedom takes the form
\begin{equation}\label{eq:Langevin}
m\ddot{u}_i+\frac{\partial V}{\partial u_i}+m\gamma\dot{u}_i=\xi_{i}\left(t\right)\,,
\end{equation}
where $\gamma$ is a damping term from laser cooling and $V$ is the total potential energy, see Eq. (\ref{eq:totalPotentialEnergy}), and  $\xi_{i}\left(t\right)$ is the stochastic force, with moments:
\begin{align}\label{eq:FluctuationDissipation}
	\ev{\xi_{i}\left(t\right)}&=0\\
	\ev{\xi_{i}\left(t\right)\xi_{j}\left(t'\right)}&=2D\delta_{ij}\delta\left(t-t'\right)\,.
\end{align}
Here, $\ev{\dots}$ indicates ensemble averaging. The diffusion coefficient $D$ of the second equation links the amplitude of the stochastic force with temperature and damping coefficients according to the fluctuation-dissipation theorem, $D=m\gamma k_B T$ \cite{Kubo2012}. In the simulations, laser cooling is treated as isotropic for all degrees of freedom.
This is a simplification with respect to the experiment. There, the damping between the transverse and axial direction is slightly different, due to the projection of the cooling laser beams on the crystal axes, with projection on $x$ and $y$ being $\cos(\SI{45}{\degree})\approx0.71$ and $z$ being $\cos(\SI{25}{\degree})\approx0.91$.

We first detail the simulation procedure and the spectral analysis. The results of the molecular dynamics simulation are then reported and discussed in Section \ref{sec:numerics:res}.

\subsection{Simulation procedure}

The ground state configuration for each trapping ratio $\alpha$ is found by simulating a crystal with $N=30$ ions choosing strong damping $m\gamma=\SI{2.5e-19}{\kilogram\per\second}$ ($\gamma=\SI{8.75e5}{\per\second}$) and with $T=0$.
The resulting equilibrium positions so obtained are the initial configuration for the simulation at finite temperature $T$. The simulation is run with lower damping $m\gamma=\SI{2.5e-21}{\kilogram\per\second}$ ($\gamma=\SI{8.75e3}{\per\second}$), comparable to experimental conditions \cite{Pyka2013}, and over a time of \SI{100}{\micro\second} to thermalize the system. The system is in a thermal state after this, which we checked via the equipartition theorem.

This result is used as a starting point for the final simulations, which run for \SI{10}{\milli\second} in total to achieve a fine resolution in the Fourier frequencies.
All simulations have an integration time step of \SI{19}{\nano\second}, which is much smaller than the expected period of the vibrational mode with the largest frequency, which is here the transverse vibration of the center of mass mode at about \SI{3}{\micro\second}. Every $100$th value is saved, resulting in time resolution of the ions evolution of \SI{1.9}{\micro\second}.

\subsection{\label{sec:sim:evaluation}Spectral analysis}

In order to extract the normal mode spectrum of the ion crystal, we carry out the Fourier transform (FT) of the trajectories of the ions' axial and transverse degrees of freedom.
Due to the simulation length and time resolution the FT has a frequency resolution of \SI{100}{\hertz} and a maximum observable frequency of about \SI{263}{\kilo\hertz}, which covers the frequency range of interest.
For our analysis, the simulation procedure described above is repeated five times, due to the stochastic nature of the thermal noise and the FTs are averaged over all simulations with identical parameters. Then the absolute value $A(\omega)_{\chi, i} = \left|\bar{F}(x_{\chi, i})\right|$ of the averaged FTs $\bar{F}$ is calculated, where $i$ is the ion index and $\chi$ is either $x$ or $z$.
We are interested in collective motions of the crystal, i.e. the normal modes, but we do not make any assumptions on possible mode vectors. Therefore, the $A(\omega_{\chi, i})$ for all ions are added together to get a signal $S(\omega)_{\chi}$:
\begin{equation}
\label{eq:S}
	S(\omega)_{\chi}=\sum_{i=1}^N A(\omega)_{\chi, i}\,,
\end{equation}
where the degrees of freedom along $x$ and $z$ are treated separately.
An example of such a signal is shown in Fig. \ref{fig:example_fft}. The width of the resonances depends on the damping $\gamma$, which is here fixed to the value $m\gamma=\SI{2.5e-21}{\kilogram\per\second}$ in order to compare the data to our experiment.

We extract the resonances and the peaks' widths from $S(\omega)$. Without prior knowledge of the complete model of the peak functions, we estimate the positions based on a peak search algorithm. It searches for local maxima that fulfill certain condition with respect to their width, absolute height and relative height to the closest base line. This method cannot treat noisy signals well. Therefore, $S$ is smoothed before starting a peak search, using a running mean over $n$ values $S_{\text{rm}}(k)=\sum_{i=0}^{n} S_{k+i}/n$. For $T=\SI{0.1}{\milli\kelvin}$ an average over $20$ points ($\approx \SI{2}{\kilo\hertz}$) and for the other shown temperatures an average over $30$  points ($\approx \SI{3}{\kilo\hertz}$) is used. The running mean of the signal is plotted as a function of the respective running mean frequency $f_\text{rm}(k)=\sum_{i=0}^n f_{k+i}/n$. In Fig. \ref{fig:example_fft} the smoothed signal $S_{\text{rm}}$ is shown as a red dotted line.
The orange line shows the peak positions and estimated half-maximum widths which we identified. Not all peaks are captured by the algo\-rithm, especially for $\omega>2\pi\cdot\SI{130}{\kilo\hertz}$,
due to the choice of selection parameters, that favor the prominent peaks at lower frequencies.
However, the high frequency peaks are outside the frequency range that we are interested in. Due to the noisy data, additional small peaks might be found close to strong resonances, e.g. the axial center of mass mode. These false positives have to be removed manually.

\subsection{\label{sec:numerics:res}Numerical Results}

\begin{figure*}
	\centering
	\includegraphics[width=0.47\textwidth]{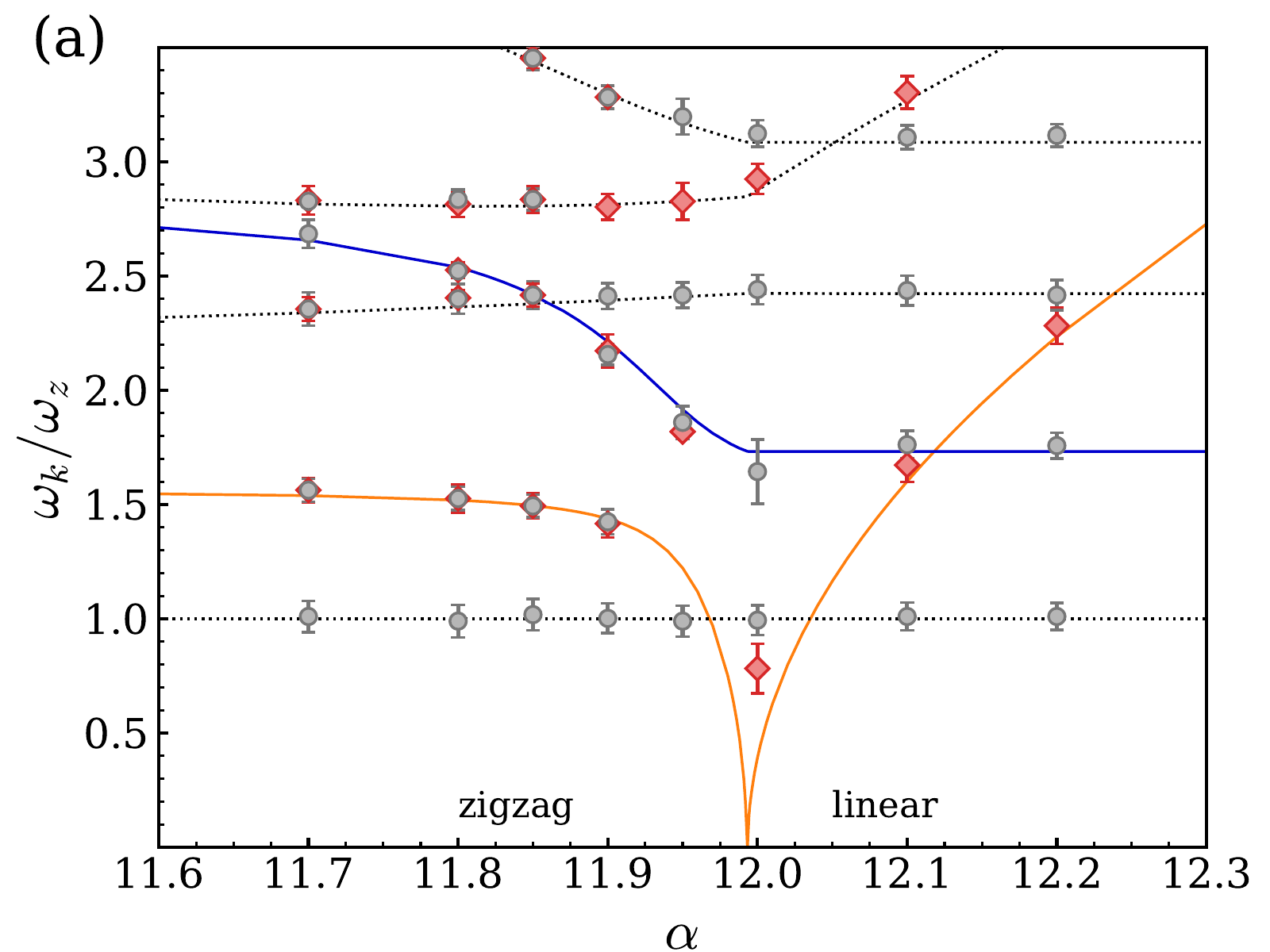}
	\includegraphics[width=0.47\textwidth]{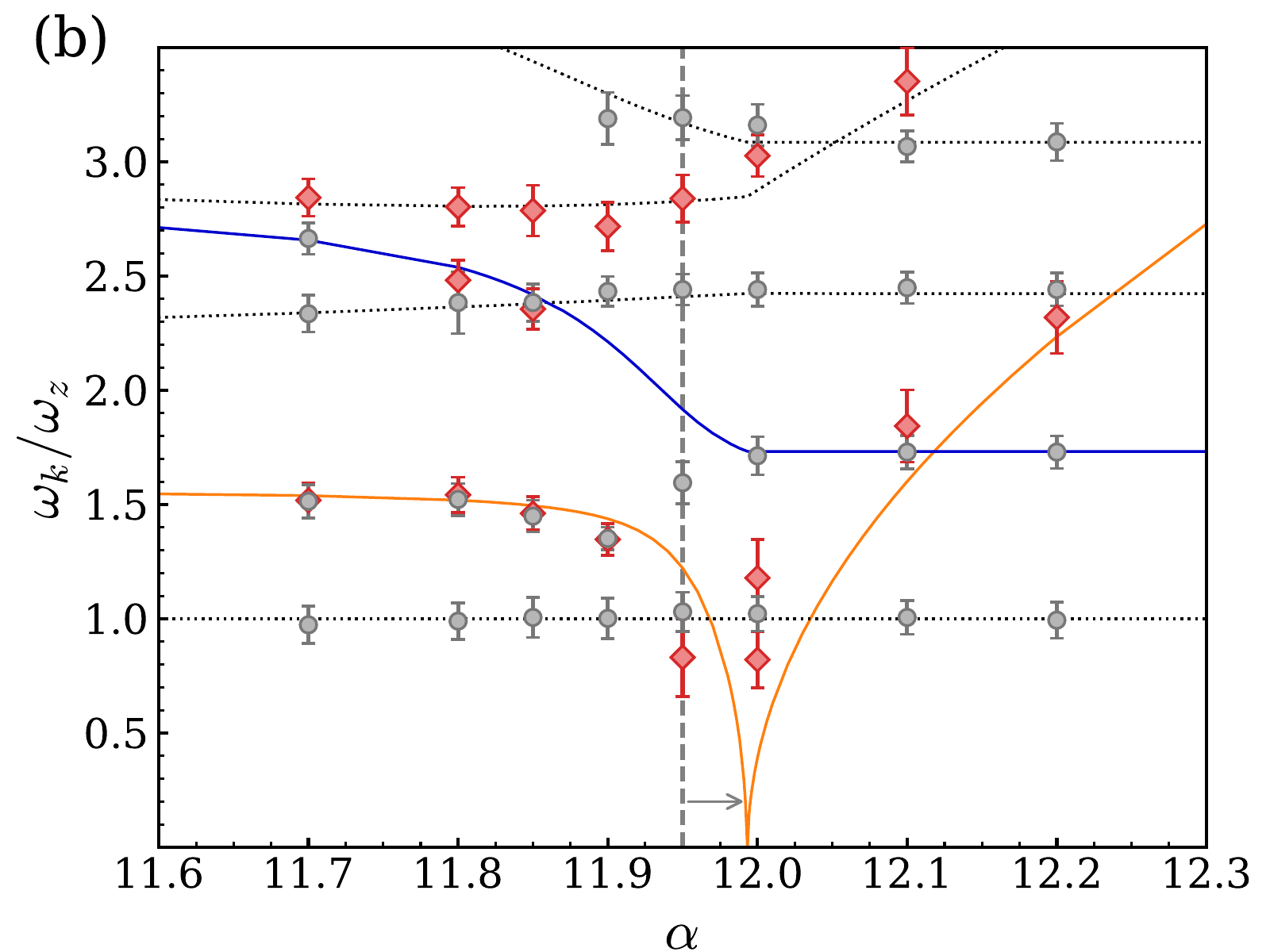}
	\vskip\baselineskip
	
	\includegraphics[width=0.47\textwidth]{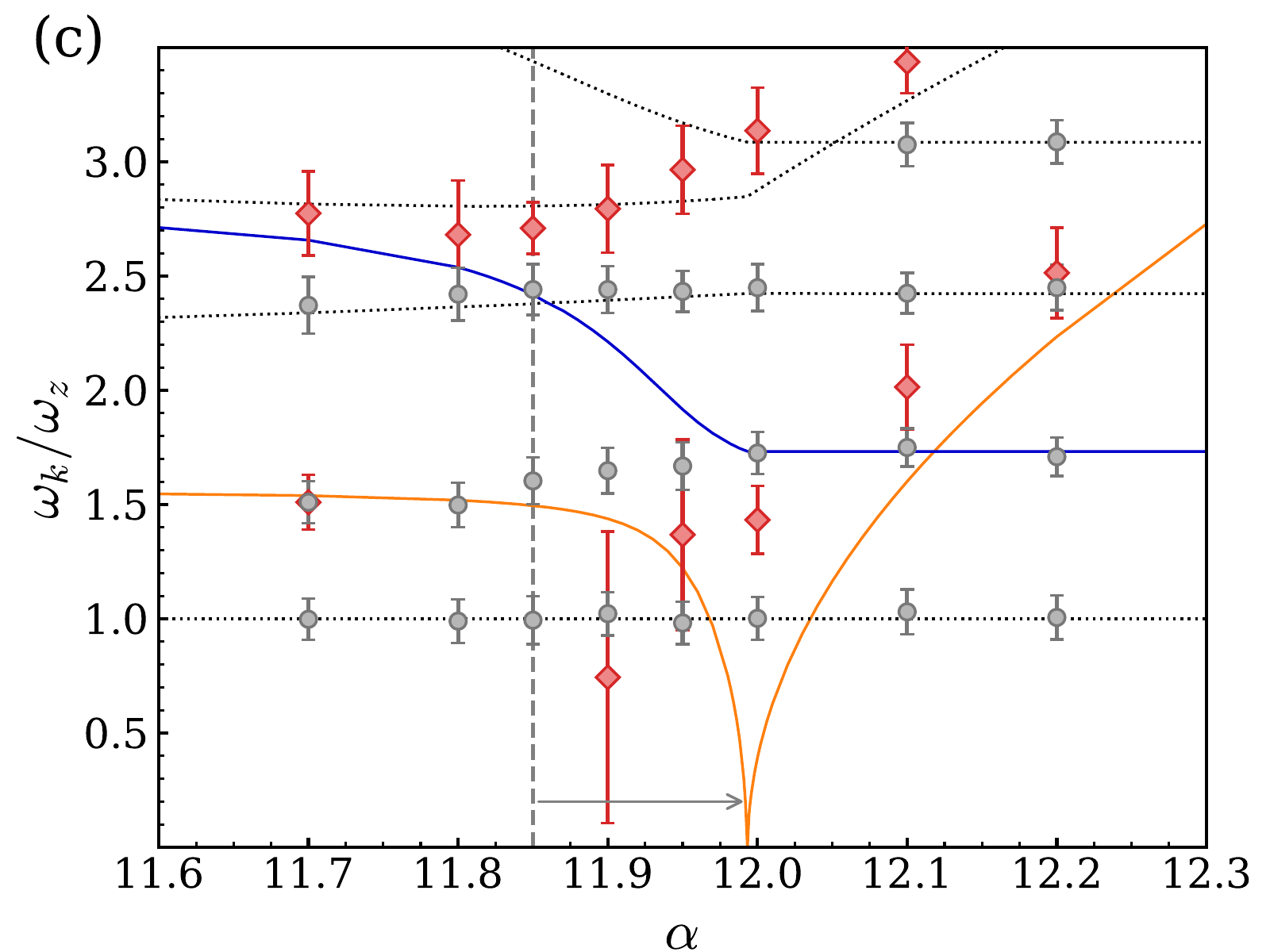}
	\includegraphics[width=0.47\textwidth]{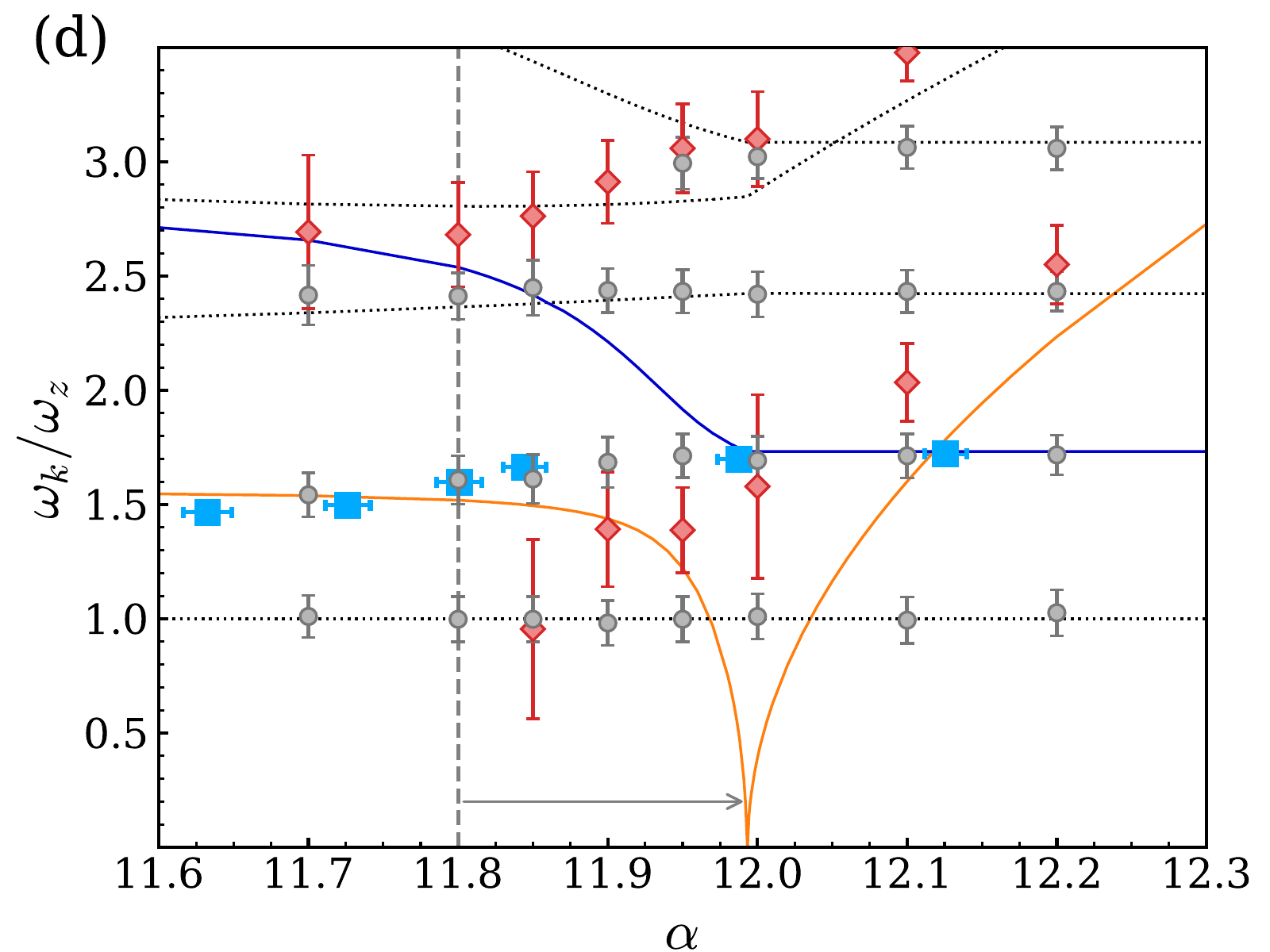}
	\caption{\label{fig:sim_measurement}Spectrum of the ion crystal vibrations as a function of $\alpha$ and for different simulation temperatures: (a) \SI{0.1}{\milli\kelvin} (b) \SI{0.5}{\milli\kelvin}. (c) \SI{2.0}{\milli\kelvin} (d) \SI{3.5}{\milli\kelvin}. Lines represent the spectrum in the absence of thermal excitations: the color code is solid orange (dark blue) for zigzag (axial breathing) mode and dotted black for other modes. The symbols are results obtained from a peak search in the sum of Fourier-transform amplitudes of the axial (gray dots) and transverse (red diamond) trajectories. The error bars indicate the estimated FWHM of the peak. Light blue squares are experimental results for measurement series (B). The gray dashed vertical line and the gray arrow mark the range of $\alpha$ at which we observe one or more jumps between the ground states per zigzag mode oscillation period $T_{zz}$.}
\end{figure*}

We evaluate the simulations for temperatures $T=[0.1, 0.5, 2.0, 3.5]\,\si{\milli\kelvin}$ and for several trapping ratios $\alpha$. Figure \ref{fig:sim_measurement} displays the estimated peak positions. The expected normal mode frequencies at $T=0$ is also shown for comparison.
We observe good agreement between the numerical results and the harmonic spectrum at low temperatures. For higher temperatures, deviations appear near the phase transition for the (1,0) and (0,$N$-1) modes, that qualitatively agree with the experimental measurement.
Meaning that the expected increase in the breathing mode frequency, when transitioning from the 1D phase into the 2D phase, is not observed, and that the zigzag mode frequency in the 2D phase seems to increase, when $\alpha$ approaches $\alpha_c$ from lower values.
The simulations for $T=\SI{3.5}{\milli\kelvin}$ match best to the measurement series (B), which are shown in Fig. \ref{fig:sim_measurement}(d) for comparison. Here, we point out, that the frequency of the zigzag mode remains finite at $\alpha_c$ and increases with the temperature, which can be seen in Fig. \ref{fig:sim_measurement}(b) to (d) in the radial points (red diamonds) around $\alpha\approx12.0$..

In the simulations we do not observe a deviation of the (2,0) mode from the harmonic approximation, as we did in the experiments. This is most likely due to the high excitation power $P_m^{(A)}$ used in series (A), as we described in Section \ref{sec:experiments:res}. The additional increase in mode amplitude leads to non-linear mode coupling on top of temperature effects.

\begin{figure}
	\centering
	\includegraphics[width=8cm]{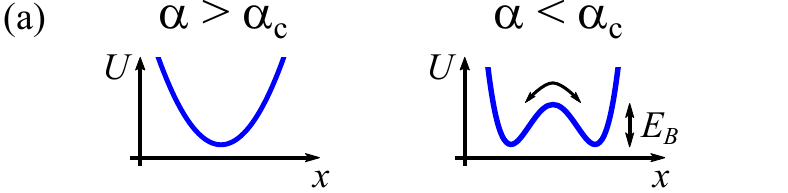}
	\includegraphics[width=0.47\textwidth]{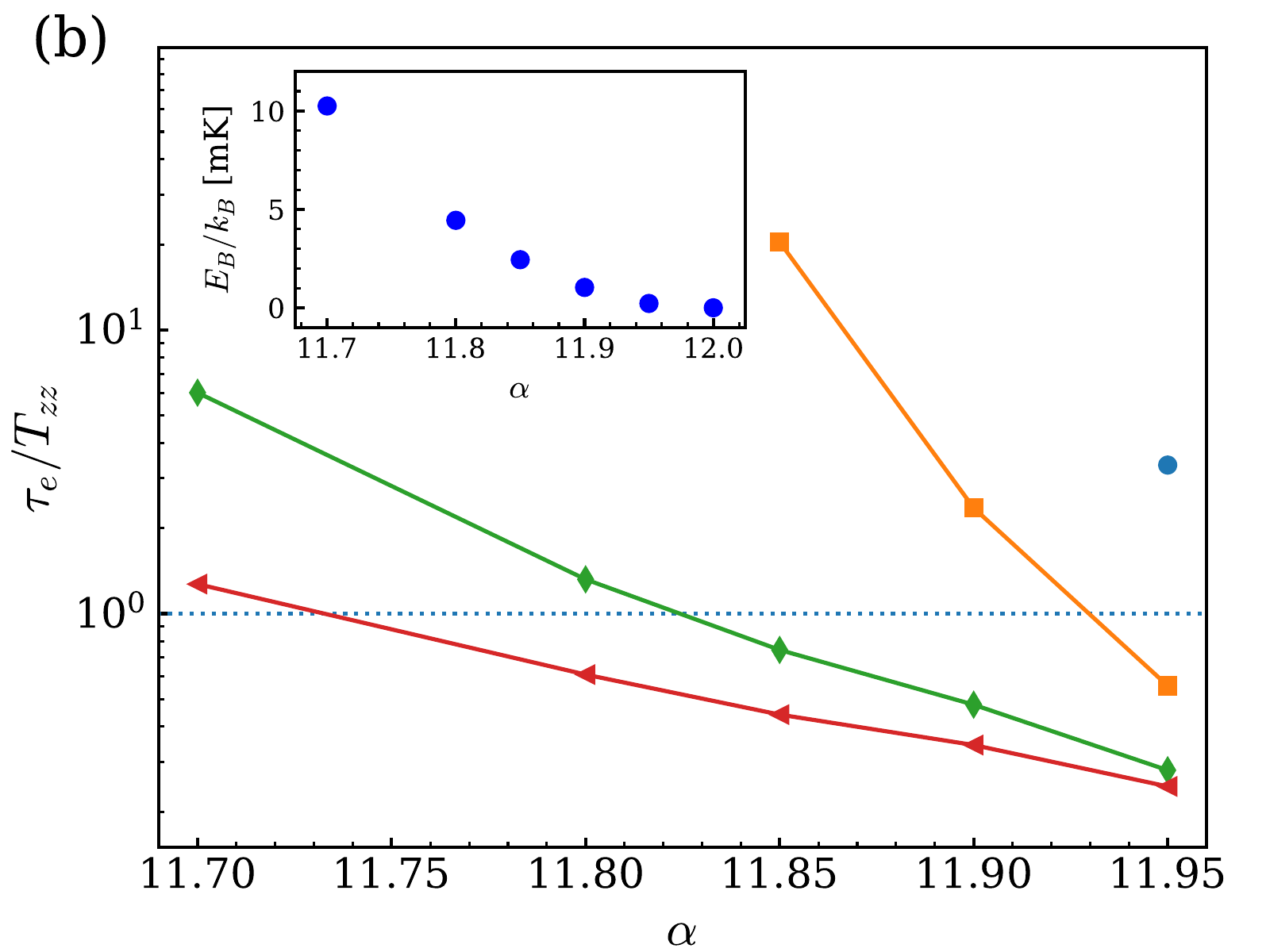}
	\caption{(a) Schematic illustration of the Landau potential $U$ in the 1D ($\alpha>\alpha_c$) and in the 2D symmetry broken phase ($\alpha<\alpha_c$). The coordinate $x$ indicates the transverse displacement of the central ion from the chain axis. Thermal excitations can overcome the barrier between the two degenerate zigzag configuration by inducing collective jumps of the crystal configuration. (b) Average dwelling time $\tau_{e}=1/k_{\rm est}$, Eq. \eqref{eq:switch}, in units of the zigzag mode period $T_{zz}$. In the simulations temperatures $T$ are \SI{0.1}{\milli\kelvin} (blue circles), \SI{0.5}{\milli\kelvin} (orange squares), \SI{2.0}{\milli\kelvin} (green diamonds), \SI{3.5}{\milli\kelvin} (red triangles). The lines are a guide for the eye. For $T=\SI{0.1}{\milli\kelvin}$ and \SI{0.5}{\milli\kelvin} missing points indicate no switches were observed during the simulation time. The dotted horizontal line indicates $\tau_{e} = T_{zz}$. Inset shows the potential barrier $E_B$ for different trapping ratios $\alpha$.}
	\label{fig:switch_time}
\end{figure}

The simulations reveal that in the 2D phase thermal effects give rise to collective jumps of the ions between the two degenerate zigzag configurations. In the Appendix \ref{sec:app:switching}, we discuss how these jumps are observed in the time evolution of the central ion, see Fig. \ref{fig:app:time_evo}. We illustrate the mechanism as thermal switching between the two minima of the Landau free-energy in the symmetry broken phase \cite{Fishman2008}, see Fig. \ref{fig:switch_time} (a). We can estimate the corresponding switching rate $k_{\text{est}}$ by counting the number of sign changes of the transverse coordinate of the central ion $P^x_{N/2}$ over the simulation length $\Delta t$:
\begin{equation}
\label{eq:switch}
	k_{\text{est}} = P^x_{N/2}/\Delta t\,.
\end{equation}
The inverse of this rate is the average dwelling time in one crystal configuration $\tau_e=k_{\text{est}}^{-1}$.
We identify two regimes with the help of $\tau_e$. In the first regime $\tau_e>T_{zz}$, with $T_{zz}$ being the characteristic period of oscillations of the {(1,$N$-1)} mode, which is the slowest oscillation contributing to the movement between the two minima, see Fig. \ref{fig:switch_time} (a).
Here the two crystalline configurations are well defined in the 2D phase and thermal noise gives rise to approximately instantaneous jumps, whose net effect is to broaden the linewidth of the resonance lines.
In the second regime, where $\tau_e\lesssim T_{zz}$, the system switches rapidly between the minima and is on average in the 1D phase. Here, non-linearities of the system are dominant and expected to modify the normal mode spectrum.
Figure \ref{fig:switch_time} (b) displays $\tau_e$ as a function of the aspect ratio and for different temperatures, the horizontal line indicates $\tau_e=T_{zz}$. We have verified that frequency deviations from the harmonic solution are observed when $\tau_e\lesssim T_{zz}$. This is visible, for instance, in Fig. \ref{fig:sim_measurement}, where the vertical dashed lines indicate the the smallest aspect ratios of the simulations at which $\tau_e\lesssim T_{zz}$.

In order to gain insight into these dynamics, we analytically estimate the switching rate using an effective potential $U$, that describes the double well structure. In the thermodynamic limit $U$ becomes the Landau free energy, see section \ref{sec:analytic}. We interpret the switching rate as the rate of thermal activation \cite{Haenggi1990} over the barrier separating the minima.
To determine $U$, we calculate the potential energy along the adiabatic path connecting the two equilibrium configurations \cite{Partner2013}.
Since the two ground states stem from the breaking of the mirror symmetry, we parameterize the path by the transverse crystal size:
\begin{equation*}
	g(\vb{u})=x_{N/2}-x_{N/2 + 1}\,,
\end{equation*}
where $x_{N/2}$ is the transverse position of the ion left of the crystal center for even $N$.
For the calculation we minimize the crystal energy using a Lagrange multiplier with a constraint for the crystal transverse size $g(\vb{u})=X$. For this, we numerically solve the following equation using Newton's method
\begin{equation*}
	\grad{\left(V(\vb{u})+\lambda\left(g(\vb{u})-X\right)\right)}=0\,,
\end{equation*}
where $V$ is given by Eq. \eqref{eq:totalPotentialEnergy}, the gradient is given by $\{\partial/\partial u_1, \dots, \partial/\partial u_{3N},\partial/\partial \lambda\}^T$ and $\lambda$ is the Lagrange multiplier.
Afterwards the total potential energy for this configuration is taken as the energy $U(X)$ of the potential at size $X$. As expected, in the 2D phase it has the shape of a double well with two minima, symmetric about $X=0$. The energy barrier $E_B$, separating the two ground states, is then given as the difference between the potential energy at $X=0$ and the minimum potential energy
\begin{equation*}
	E_B = U(0) - \min(U(X))\,.
\end{equation*}
Sufficiently close to the transition the energy barrier increases with $\left|\alpha-\alpha_c\right|^2$, see inset in Fig. \ref{fig:switch_time} (b), in agreement with the predictions of Ref. \cite{Fishman2008} and with the numerical simulations of the linear to zigzag transition in clusters of metallic beads \cite{Delfau2013}.

The trajectory of the collective coordinate of the crystal that jumps between the minima of the bistable potential results from the interplay of driving, damping, and noise. Quantitatively accounting for the prefactors in the Kramer's escape formula \cite{Kramers1940} is beyond the scope of the current work. Here, we perform an estimate using transition-state theory \cite{Haenggi1990}:
\begin{equation*}
	k_{\text{TST}}\approx\omega_a/(2\pi)\exp{-E_B/(k_BT)},
\end{equation*}
with  $\omega_a=\sqrt{U''(X_{\text{min}})/m}$, where $U''$ is the second derivative with respect to $X$ and $X_{\text{min}}$ is the transverse crystal size in equilibrium.
For $\alpha=11.8$ and $T=\SI{2.0}{\milli\kelvin}$ the transition-state theory predicts a rate about \SI{16000}{\per\second}. Taking into account, that in the simulations the particles can also return to each minima the escape rate in the simulation from one minimum is about \SI{14000}{\per\second}.
In Appendix \ref{sec:app:switching} we compare the rates extracted from the simulations with the predictions of transition-state theory rates over a range of parameters. We find agreement within a factor of 2.

The molecular dynamics simulation validate the thermal fluctuations as the source of the observed frequency deviations. Moreover, they supply a deeper insight into the exact dynamics behind the non-linear mechanism at hand, i.e. the frequent crossing of the potential barrier between the two degenerate ground states of the 2D phase.
\FloatBarrier

\section{\label{sec:analytic}Effective model for the modes dynamics}

In this Section we use a simplified model in order to determine the temperature dependence of the mode spectrum close to the linear to zigzag instability. For this purpose we use a complementary approach to the one based on dwelling times and consider the normal mode expansion around the linear chain for aspect ratios at which the linear chain is mechanically unstable. We then evaluate the average effect of the higher modes on the lowest part of the spectrum using a time-scale separation ansatz and determine the resulting spectrum as a function of the temperature. The resulting normal mode spectrum agrees with the numerical results close to the transition point, as we discuss below and summarize in Fig. \ref{fig:sim_analytical_lin}.

\subsubsection{Normal modes at the instability}
We first review the normal modes of the linear chain and the equations for the structural instability in the absence of damping and noise. Close to the linear to zigzag instability we expand the total potential energy $V$ of Eq. \eqref{eq:totalPotentialEnergy}   to fourth order around the equilibrium positions of the linear chain,
\begin{multline}\label{eq:4thOrderPotential}
	V_4'=\frac{1}{2!}\sum_{ij=1}^{3N} K^{\prime}_{ij} q_i q_j +
	\frac{1}{3!}\sum_{ijk=1}^{3N} L^{\prime}_{ijk} q_i q_j q_k + \\
	\frac{1}{4!}\sum_{ijkl=1}^{3N} M^{\prime}_{ijkl} q_i q_j q_k q_l\,,
\end{multline}
where $q_i$ are the displacements around the equilibrium positions $u_i(0)$, $q_i=u_i-u_i(0)$, and the tensors $L^{\prime}$ and $M^{\prime}$ are given by the expressions:
\begin{align}
	L^{\prime}_{ijk}=&\eval{\frac{\partial^3 V}{\partial u_i \partial u_j \partial u_k}}_{\vb{u}(0)}\\
	M^{\prime}_{ijkl}=&\eval{\frac{\partial^4 V}{\partial u_i \partial u_j \partial u_k \partial u_l}}_{\vb{u}(0)}\,,
\end{align}
and $K^{\prime}$ is given by Eq. \eqref{eq:HessianMatrix}.
Note that $V_4$ approximates the total potential $V$, Eq. \eqref{eq:totalPotentialEnergy}, in the limit in which the displacements around the equilibrium positions are much smaller than the interparticle distance at equilibrium. 

As introduced in section \ref{sec:LinZZ}$, \Theta_j$ denote the normal modes of the linear chain, which diagonalize the matrix $K'$ and have eigenvalues $m\omega_j^2$. The linear chain is stable provided that all eigenvalues are positive. In this regime the $\omega_j$ are real and correspond to the normal mode frequencies. The condition $\min_j\omega_j=0$ identifies the classical transition point of the linear to zigzag instability. Potential \eqref{eq:4thOrderPotential} is cast in terms of the normal mode by means of the dynamical matrix $\lambda_{ij}$ such that $q_i=\sum_j\lambda_{ij}\Theta_j$ and takes the form
\begin{multline}\label{eq:potentialNormalModes}
	V_4=\frac{1}{2!}\sum_{i=1}^{3N} m\omega_i^2 \Theta_i^2 +
	\frac{1}{3!}\sum_{ijk=1}^{3N} L_{ijk} \Theta_i\Theta_j\Theta_k + \\
	\frac{1}{4!}\sum_{ijkl=1}^{3N} M_{ijkl}\Theta_i\Theta_j\Theta_k\Theta_l
\end{multline}
where now the tensors $L$ and $M$ are related to the tensors $L'$ and $M'$ by the relations:
\begin{align}\label{eq:convertToModePicture}
	L_{ijk}&=\sum_{mns=1}^{3N}L_{mns}^{\prime}\lambda_{mi}\lambda_{nj}\lambda_{sk}\\
	M_{ijkl}&=\sum_{mnst=1}^{3N}M_{mnst}^{\prime}\lambda_{mi}\lambda_{nj}\lambda_{sk}\lambda_{tl}.
\end{align}
The total Lagrangian for the normal modes takes the form $L=\frac{1}{2}m\sum_{i=1}^{3N}\dot{\Theta}_i^2-V_4$.

In an appropriately defined thermodynamic limit, for which the critical aspect ratio converges to a finite value as $N\to\infty$, the linear-zigzag transition can be cast in terms of the Landau potential:
\begin{equation}
	\label{Landau}
	U_{LG}=\mathcal V \Theta_{zz}^2+A\Theta_{zz}^4\,,
\end{equation}
where $\Theta_{zz}$ is the amplitude of the zigzag mode in the linear chain, $A>0$ and $\mathcal V\propto (\alpha^2-\alpha_c^2)$.
This potential is determined from potential $V_4$ in lowest order in a gradient expansion \cite{Fishman2008}.
The one dimensional model strictly exhibits a phase transition at zero temperature, where a quantum description becomes appropriate \cite{Retzker2008,Shimshoni2011,Silvi2014,Podolsky2014}. In what follows, instead, we consider a finite system and do not scale the physical parameters with $N$.

\subsubsection{Thermal effects}
We now discuss the low frequency spectrum of the linear chain across the linear to zigzag instability and in the presence of laser cooling. We  consider a finite chain and, starting from the Fokker-Planck equation \cite{Morigi2001}, we model the dynamics of laser Doppler cooling in terms of Langevin equations. We denote the damping (cooling) rates of the normal modes by $\gamma_i$ and write the corresponding Langevin equations as \cite{DeChiara2010}
\begin{equation}\label{Lars}
	\ddot\Theta_i=-\frac{1}{m}\frac{\partial V_4}{\partial \Theta_i}-\gamma_i\Theta_i+\Xi_i(t)\,,
\end{equation}
where $\Xi_i(t)$ is the Langevin force for the normal mode $\Theta_i$, with $\langle \Xi_i(t)\rangle=0$, $\langle \Xi_i(t)\Xi_j(t')\rangle=2\gamma_i(k_BT/m)\delta_{ij}\delta(t-t')$ and we neglect here mode-mode correlations due to the dissipative dynamics.

For finite chains and in the 2D phase the lowest frequency mode is a superposition of  the zigzag mode and of the axial breathing mode of the linear chain as seen in Fig. \ref{fig:Modevec}c). The gap between the soft mode and all other normal modes remains finite. Thus, whenever thermal excitations and the line broadening are smaller than the gap, normal-mode spectroscopy of the chain shall provide in first approximation the mode spectrum obtained by diagonalizing the quadratic term of potential \eqref{eq:totalPotentialEnergy} about the stable equilibrium positions. 

\begin{figure*}[!bth]
	\centering
	\includegraphics[width=0.47\textwidth]{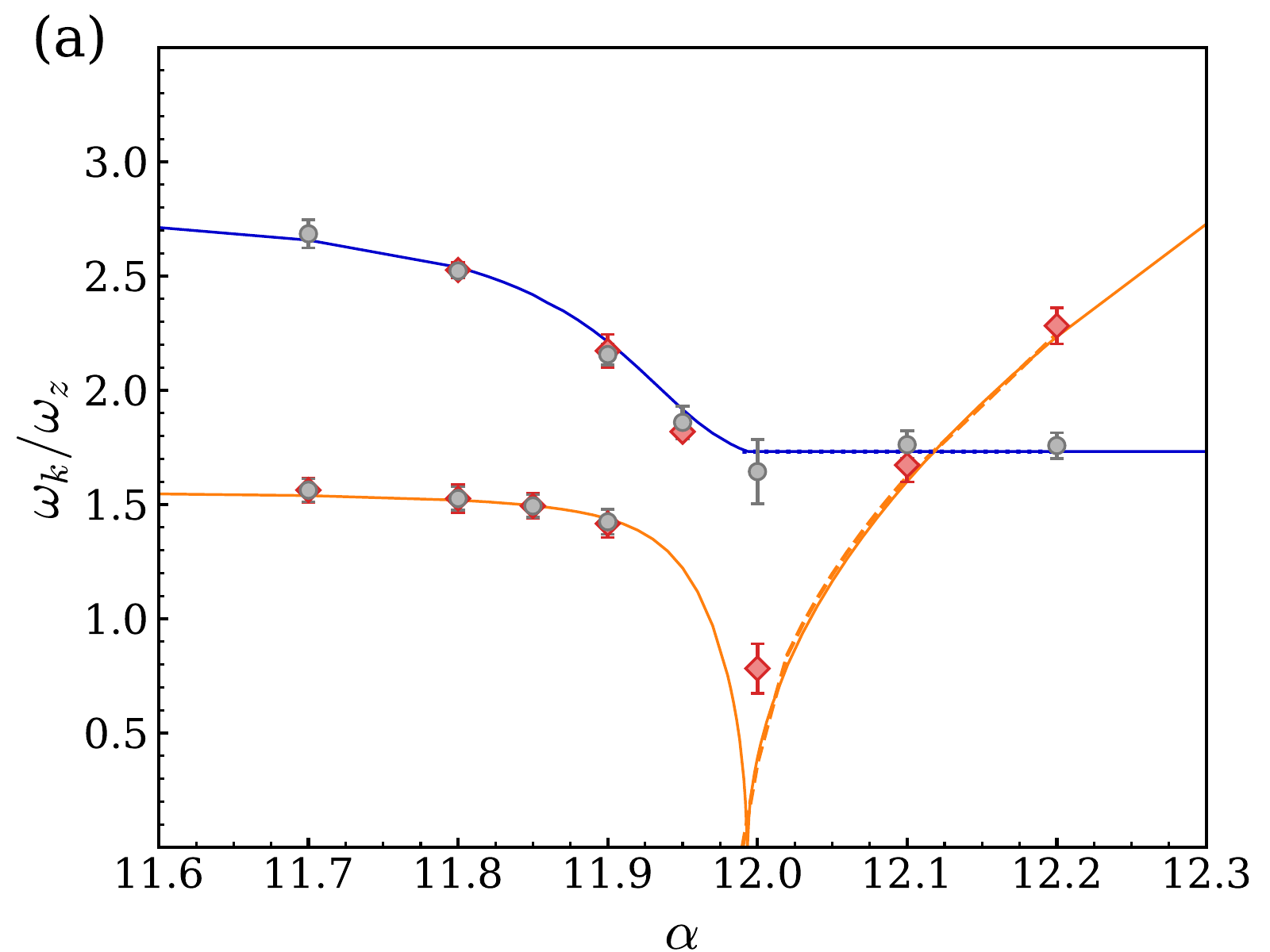}
	\includegraphics[width=0.47\textwidth]{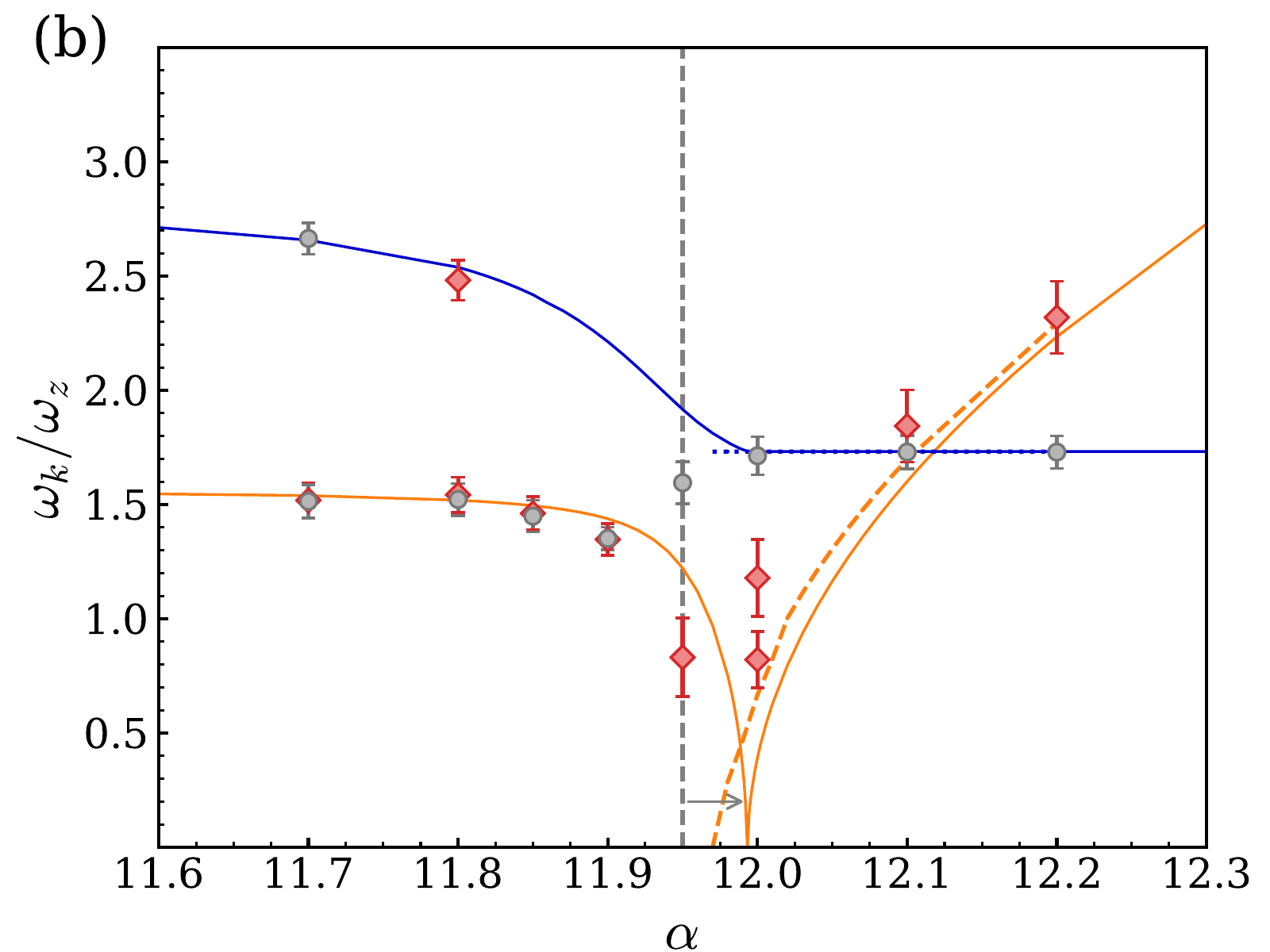}
	\vskip\baselineskip
	\includegraphics[width=0.47\textwidth]{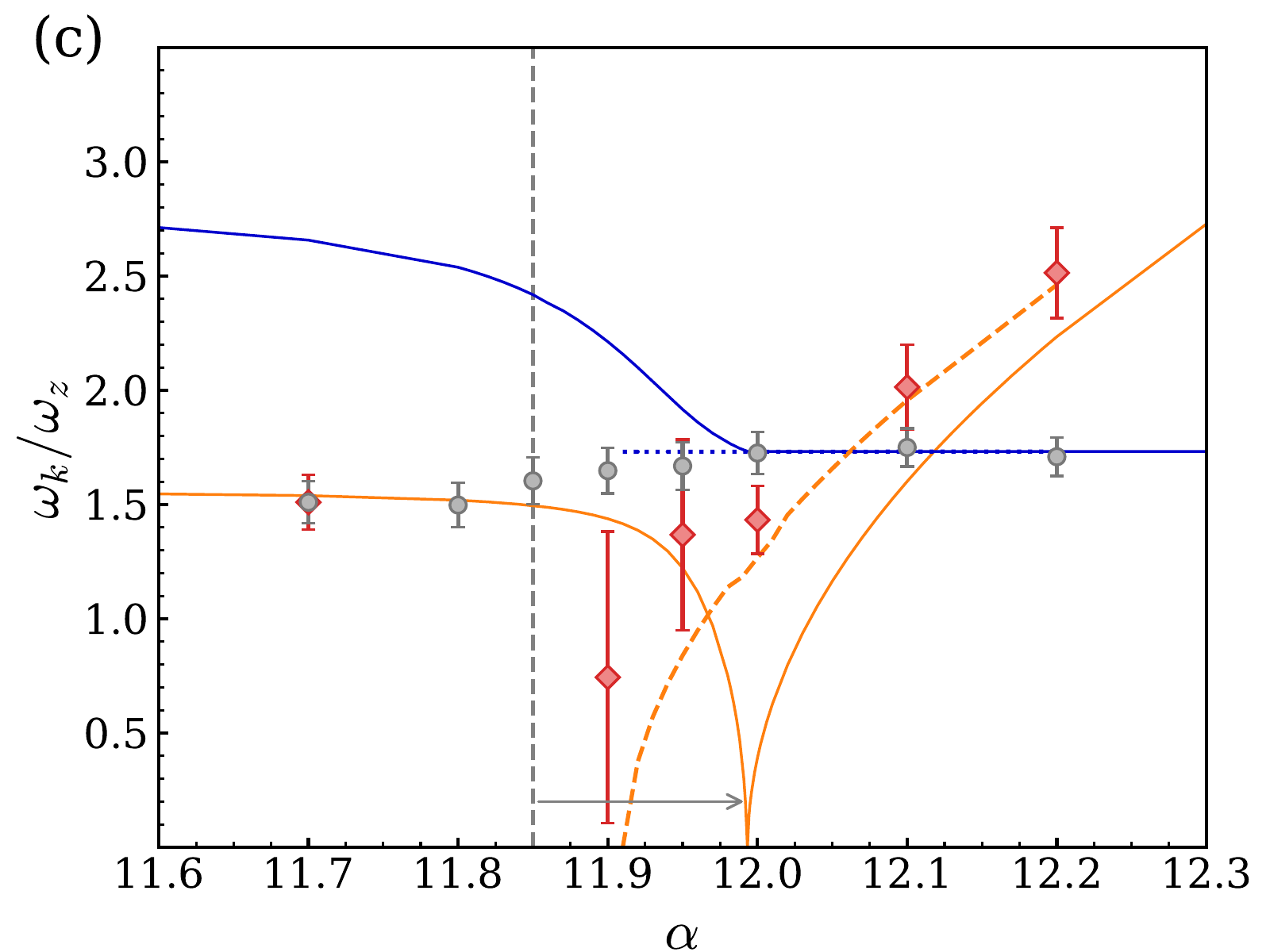}
	\includegraphics[width=0.47\textwidth]{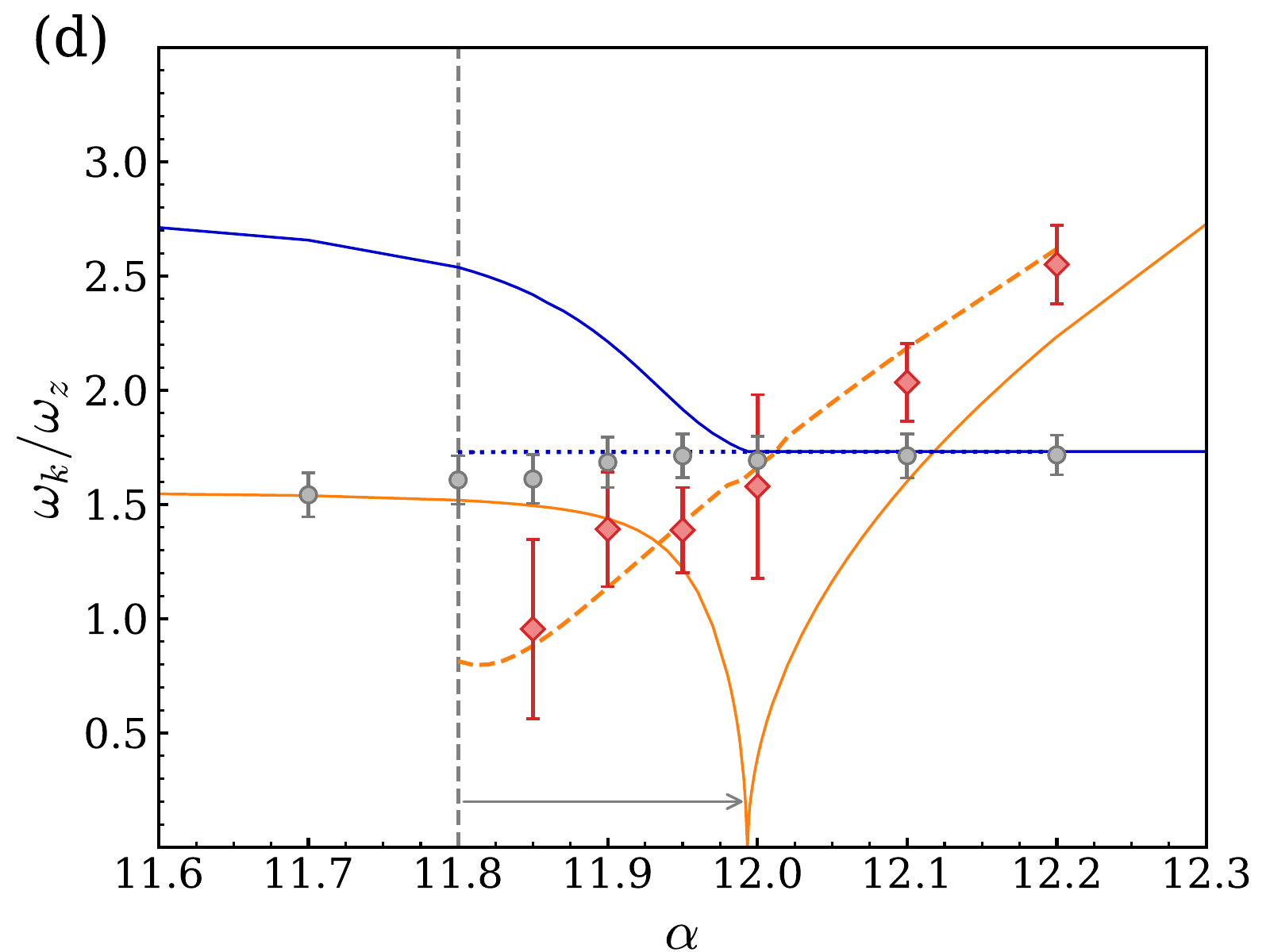}
	\caption{\label{fig:sim_analytical_lin} The zigzag and breathing mode frequencies as a function of the aspect ratio $\alpha$. The parameters and legend are the same Fig. \ref{fig:sim_measurement}. Moreover, the dashed and dotted lines correspond to the predictions of the analytical model: $\tilde{\omega}_2$ (dark blue, dotted) and $\tilde{\omega}_1$ (orange, dashed). The analytical solutions are shown until $\tilde{\omega}_1$ becomes imaginary or until less than one switch per zigzag mode oscillation period $T_{zz}$ is observed, i.e. for values of $\alpha$ below the range marked by the gray dashed vertical lines and the gray arrows.}
\end{figure*}

We now determine the effects of thermal excitation on the lowest energy spectrum by considering the equa\-tions of the lowest energy mode, here (0,$N$-1), which we label by $\Theta_1$, and the mode which is closest in frequency and to which it couples. This mode is labeled by $\Theta_2$ and is according to our notation  (1,0).
We then make the simplifying assumption that  $\omega_1,\omega_2\ll\omega_\ell$, where $\omega_{\ell}$ are here the frequencies of the modes $\Theta_\ell$ to which $\Theta_1$ and $\Theta_2$ appreciably couple through the anharmonicities. In this regime we can identify the time scale $\delta t$ for which $\omega_1\delta t,\omega_2\delta t\ll 1$ and $\omega_\ell\delta t\gg 1$. Moreover, we assume that the modes $\Theta_\ell$ are at thermal equilibrium. We now perform the time average of Eq. \eqref{Lars} over the grid with step $\delta t$. For convenience we introduce the notation
\begin{equation}\label{Lars:1}
	\frac{1}{\delta t}\int_t^{t+\delta t}{\rm d}\tau\,f(\tau)\equiv \langle f(t)\rangle_{\delta t}\,,
\end{equation}
where $f(t)$ is a function of time. Since $\omega_1\delta t,\omega_2\delta t\ll 1$, then $\langle \Theta_{1,2}(t)\rangle_{\delta t}\approx \Theta_{1,2}(t)$. Moreover, to provide an example, the contribution of the fourth order term of Eq. \eqref{eq:potentialNormalModes}, which de denote by $W_4$, takes the form
$$\left\langle \frac{\partial W_4}{\partial \Theta_{1}}\right\rangle_{\delta t}\approx \sum_\ell \left(\frac{1}{2}M_{11\ell\ell}\langle\Theta_\ell^2\rangle_{\delta t}\Theta_{1}+\frac{1}{2}M_{12\ell\ell}\langle\Theta_\ell^2\rangle_{\delta t}\Theta_2\right)\,,$$
where the equation for $\Theta_2$ is found by replacing $1\to 2$ and we used that for  $\ell\neq 1,2$ the eigenmodes are at thermal equilibrium, thus $\langle\Theta_\ell\rangle_{\delta t}=0$ and $\langle\Theta_\ell\Theta_{\ell'}\rangle_{\delta t}=\delta_{\ell,\ell'}\langle\Theta_\ell^2\rangle_{\delta t}$. Finally, assuming ergodicity we obtain $\langle\Theta_\ell^2\rangle_{\delta t}=k_BT/(m\omega_\ell^2)$ from the classical equipartition theorem.
This procedure leads to the two coupled equations:
\begin{eqnarray}
	\label{Langevin:red}
	&&\ddot\Theta_1=-\tilde\omega_1^2\Theta_1-\frac{1}{2}\nu_{12}^2\Theta_2+\eta_1-\gamma_1\Theta_1+\Xi_1\\
	\label{Langevin:red2}
	&&\ddot\Theta_2=-\tilde\omega_2^2\Theta_2-\frac{1}{2}\nu_{12}^2\Theta_1+\eta_2-\gamma_2\Theta_2+\Xi_2\,,
\end{eqnarray}
where $\tilde\omega_i$, $\nu_{12}$ and $\eta_i$ are explicitly dependent on the temperature. In particular, the frequency squared $\tilde{\omega}_i^2$ now reads
\begin{equation}\label{eq:freq_shift}
\tilde\omega_i(T)^2=\omega_i^2+\nu_i(T)^2 = \omega_i^2+\nu_{\text{eff},i}^2 T\,,
\end{equation}
and it contains a shift proportional to the temperature with proportionality constant
\begin{align}\label{eq:nu_i_eff}
	\nu_{\text{eff},i}^2=\frac{1}{2m}\sum_{k\ne1,2}M_{iikk}\frac{k_B}{m\omega_k^2}\,.
\end{align}
The second and third terms on the right-hand side of Eqs. \eqref{Langevin:red}-\eqref{Langevin:red2} describe an effective coupling between the two modes and  a mean displacement force, respectively, with
\begin{align}\label{eq:nu_12_xi_eff}
	\nu_{12}(T)^2&=\frac{1}{m}\sum_{k\ne1,2}M_{12kk}\frac{k_BT}{m\omega_k^2}\equiv\nu_{\text{eff},12}^2\cdot T\\
	\label{eq:xi_eff}
	\eta_i&=-\frac{1}{2}\sum_{k\ne1,2}L_{ikk}\frac{k_BT}{m\omega_k^2}\equiv-\eta_{\text{eff},i}\cdot T\,.
\end{align}
The effective, temperature-independent constants $\nu_{\text{eff},i}^2$, $\nu_{\text{eff},12}^2$ and $\eta_{\text{eff},i}$ for the two modes $i=1,2$ are determined by carrying out the summation in Eqs. \eqref{eq:nu_i_eff}-\eqref{eq:xi_eff} over all other modes $k$. We remark that the shifts depend on $\alpha$ through the coefficients of the expansion of $V_4$.

Equations \eqref{Langevin:red} and \eqref{Langevin:red2} describe mode mixing and frequency shifts induced by the thermal excitation of the chain.
Within this classical model these terms are directly proportional to the temperature. We can now determine the resulting normal mode frequencies. For this purpose we note that the term $\nu_{12}=0$ for the expansion about the linear chain equilibrium positions, see Table \ref{tab:app:eff_consts_lin} of Appendix \ref{sec:app:eff_constants}.
This is a consequence of the fact the breathing mode is an exact eigenmode of the linear chain \cite{Morigi2004}.
In the underdamped limit, corresponding to $\gamma_i\ll\omega_i$, the characteristic frequencies are now given by Eq. \eqref{eq:freq_shift} and Eq. \eqref{eq:nu_i_eff}.
Figure \ref{fig:sim_analytical_lin} displays the frequencies $\tilde{\omega}_{i}$ as a function of the aspect ratio $\alpha$ and for four increasing values of the temperature, ranging between \SI{0.1}{\milli\kelvin} and \SI{3.5}{\milli\kelvin}.
For comparison, the results of the numerical simulation of Eq. \eqref{eq:Langevin} are reported, which agreed well with the experimental measurements presented in section \ref{sec:experiments} for $T=\SI{3.5}{\milli\kelvin}$.
The prediction of the analytical model and the result of the numerical simulation agree for aspect ratios close to the transition point $\alpha_c$: this is the regime where our model is plausible since the truncation of the Taylor expansion is justified. We note that, even though these analytical arguments have been applied to a finite chain, the considerations of our theoretical model are also valid in the thermodynamic limit and show that at finite temperatures the coupling with the high frequency modes significantly modifies the properties at the instability. Loosely speaking, the thermal effects stabilize the linear chain also for aspect ratios beyond the critical point.
This behavior might be interpreted as a shift of the transition point \cite{Gong2010}, in the sense that a measurement of the mean transverse displacement will give zero in the regime where the ions dynamics consists of thermally activated jumps between the two zigzag configurations. Thus, a linear chain will be effectively detected for aspect ratios $\alpha$ below but close to $\alpha_c$. Nevertheless, in the classical regime this is the manifestation of a non-universal crossover dynamics.

\FloatBarrier

\section{Conclusion}
\label{sec:conclusions}

In this work we investigated experimentally and theoretically the effect of thermal noise on the low-frequency spectrum of an ion chain near the symmetry-breaking linear to zigzag transition.

In the experiment we employed resonant light force modulation with an amplitude modulated laser beam to excite collective oscillations in a crystal.
The method is simple to implement and can also be used to measure trapping frequencies, replacing established excitation methods, such as modulation of the trapping potentials \cite{Ibaraki2011}.
This allows for stronger filters in the RF and DC electronics of the Paul trap \cite{Pyka2014,Keller2019}, reducing the heating by electrical noise of the trapped ion crystals.
While we used an excitation beam profile encompassing multiple ions, a more focused beam or a spatially engineered beam profile, e.g. generated by an spatial light modulator \cite{Bergamini2004,Zupancic2016}, would allow for arbitrary mode excitations.

The experimental measurements did not show the softening of the zigzag mode that is predicted at the structural phase transition. Also the frequency of the breathing mode was nearly constant and independent of $\alpha$ when sweeping into the 2D phase, instead of increasing as expected in the absence of thermal noise.

With the help of molecular dynamics simulations we could reproduce the experimental observations within the uncertainties, thus confirming that this behavior is majorly due to thermal excitations. In particular,
inspection of the trajectories show that finite temperature effects induce collective jumps of the ions between the two degenerate zigzag configurations. This microscopic picture is at the basis of the expected crossover behavior at finite temperatures.

We developed a simple analytical model that builds on these findings and predicts the experimentally observed frequency spectrum. This model shows that the temperature dependent shift of the zigzag mode at the transition point is due to anharmonic coupling with high frequency modes, which act as an effective phonon environment. Separation of timescales between the low frequency soft mode and the higher frequency modes allows taking the averaged higher frequency modes as an effective potential that influences the soft mode. 
Note, that the thermally excited phonon environment in our model could be replaced by non-thermal excitations. Single quanta excitations with coherent interaction in third order have been investigated in \cite{Marquet2003}. In future theoretical works their method could be extended in order to describe the effects of finite temperatures.
While we do not include the effects of micromotion induced by the RF of the trap in our theoretical models, it could be treated analytically following \cite{Kaufmann2012}.
From our model it also follows, that the observation of a low-frequency zigzag mode near the linear-zigzag transition at Doppler temperature of \Yb\, is unlikely. However, it might be observed in a crystal with $T\approx$\SI{100}{\micro\kelvin}. Therefore, in future experiments methods of (near) ground state cooling, that can cool several modes in an ion Coulomb crystal, such as Sisyphus cooling \cite{Wineland1992,Ejtemaee2017,Joshi2020} or electromagnetically-induced-transparency cooling \cite{Morigi2000,Morigi2003,Lechner2016,Scharnhorst2018,Jordan2019}, need to be considered.

Our results suggest that a similar model can be developed in order to describe the  experimental measurements of the Aubry-type transition in ion Coulomb crystals. Here the soft mode of the pinning to sliding transition exhibited a finite frequency at the critical point, when a finite temperature allowed the system to switch between different minima of the Peierls-Nabarro potential \cite{Kiethe2017,Kiethe2018}.

Our work is relevant for experiments operating close to phase transitions in ion chains such as studies of energy transport \cite{Timm2020} or quantum information using the gapped topological defect mode \cite{Landa2010,Landa2014}.
According to the results presented the cooling of high-frequency modes is crucial to avoid the heating of the soft mode due to higher-order coupling, showing the experimental complexity of these plans. Similar challenges were recently discussed for laser cooling a 2D planar crystal confined in a Penning trap \cite{Shankar2020}.

\begin{acknowledgments}
We gratefully thank J. Keller for fruitful dis\-cus\-sions. This project has been funded
by the Deutsche Forschungsgemeinschaft (DFG, German Research Foundation) through
Grant No. CRC 1227 (DQ-mat, project A07) and under Germany’s Excellence Strategy –EXC-2123 QuantumFrontiers –390837967.
This project has received funding from the European Metrology Programme for Innovation and Research (EMPIR) co-financed by the Participating States and from the European Unions Horizon 2020
research and innovation programme. It was funded under Project No. 17FUN07 CC4C.
G. M. acknowledges the support by the Deutsche Forschungsgemeinschaft (DFG, German Research Foundation) through the grant No. CRC TRR 306 QuCoLiMa ("Quantum Cooperativity of Light and Matter'') and by the German Ministry of Education and Research (BMBF) via the QuantERA project NAQUAS. Project NAQUAS has received funding from the QuantERA ERA-NET Cofund in Quantum Technologies implemented within the European Union's Horizon2020 program.
\end{acknowledgments}

\appendix

\section{\label{sec:app:eff_constants}Effective analytical constants}
\begin{table}[hbt]
	\begin{tabular}{c|c|c|c|c|c}
		$\alpha$ & $\nu_{\text{eff}, 1}^2$ & $\nu_{\text{eff}, 2}^2$ & $\nu_{\text{eff}, 12}^2$ & $\eta_{\text{eff}, 1}$ & $\eta_{\text{eff}, 2}$ \\
		\hline
		11.70 & 3767.3904 & -6.4896 & -0.0039 & 0.0003 & -26.1177 \\
		\hline
		11.80 & 1505.2251 & -2.7630 & -0.0021 & 0.0001 & -11.1188 \\
		\hline
		11.85 & 1200.0101 & -2.2324 & 0.0020 & -0.0001 & -8.9835 \\
		\hline
		11.90 & 1007.8599 & -1.8871 & 0.0021 & -0.0001 & -7.5937 \\
		\hline
		11.91 & 977.3964 & -1.8312 & -0.0021 & 0.0001 & -7.3688 \\
		\hline
		11.92 & 948.9239 & -1.7786 & 0.0022 & -0.0001 & -7.1572 \\
		\hline
		11.93 & 922.2375 & -1.7290 & -0.0022 & 0.0001 & -6.9577 \\
		\hline
		11.94 & 897.1609 & -1.6821 & 0.0023 & -0.0001 & -6.7689 \\
		\hline
		11.95 & 873.5402 & -1.6377 & 0.0023 & -0.0001 & -6.5900 \\
		\hline
		11.96 & 851.2413 & -1.5955 & -0.0024 & 0.0001 & -6.4201 \\
		\hline
		11.97 & 830.1467 & -1.5553 & -0.0025 & 0.0001 & -6.2584 \\
		\hline
		11.98 & 810.1524 & -1.5169 & -0.0027 & 0.0001 & -6.1041 \\
		\hline
		11.99 & 791.1669 & -1.4803 & -0.0028 & 0.0001 & -5.9565 \\
		\hline
		12.00 & 773.1083 & -1.4453 & -0.0030 & -0.0001 & 5.8158 \\
		\hline
		12.05 & 694.4609 & -1.2899 & 0.0000 & 0.0000 & 5.1913 \\
		\hline
		12.10 & 630.6453 & -1.1606 & 0.0000 & 0.0000 & 4.6702 \\
		\hline
		12.20 & 532.5320 & -0.9539 & 0.0000 & 0.0000 & 3.8383 \\
	\end{tabular}
	\caption{Effective higher order constants for a $N=30$ ion Coulomb crystal near the linear to zigzag phase transition for the zigzag mode and breathing mode. Crystal expanded around a linear chain. $\nu^2_{\text{eff},i}$ and $\nu^2_{\text{eff}, 12}$ are given in units of the squared axial frequency $\omega_{z}^2$. The constants $\eta_{\text{eff}, i}$ are given in units of $m\omega_z^2l_c$, where $l_c=\left[e^2/(4\pi\epsilon_0m\omega_z^2)\right]^{1/3}$ is the length constant of a trapped ion Coulomb crystal.}\label{tab:app:eff_consts_lin}
\end{table}
\FloatBarrier

\section{\label{sec:app:switching}Switching rates: Simulation and Transition-state Theory}
\begin{figure}
	\centering
	\includegraphics[width=0.45\textwidth]{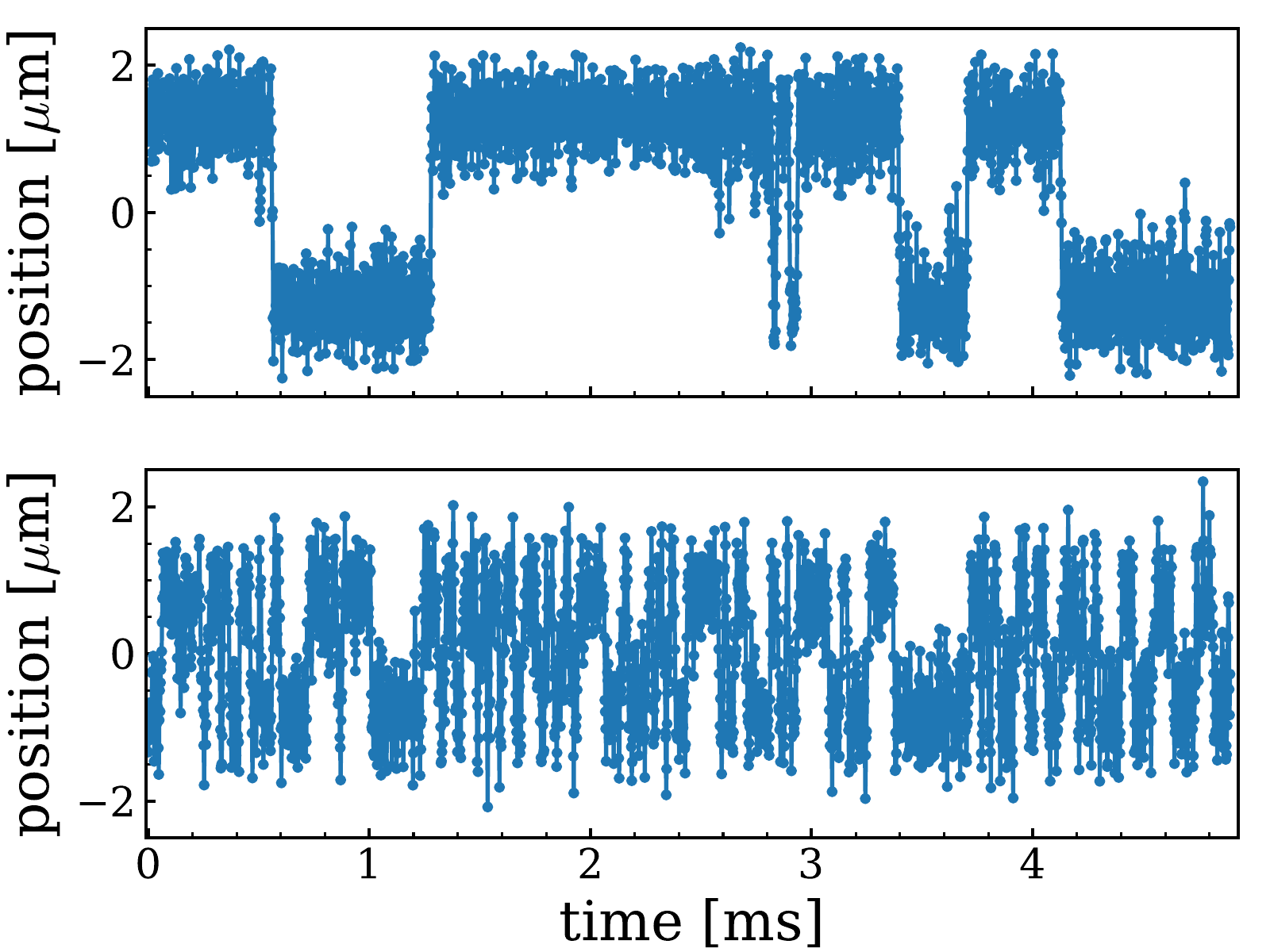}
	\caption{Time evolution of transverse coordinate of ion 15. The top row is for $\alpha=11.7$ and the bottom row for $\alpha=11.85$. Closer to the transition at $\alpha_c\approx 12.0$ the ion changes more often between the two ground state configurations, about which the crystal oscillates.}
	\label{fig:app:time_evo}
\end{figure}

\begin{figure}
	\centering
	\includegraphics[width=0.45\textwidth]{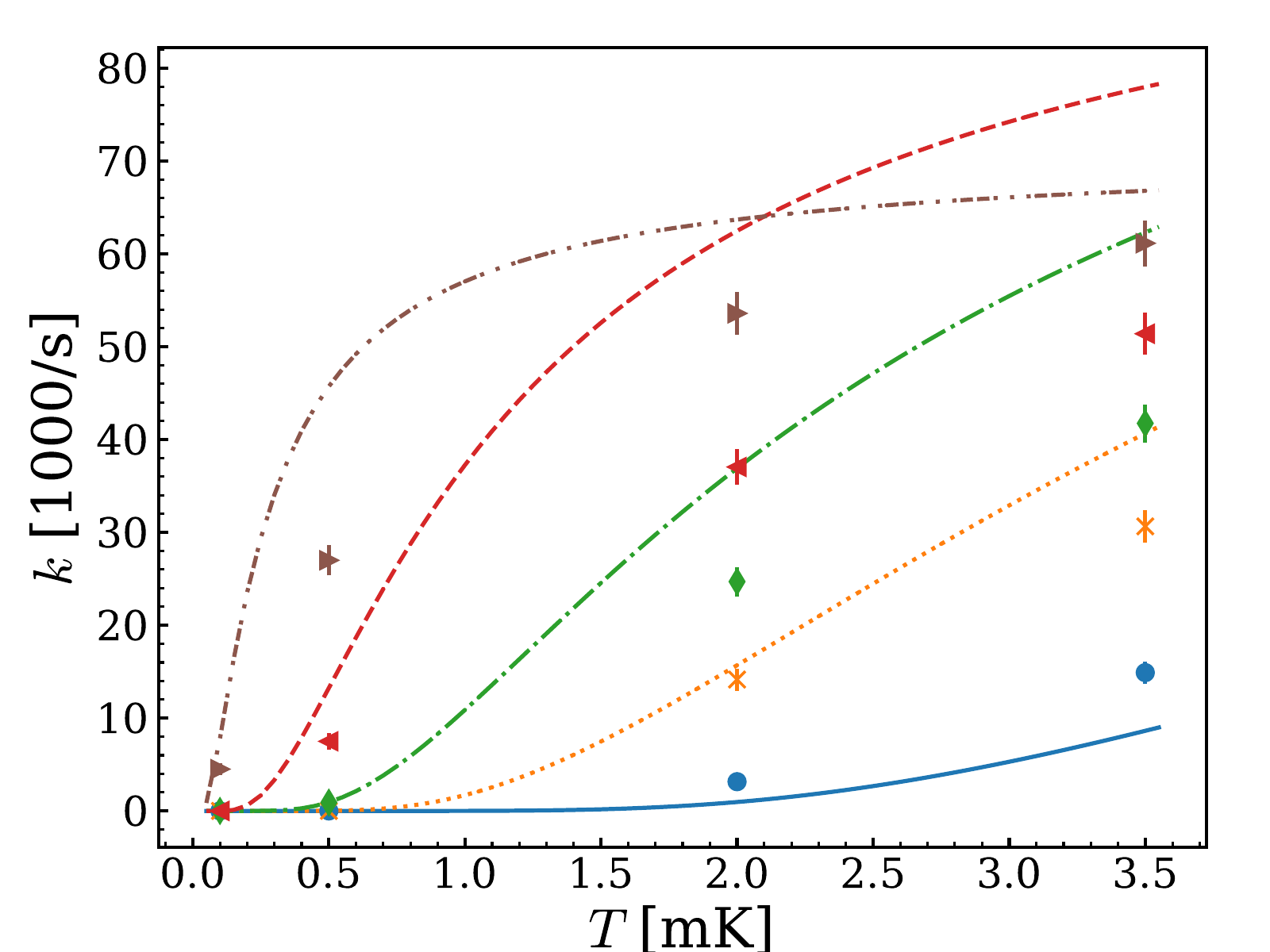}
	\caption{Escape rate from transition-state theory $k_{\text{TST}}$ (lines) and estimate rates from simulations $k_{\text{est}}/2$ (symbols). Colors and markers are: $\alpha=11.7$: blue circles \& solid line, $\alpha=11.8$: orange crosses \& dotted line, $\alpha=11.85$: green diamonds \& dash-dotted line, $\alpha=11.9$: red left triangles \& dashed line, $\alpha=11.95$: brown right triangles \& double-dash-dotted line.}
	\label{fig:app:escape_rate}
\end{figure}

In Fig. \ref{fig:app:time_evo} we show the time evolution of the transverse coordinate of the 15th ion for two different $\alpha$ and for $T=\SI{2.0}{\milli\kelvin}$, to illustrate the switching of the crystal between the two ground state configurations.
In Fig. \ref{fig:app:escape_rate} we show the comparison between the escape rate obtained by transition-state theory from the double well potentials calculated in Section \ref{sec:numerics:res} and the rates estimated from the simulation results. The latter have been corrected by the factor two for the comparison to account for the possibility to come back to a potential minimum.

\section{\label{sec:app:power_dep}Power dependency of experimental signal}
\begin{figure}[htb]
	\includegraphics[width=0.45\textwidth]{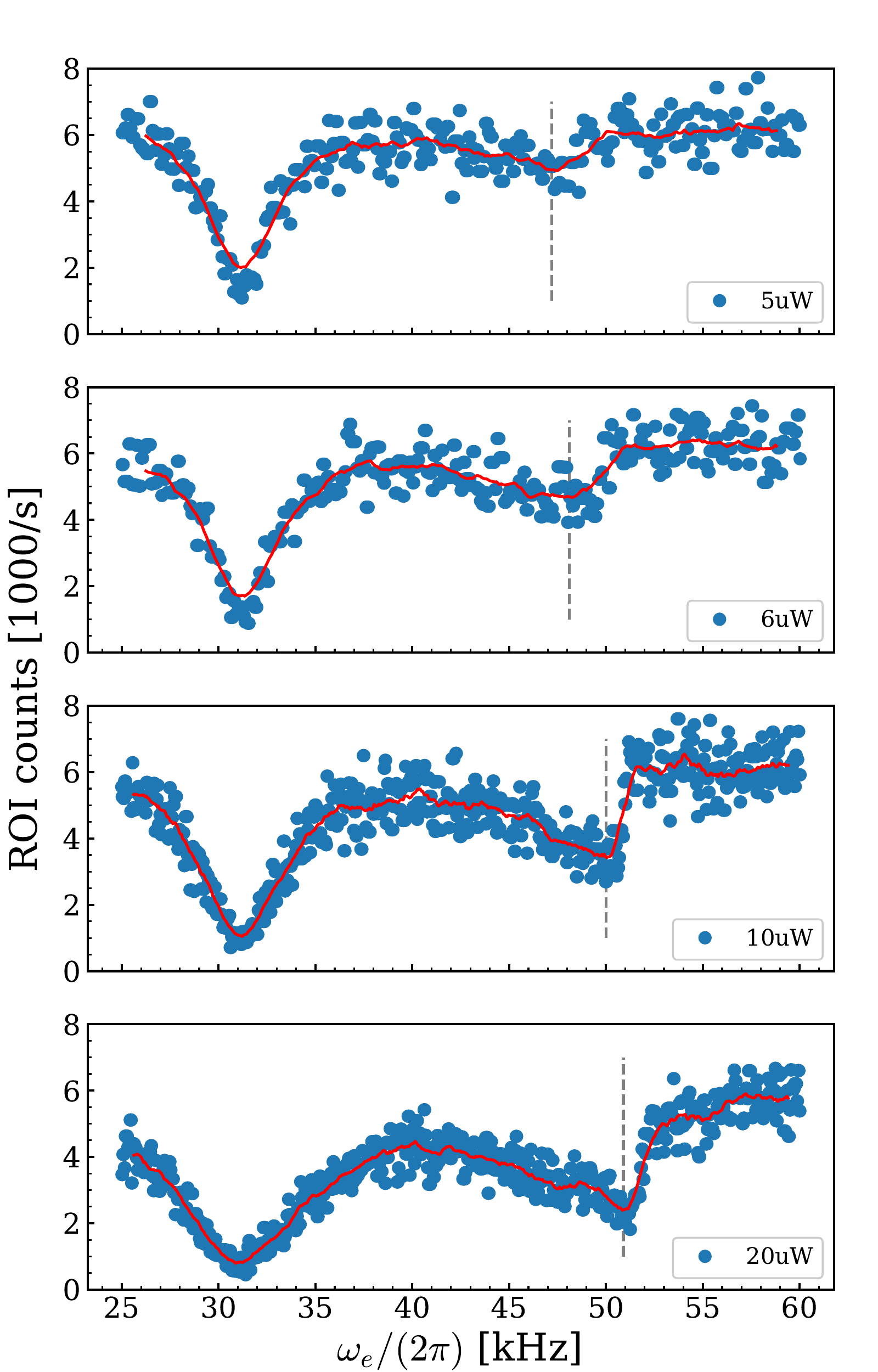}
	\caption{\label{fig:app:powersweeps}Measured ROI fluorescence (blue dots) against modulation frequency $\omega_e$ at $\alpha\approx 11.72$. The red line is a running mean over 50 points, acting as a guide to the eye. The dashed gray vertical lines indicate estimated maximum excitation for the breathing mode.}
\end{figure}
Driving the intrinsically non-linear Coulomb system can lead to additional frequency shifts due to further increased amplitudes.
The excitation method employed in this paper can lead to such shifts, depending on the laser power of the modulated laser $P_0$.

In Fig. \ref{fig:app:powersweeps} we show the power dependence of recorded resonance features of measurement series (B), when sweeping over the axial center of mass and the breathing mode at $\alpha\approx11.72$.
It can be clearly seen, that the position of maximum excitation shifts, when increasing the driving force. Additionally, the resonances become increasingly asymmetric with higher forces. The center of mass mode resonance around \SI{31}{\kilo\hertz} increases in width with higher forces.

Measurements presented in Section \ref{sec:experiments:res}, were carried out with $P_m=\SI{20}{\micro\watt}$ for series (A) and with $P_m=\SI{6}{\micro\watt}$ for series (B).
\FloatBarrier


\begin{thebibliography}{80}%
	\makeatletter
	\providecommand \@ifxundefined [1]{%
		\@ifx{#1\undefined}
	}%
	\providecommand \@ifnum [1]{%
		\ifnum #1\expandafter \@firstoftwo
		\else \expandafter \@secondoftwo
		\fi
	}%
	\providecommand \@ifx [1]{%
		\ifx #1\expandafter \@firstoftwo
		\else \expandafter \@secondoftwo
		\fi
	}%
	\providecommand \natexlab [1]{#1}%
	\providecommand \enquote  [1]{``#1''}%
	\providecommand \bibnamefont  [1]{#1}%
	\providecommand \bibfnamefont [1]{#1}%
	\providecommand \citenamefont [1]{#1}%
	\providecommand \href@noop [0]{\@secondoftwo}%
	\providecommand \href [0]{\begingroup \@sanitize@url \@href}%
	\providecommand \@href[1]{\@@startlink{#1}\@@href}%
	\providecommand \@@href[1]{\endgroup#1\@@endlink}%
	\providecommand \@sanitize@url [0]{\catcode `\\12\catcode `\$12\catcode
		`\&12\catcode `\#12\catcode `\^12\catcode `\_12\catcode `\%12\relax}%
	\providecommand \@@startlink[1]{}%
	\providecommand \@@endlink[0]{}%
	\providecommand \url  [0]{\begingroup\@sanitize@url \@url }%
	\providecommand \@url [1]{\endgroup\@href {#1}{\urlprefix }}%
	\providecommand \urlprefix  [0]{URL }%
	\providecommand \Eprint [0]{\href }%
	\providecommand \doibase [0]{https://doi.org/}%
	\providecommand \selectlanguage [0]{\@gobble}%
	\providecommand \bibinfo  [0]{\@secondoftwo}%
	\providecommand \bibfield  [0]{\@secondoftwo}%
	\providecommand \translation [1]{[#1]}%
	\providecommand \BibitemOpen [0]{}%
	\providecommand \bibitemStop [0]{}%
	\providecommand \bibitemNoStop [0]{.\EOS\space}%
	\providecommand \EOS [0]{\spacefactor3000\relax}%
	\providecommand \BibitemShut  [1]{\csname bibitem#1\endcsname}%
	\let\auto@bib@innerbib\@empty
	\bibitem [{\citenamefont {Dubin}\ and\ \citenamefont
		{O'Neil}(1999)}]{Dubin1999}%
	\BibitemOpen
	\bibfield  {author} {\bibinfo {author} {\bibfnamefont {D.~H.~E.}\
			\bibnamefont {Dubin}}\ and\ \bibinfo {author} {\bibfnamefont
			{T.}~\bibnamefont {O'Neil}},\ }\bibfield  {title} {\bibinfo {title} {Trapped
			nonneutral plasmas, liquids, and crystals (the thermal equilibrium states)},\
	}\href {https://doi.org/10.1103/RevModPhys.71.87} {\bibfield  {journal}
		{\bibinfo  {journal} {Rev. Mod. Phys.}\ }\textbf {\bibinfo {volume} {71}},\
		\bibinfo {pages} {87} (\bibinfo {year} {1999})}\BibitemShut {NoStop}%
	\bibitem [{\citenamefont {Leibfried}\ \emph
		{et~al.}(2003{\natexlab{a}})\citenamefont {Leibfried}, \citenamefont {Blatt},
		\citenamefont {Monroe},\ and\ \citenamefont {Wineland}}]{Leibfried2003}%
	\BibitemOpen
	\bibfield  {author} {\bibinfo {author} {\bibfnamefont {D.}~\bibnamefont
			{Leibfried}}, \bibinfo {author} {\bibfnamefont {R.}~\bibnamefont {Blatt}},
		\bibinfo {author} {\bibfnamefont {C.}~\bibnamefont {Monroe}},\ and\ \bibinfo
		{author} {\bibfnamefont {D.}~\bibnamefont {Wineland}},\ }\bibfield  {title}
	{\bibinfo {title} {Quantum dynamics of single trapped ions},\ }\href
	{https://doi.org/10.1103/RevModPhys.75.281} {\bibfield  {journal} {\bibinfo
			{journal} {Rev. Mod. Phys.}\ }\textbf {\bibinfo {volume} {75}},\ \bibinfo
		{pages} {281} (\bibinfo {year} {2003}{\natexlab{a}})}\BibitemShut {NoStop}%
	\bibitem [{\citenamefont {Cirac}\ and\ \citenamefont
		{Zoller}(1995)}]{Cirac1995}%
	\BibitemOpen
	\bibfield  {author} {\bibinfo {author} {\bibfnamefont {J.~I.}\ \bibnamefont
			{Cirac}}\ and\ \bibinfo {author} {\bibfnamefont {P.}~\bibnamefont {Zoller}},\
	}\bibfield  {title} {\bibinfo {title} {Quantum computations with cold trapped
			ions},\ }\href {https://doi.org/10.1103/PhysRevLett.74.4091} {\bibfield
		{journal} {\bibinfo  {journal} {Phys. Rev. Lett.}\ }\textbf {\bibinfo
			{volume} {74}},\ \bibinfo {pages} {4091} (\bibinfo {year}
		{1995})}\BibitemShut {NoStop}%
	\bibitem [{\citenamefont {Leibfried}\ \emph
		{et~al.}(2003{\natexlab{b}})\citenamefont {Leibfried}, \citenamefont
		{DeMarco}, \citenamefont {Meyer}, \citenamefont {Lucas}, \citenamefont
		{Barrett}, \citenamefont {Britton}, \citenamefont {Itano}, \citenamefont
		{Jelenković}, \citenamefont {Langer}, \citenamefont {Rosenband},\ and\
		\citenamefont {Wineland}}]{Leibfried2003a}%
	\BibitemOpen
	\bibfield  {author} {\bibinfo {author} {\bibfnamefont {D.}~\bibnamefont
			{Leibfried}}, \bibinfo {author} {\bibfnamefont {B.}~\bibnamefont {DeMarco}},
		\bibinfo {author} {\bibfnamefont {V.}~\bibnamefont {Meyer}}, \bibinfo
		{author} {\bibfnamefont {D.}~\bibnamefont {Lucas}}, \bibinfo {author}
		{\bibfnamefont {M.}~\bibnamefont {Barrett}}, \bibinfo {author} {\bibfnamefont
			{J.}~\bibnamefont {Britton}}, \bibinfo {author} {\bibfnamefont {W.~M.}\
			\bibnamefont {Itano}}, \bibinfo {author} {\bibfnamefont {B.}~\bibnamefont
			{Jelenković}}, \bibinfo {author} {\bibfnamefont {C.}~\bibnamefont {Langer}},
		\bibinfo {author} {\bibfnamefont {T.}~\bibnamefont {Rosenband}},\ and\
		\bibinfo {author} {\bibfnamefont {D.~J.}\ \bibnamefont {Wineland}},\
	}\bibfield  {title} {\bibinfo {title} {Experimental demonstration of a
			robust, high-fidelity geometric two ion-qubit phase gate},\ }\href
	{https://doi.org/10.1038/nature01492} {\bibfield  {journal} {\bibinfo
			{journal} {Nature}\ }\textbf {\bibinfo {volume} {422}},\ \bibinfo {pages}
		{412} (\bibinfo {year} {2003}{\natexlab{b}})}\BibitemShut {NoStop}%
	\bibitem [{\citenamefont {Schmidt-Kaler}\ \emph {et~al.}(2003)\citenamefont
		{Schmidt-Kaler}, \citenamefont {H{\"a}ffner}, \citenamefont {Riebe},
		\citenamefont {Gulde}, \citenamefont {Lancaster}, \citenamefont {Deuschle},
		\citenamefont {Becher}, \citenamefont {Roos}, \citenamefont {Eschner},\ and\
		\citenamefont {Blatt}}]{SchmidtKaler2003}%
	\BibitemOpen
	\bibfield  {author} {\bibinfo {author} {\bibfnamefont {F.}~\bibnamefont
			{Schmidt-Kaler}}, \bibinfo {author} {\bibfnamefont {H.}~\bibnamefont
			{H{\"a}ffner}}, \bibinfo {author} {\bibfnamefont {M.}~\bibnamefont {Riebe}},
		\bibinfo {author} {\bibfnamefont {S.}~\bibnamefont {Gulde}}, \bibinfo
		{author} {\bibfnamefont {G.~P.~T.}\ \bibnamefont {Lancaster}}, \bibinfo
		{author} {\bibfnamefont {T.}~\bibnamefont {Deuschle}}, \bibinfo {author}
		{\bibfnamefont {C.}~\bibnamefont {Becher}}, \bibinfo {author} {\bibfnamefont
			{C.~F.}\ \bibnamefont {Roos}}, \bibinfo {author} {\bibfnamefont
			{J.}~\bibnamefont {Eschner}},\ and\ \bibinfo {author} {\bibfnamefont
			{R.}~\bibnamefont {Blatt}},\ }\bibfield  {title} {\bibinfo {title}
		{Realization of the {Cirac--Zoller} controlled-{NOT} quantum gate},\ }\href
	{https://doi.org/10.1038/nature01494} {\bibfield  {journal} {\bibinfo
			{journal} {Nature}\ }\textbf {\bibinfo {volume} {422}},\ \bibinfo {pages}
		{408} (\bibinfo {year} {2003})}\BibitemShut {NoStop}%
	\bibitem [{\citenamefont {Schindler}\ \emph {et~al.}(2013)\citenamefont
		{Schindler}, \citenamefont {Nigg}, \citenamefont {Monz}, \citenamefont
		{Barreiro}, \citenamefont {Martinez}, \citenamefont {Wang}, \citenamefont
		{Quint}, \citenamefont {Brandl}, \citenamefont {Nebendahl}, \citenamefont
		{Roos}, \citenamefont {Chwalla}, \citenamefont {Hennrich},\ and\
		\citenamefont {Blatt}}]{Schindler2013}%
	\BibitemOpen
	\bibfield  {author} {\bibinfo {author} {\bibfnamefont {P.}~\bibnamefont
			{Schindler}}, \bibinfo {author} {\bibfnamefont {D.}~\bibnamefont {Nigg}},
		\bibinfo {author} {\bibfnamefont {T.}~\bibnamefont {Monz}}, \bibinfo {author}
		{\bibfnamefont {J.~T.}\ \bibnamefont {Barreiro}}, \bibinfo {author}
		{\bibfnamefont {E.}~\bibnamefont {Martinez}}, \bibinfo {author}
		{\bibfnamefont {S.~X.}\ \bibnamefont {Wang}}, \bibinfo {author}
		{\bibfnamefont {S.}~\bibnamefont {Quint}}, \bibinfo {author} {\bibfnamefont
			{M.~F.}\ \bibnamefont {Brandl}}, \bibinfo {author} {\bibfnamefont
			{V.}~\bibnamefont {Nebendahl}}, \bibinfo {author} {\bibfnamefont {C.~F.}\
			\bibnamefont {Roos}}, \bibinfo {author} {\bibfnamefont {M.}~\bibnamefont
			{Chwalla}}, \bibinfo {author} {\bibfnamefont {M.}~\bibnamefont {Hennrich}},\
		and\ \bibinfo {author} {\bibfnamefont {R.}~\bibnamefont {Blatt}},\ }\bibfield
	{title} {\bibinfo {title} {A quantum information processor with trapped
			ions},\ }\href {https://doi.org/10.1088/1367-2630/15/12/123012} {\bibfield
		{journal} {\bibinfo  {journal} {New J. Phys.}\ }\textbf {\bibinfo {volume}
			{15}},\ \bibinfo {pages} {123012} (\bibinfo {year} {2013})}\BibitemShut
	{NoStop}%
	\bibitem [{\citenamefont {Monroe}\ \emph {et~al.}(2014)\citenamefont {Monroe},
		\citenamefont {Raussendorf}, \citenamefont {Ruthven}, \citenamefont {Brown},
		\citenamefont {Maunz}, \citenamefont {Duan},\ and\ \citenamefont
		{Kim}}]{Monroe2014}%
	\BibitemOpen
	\bibfield  {author} {\bibinfo {author} {\bibfnamefont {C.}~\bibnamefont
			{Monroe}}, \bibinfo {author} {\bibfnamefont {R.}~\bibnamefont {Raussendorf}},
		\bibinfo {author} {\bibfnamefont {A.}~\bibnamefont {Ruthven}}, \bibinfo
		{author} {\bibfnamefont {K.~R.}\ \bibnamefont {Brown}}, \bibinfo {author}
		{\bibfnamefont {P.}~\bibnamefont {Maunz}}, \bibinfo {author} {\bibfnamefont
			{L.-M.}\ \bibnamefont {Duan}},\ and\ \bibinfo {author} {\bibfnamefont
			{J.}~\bibnamefont {Kim}},\ }\bibfield  {title} {\bibinfo {title} {Large-scale
			modular quantum-computer architecture with atomic memory and photonic
			interconnects},\ }\href {https://doi.org/10.1103/PhysRevA.89.022317}
	{\bibfield  {journal} {\bibinfo  {journal} {Phys. Rev. A}\ }\textbf {\bibinfo
			{volume} {89}},\ \bibinfo {pages} {022317} (\bibinfo {year}
		{2014})}\BibitemShut {NoStop}%
	\bibitem [{\citenamefont {Harty}\ \emph {et~al.}(2014)\citenamefont {Harty},
		\citenamefont {Allcock}, \citenamefont {Ballance}, \citenamefont {Guidoni},
		\citenamefont {Janacek}, \citenamefont {Linke}, \citenamefont {Stacey},\ and\
		\citenamefont {Lucas}}]{Harty2014}%
	\BibitemOpen
	\bibfield  {author} {\bibinfo {author} {\bibfnamefont {T.~P.}\ \bibnamefont
			{Harty}}, \bibinfo {author} {\bibfnamefont {D.~T.~C.}\ \bibnamefont
			{Allcock}}, \bibinfo {author} {\bibfnamefont {C.~J.}\ \bibnamefont
			{Ballance}}, \bibinfo {author} {\bibfnamefont {L.}~\bibnamefont {Guidoni}},
		\bibinfo {author} {\bibfnamefont {H.~A.}\ \bibnamefont {Janacek}}, \bibinfo
		{author} {\bibfnamefont {N.~M.}\ \bibnamefont {Linke}}, \bibinfo {author}
		{\bibfnamefont {D.~N.}\ \bibnamefont {Stacey}},\ and\ \bibinfo {author}
		{\bibfnamefont {D.~M.}\ \bibnamefont {Lucas}},\ }\bibfield  {title} {\bibinfo
		{title} {High-fidelity preparation, gates, memory, and readout of a
			trapped-ion quantum bit},\ }\href
	{https://doi.org/10.1103/PhysRevLett.113.220501} {\bibfield  {journal}
		{\bibinfo  {journal} {Phys. Rev. Lett.}\ }\textbf {\bibinfo {volume} {113}},\
		\bibinfo {pages} {220501} (\bibinfo {year} {2014})}\BibitemShut {NoStop}%
	\bibitem [{\citenamefont {Monz}\ \emph {et~al.}(2016)\citenamefont {Monz},
		\citenamefont {Nigg}, \citenamefont {Martinez}, \citenamefont {Brandl},
		\citenamefont {Schindler}, \citenamefont {Rines}, \citenamefont {Wang},
		\citenamefont {Chuang},\ and\ \citenamefont {Blatt}}]{Monz2016}%
	\BibitemOpen
	\bibfield  {author} {\bibinfo {author} {\bibfnamefont {T.}~\bibnamefont
			{Monz}}, \bibinfo {author} {\bibfnamefont {D.}~\bibnamefont {Nigg}}, \bibinfo
		{author} {\bibfnamefont {E.~A.}\ \bibnamefont {Martinez}}, \bibinfo {author}
		{\bibfnamefont {M.~F.}\ \bibnamefont {Brandl}}, \bibinfo {author}
		{\bibfnamefont {P.}~\bibnamefont {Schindler}}, \bibinfo {author}
		{\bibfnamefont {R.}~\bibnamefont {Rines}}, \bibinfo {author} {\bibfnamefont
			{S.~X.}\ \bibnamefont {Wang}}, \bibinfo {author} {\bibfnamefont {I.~L.}\
			\bibnamefont {Chuang}},\ and\ \bibinfo {author} {\bibfnamefont
			{R.}~\bibnamefont {Blatt}},\ }\bibfield  {title} {\bibinfo {title}
		{Realization of a scalable {Shor} algorithm},\ }\href
	{https://doi.org/10.1126/science.aad9480} {\bibfield  {journal} {\bibinfo
			{journal} {Science}\ }\textbf {\bibinfo {volume} {351}},\ \bibinfo {pages}
		{1068} (\bibinfo {year} {2016})}\BibitemShut {NoStop}%
	\bibitem [{\citenamefont {Wright}\ \emph {et~al.}(2019)\citenamefont {Wright},
		\citenamefont {Beck}, \citenamefont {Debnath}, \citenamefont {Amini},
		\citenamefont {Nam}, \citenamefont {Grzesiak}, \citenamefont {Chen},
		\citenamefont {Pisenti}, \citenamefont {Chmielewski}, \citenamefont
		{Collins}, \citenamefont {Hudek}, \citenamefont {Mizrahi}, \citenamefont
		{Wong-Campos}, \citenamefont {Allen}, \citenamefont {Apisdorf}, \citenamefont
		{Solomon}, \citenamefont {Williams}, \citenamefont {Ducore}, \citenamefont
		{Blinov}, \citenamefont {Kreikemeier}, \citenamefont {Chaplin}, \citenamefont
		{Keesan}, \citenamefont {Monroe},\ and\ \citenamefont {Kim}}]{Wright2019}%
	\BibitemOpen
	\bibfield  {author} {\bibinfo {author} {\bibfnamefont {K.}~\bibnamefont
			{Wright}}, \bibinfo {author} {\bibfnamefont {K.~M.}\ \bibnamefont {Beck}},
		\bibinfo {author} {\bibfnamefont {S.}~\bibnamefont {Debnath}}, \bibinfo
		{author} {\bibfnamefont {J.~M.}\ \bibnamefont {Amini}}, \bibinfo {author}
		{\bibfnamefont {Y.}~\bibnamefont {Nam}}, \bibinfo {author} {\bibfnamefont
			{N.}~\bibnamefont {Grzesiak}}, \bibinfo {author} {\bibfnamefont {J.-S.}\
			\bibnamefont {Chen}}, \bibinfo {author} {\bibfnamefont {N.~C.}\ \bibnamefont
			{Pisenti}}, \bibinfo {author} {\bibfnamefont {M.}~\bibnamefont
			{Chmielewski}}, \bibinfo {author} {\bibfnamefont {C.}~\bibnamefont
			{Collins}}, \bibinfo {author} {\bibfnamefont {K.~M.}\ \bibnamefont {Hudek}},
		\bibinfo {author} {\bibfnamefont {J.}~\bibnamefont {Mizrahi}}, \bibinfo
		{author} {\bibfnamefont {J.~D.}\ \bibnamefont {Wong-Campos}}, \bibinfo
		{author} {\bibfnamefont {S.}~\bibnamefont {Allen}}, \bibinfo {author}
		{\bibfnamefont {J.}~\bibnamefont {Apisdorf}}, \bibinfo {author}
		{\bibfnamefont {P.}~\bibnamefont {Solomon}}, \bibinfo {author} {\bibfnamefont
			{M.}~\bibnamefont {Williams}}, \bibinfo {author} {\bibfnamefont {A.~M.}\
			\bibnamefont {Ducore}}, \bibinfo {author} {\bibfnamefont {A.}~\bibnamefont
			{Blinov}}, \bibinfo {author} {\bibfnamefont {S.~M.}\ \bibnamefont
			{Kreikemeier}}, \bibinfo {author} {\bibfnamefont {V.}~\bibnamefont
			{Chaplin}}, \bibinfo {author} {\bibfnamefont {M.}~\bibnamefont {Keesan}},
		\bibinfo {author} {\bibfnamefont {C.}~\bibnamefont {Monroe}},\ and\ \bibinfo
		{author} {\bibfnamefont {J.}~\bibnamefont {Kim}},\ }\bibfield  {title}
	{\bibinfo {title} {Benchmarking an 11-qubit quantum computer},\ }\href
	{https://doi.org/10.1038/s41467-019-13534-2} {\bibfield  {journal} {\bibinfo
			{journal} {Nat. Commun.}\ }\textbf {\bibinfo {volume} {10}},\ \bibinfo
		{pages} {5464} (\bibinfo {year} {2019})}\BibitemShut {NoStop}%
	\bibitem [{\citenamefont {Bylinskii}\ \emph {et~al.}(2015)\citenamefont
		{Bylinskii}, \citenamefont {Gangloff},\ and\ \citenamefont
		{Vuleti\'{c}}}]{Bylinskii2015}%
	\BibitemOpen
	\bibfield  {author} {\bibinfo {author} {\bibfnamefont {A.}~\bibnamefont
			{Bylinskii}}, \bibinfo {author} {\bibfnamefont {D.}~\bibnamefont
			{Gangloff}},\ and\ \bibinfo {author} {\bibfnamefont {V.}~\bibnamefont
			{Vuleti\'{c}}},\ }\bibfield  {title} {\bibinfo {title} {Tuning friction
			atom-by-atom in an ion-crystal simulator},\ }\href
	{https://doi.org/10.1126/science.1261422} {\bibfield  {journal} {\bibinfo
			{journal} {Science}\ }\textbf {\bibinfo {volume} {348}},\ \bibinfo {pages}
		{1115} (\bibinfo {year} {2015})}\BibitemShut {NoStop}%
	\bibitem [{\citenamefont {Kiethe}\ \emph {et~al.}(2017)\citenamefont {Kiethe},
		\citenamefont {Nigmatullin}, \citenamefont {Kalincev}, \citenamefont
		{Schmirander},\ and\ \citenamefont {Mehlst{\"a}ubler}}]{Kiethe2017}%
	\BibitemOpen
	\bibfield  {author} {\bibinfo {author} {\bibfnamefont {J.}~\bibnamefont
			{Kiethe}}, \bibinfo {author} {\bibfnamefont {R.}~\bibnamefont {Nigmatullin}},
		\bibinfo {author} {\bibfnamefont {D.}~\bibnamefont {Kalincev}}, \bibinfo
		{author} {\bibfnamefont {T.}~\bibnamefont {Schmirander}},\ and\ \bibinfo
		{author} {\bibfnamefont {T.~E.}\ \bibnamefont {Mehlst{\"a}ubler}},\
	}\bibfield  {title} {\bibinfo {title} {{Probing nanofriction and Aubry-type
				signatures in a finite self-organized system}},\ }\href
	{https://doi.org/10.1038/ncomms15364} {\bibfield  {journal} {\bibinfo
			{journal} {Nat. Commun.}\ }\textbf {\bibinfo {volume} {8}},\ \bibinfo {pages}
		{15364} (\bibinfo {year} {2017})}\BibitemShut {NoStop}%
	\bibitem [{\citenamefont {Matjeschk}\ \emph {et~al.}(2012)\citenamefont
		{Matjeschk}, \citenamefont {Schneider}, \citenamefont {Enderlein},
		\citenamefont {Huber}, \citenamefont {Schmitz}, \citenamefont {Glueckert},\
		and\ \citenamefont {Schaetz}}]{Matjeschk2012}%
	\BibitemOpen
	\bibfield  {author} {\bibinfo {author} {\bibfnamefont {R.}~\bibnamefont
			{Matjeschk}}, \bibinfo {author} {\bibfnamefont {C.}~\bibnamefont
			{Schneider}}, \bibinfo {author} {\bibfnamefont {M.}~\bibnamefont
			{Enderlein}}, \bibinfo {author} {\bibfnamefont {T.}~\bibnamefont {Huber}},
		\bibinfo {author} {\bibfnamefont {H.}~\bibnamefont {Schmitz}}, \bibinfo
		{author} {\bibfnamefont {J.}~\bibnamefont {Glueckert}},\ and\ \bibinfo
		{author} {\bibfnamefont {T.}~\bibnamefont {Schaetz}},\ }\bibfield  {title}
	{\bibinfo {title} {Experimental simulation and limitations of quantum walks
			with trapped ions},\ }\href {https://doi.org/10.1088/1367-2630/14/3/035012}
	{\bibfield  {journal} {\bibinfo  {journal} {New J. Phys.}\ }\textbf {\bibinfo
			{volume} {14}},\ \bibinfo {pages} {035012} (\bibinfo {year}
		{2012})}\BibitemShut {NoStop}%
	\bibitem [{\citenamefont {Toyoda}\ \emph {et~al.}(2013)\citenamefont {Toyoda},
		\citenamefont {Matsuno}, \citenamefont {Noguchi}, \citenamefont {Haze},\ and\
		\citenamefont {Urabe}}]{Toyoda2013}%
	\BibitemOpen
	\bibfield  {author} {\bibinfo {author} {\bibfnamefont {K.}~\bibnamefont
			{Toyoda}}, \bibinfo {author} {\bibfnamefont {Y.}~\bibnamefont {Matsuno}},
		\bibinfo {author} {\bibfnamefont {A.}~\bibnamefont {Noguchi}}, \bibinfo
		{author} {\bibfnamefont {S.}~\bibnamefont {Haze}},\ and\ \bibinfo {author}
		{\bibfnamefont {S.}~\bibnamefont {Urabe}},\ }\bibfield  {title} {\bibinfo
		{title} {Experimental realization of a quantum phase transition of
			polaritonic excitations},\ }\href
	{https://doi.org/10.1103/PhysRevLett.111.160501} {\bibfield  {journal}
		{\bibinfo  {journal} {Phys. Rev. Lett.}\ }\textbf {\bibinfo {volume} {111}},\
		\bibinfo {pages} {160501} (\bibinfo {year} {2013})}\BibitemShut {NoStop}%
	\bibitem [{\citenamefont {Martinez}\ \emph {et~al.}(2016)\citenamefont
		{Martinez}, \citenamefont {Muschik}, \citenamefont {Schindler}, \citenamefont
		{Nigg}, \citenamefont {Erhard}, \citenamefont {Heyl}, \citenamefont {Hauke},
		\citenamefont {Dalmonte}, \citenamefont {Monz}, \citenamefont {Zoller},\ and\
		\citenamefont {Blatt}}]{Martinez2016}%
	\BibitemOpen
	\bibfield  {author} {\bibinfo {author} {\bibfnamefont {E.~A.}\ \bibnamefont
			{Martinez}}, \bibinfo {author} {\bibfnamefont {C.~A.}\ \bibnamefont
			{Muschik}}, \bibinfo {author} {\bibfnamefont {P.}~\bibnamefont {Schindler}},
		\bibinfo {author} {\bibfnamefont {D.}~\bibnamefont {Nigg}}, \bibinfo {author}
		{\bibfnamefont {A.}~\bibnamefont {Erhard}}, \bibinfo {author} {\bibfnamefont
			{M.}~\bibnamefont {Heyl}}, \bibinfo {author} {\bibfnamefont {P.}~\bibnamefont
			{Hauke}}, \bibinfo {author} {\bibfnamefont {M.}~\bibnamefont {Dalmonte}},
		\bibinfo {author} {\bibfnamefont {T.}~\bibnamefont {Monz}}, \bibinfo {author}
		{\bibfnamefont {P.}~\bibnamefont {Zoller}},\ and\ \bibinfo {author}
		{\bibfnamefont {R.}~\bibnamefont {Blatt}},\ }\bibfield  {title} {\bibinfo
		{title} {Real-time dynamics of lattice gauge theories with a few-qubit
			quantum computer},\ }\href {https://doi.org/10.1038/nature18318} {\bibfield
		{journal} {\bibinfo  {journal} {Nature}\ }\textbf {\bibinfo {volume} {534}},\
		\bibinfo {pages} {516} (\bibinfo {year} {2016})}\BibitemShut {NoStop}%
	\bibitem [{\citenamefont {Zhang}\ \emph {et~al.}(2017)\citenamefont {Zhang},
		\citenamefont {Pagano}, \citenamefont {Hess}, \citenamefont {Kyprianidis},
		\citenamefont {Becker}, \citenamefont {Kaplan}, \citenamefont {Gorshkov},
		\citenamefont {Gong},\ and\ \citenamefont {Monroe}}]{Zhang2017}%
	\BibitemOpen
	\bibfield  {author} {\bibinfo {author} {\bibfnamefont {J.}~\bibnamefont
			{Zhang}}, \bibinfo {author} {\bibfnamefont {G.}~\bibnamefont {Pagano}},
		\bibinfo {author} {\bibfnamefont {P.~W.}\ \bibnamefont {Hess}}, \bibinfo
		{author} {\bibfnamefont {A.}~\bibnamefont {Kyprianidis}}, \bibinfo {author}
		{\bibfnamefont {P.}~\bibnamefont {Becker}}, \bibinfo {author} {\bibfnamefont
			{H.}~\bibnamefont {Kaplan}}, \bibinfo {author} {\bibfnamefont {A.~V.}\
			\bibnamefont {Gorshkov}}, \bibinfo {author} {\bibfnamefont {Z.-X.}\
			\bibnamefont {Gong}},\ and\ \bibinfo {author} {\bibfnamefont
			{C.}~\bibnamefont {Monroe}},\ }\bibfield  {title} {\bibinfo {title}
		{Observation of a many-body dynamical phase transition with a 53-qubit
			quantum simulator},\ }\href {https://doi.org/10.1038/nature24654} {\bibfield
		{journal} {\bibinfo  {journal} {Nature}\ }\textbf {\bibinfo {volume} {551}},\
		\bibinfo {pages} {601} (\bibinfo {year} {2017})}\BibitemShut {NoStop}%
	\bibitem [{\citenamefont {G{\"a}rttner}\ \emph {et~al.}(2017)\citenamefont
		{G{\"a}rttner}, \citenamefont {Bohnet}, \citenamefont {Safavi-Naini},
		\citenamefont {Wall}, \citenamefont {Bollinger},\ and\ \citenamefont
		{Rey}}]{Gaerttner2017}%
	\BibitemOpen
	\bibfield  {author} {\bibinfo {author} {\bibfnamefont {M.}~\bibnamefont
			{G{\"a}rttner}}, \bibinfo {author} {\bibfnamefont {J.~G.}\ \bibnamefont
			{Bohnet}}, \bibinfo {author} {\bibfnamefont {A.}~\bibnamefont
			{Safavi-Naini}}, \bibinfo {author} {\bibfnamefont {M.~L.}\ \bibnamefont
			{Wall}}, \bibinfo {author} {\bibfnamefont {J.~J.}\ \bibnamefont
			{Bollinger}},\ and\ \bibinfo {author} {\bibfnamefont {A.~M.}\ \bibnamefont
			{Rey}},\ }\bibfield  {title} {\bibinfo {title} {Measuring out-of-time-order
			correlations and multiple quantum spectra in a trapped-ion quantum magnet},\
	}\href {https://doi.org/10.1038/nphys4119} {\bibfield  {journal} {\bibinfo
			{journal} {Nat. Phys.}\ }\textbf {\bibinfo {volume} {13}},\ \bibinfo {pages}
		{781} (\bibinfo {year} {2017})}\BibitemShut {NoStop}%
	\bibitem [{\citenamefont {Brox}\ \emph {et~al.}(2017)\citenamefont {Brox},
		\citenamefont {Kiefer}, \citenamefont {Bujak}, \citenamefont {Schaetz},\ and\
		\citenamefont {Landa}}]{Brox2017}%
	\BibitemOpen
	\bibfield  {author} {\bibinfo {author} {\bibfnamefont {J.}~\bibnamefont
			{Brox}}, \bibinfo {author} {\bibfnamefont {P.}~\bibnamefont {Kiefer}},
		\bibinfo {author} {\bibfnamefont {M.}~\bibnamefont {Bujak}}, \bibinfo
		{author} {\bibfnamefont {T.}~\bibnamefont {Schaetz}},\ and\ \bibinfo {author}
		{\bibfnamefont {H.}~\bibnamefont {Landa}},\ }\bibfield  {title} {\bibinfo
		{title} {Spectroscopy and directed transport of topological solitons in
			crystals of trapped ions},\ }\href
	{https://doi.org/10.1103/physrevlett.119.153602} {\bibfield  {journal}
		{\bibinfo  {journal} {Phys. Rev. Lett.}\ }\textbf {\bibinfo {volume} {119}},\
		\bibinfo {pages} {153602} (\bibinfo {year} {2017})}\BibitemShut {NoStop}%
	\bibitem [{\citenamefont {Gorman}\ \emph {et~al.}(2018)\citenamefont {Gorman},
		\citenamefont {Hemmerling}, \citenamefont {Megidish}, \citenamefont
		{Moeller}, \citenamefont {Schindler}, \citenamefont {Sarovar},\ and\
		\citenamefont {Haeffner}}]{Gorman2018}%
	\BibitemOpen
	\bibfield  {author} {\bibinfo {author} {\bibfnamefont {D.~J.}\ \bibnamefont
			{Gorman}}, \bibinfo {author} {\bibfnamefont {B.}~\bibnamefont {Hemmerling}},
		\bibinfo {author} {\bibfnamefont {E.}~\bibnamefont {Megidish}}, \bibinfo
		{author} {\bibfnamefont {S.~A.}\ \bibnamefont {Moeller}}, \bibinfo {author}
		{\bibfnamefont {P.}~\bibnamefont {Schindler}}, \bibinfo {author}
		{\bibfnamefont {M.}~\bibnamefont {Sarovar}},\ and\ \bibinfo {author}
		{\bibfnamefont {H.}~\bibnamefont {Haeffner}},\ }\bibfield  {title} {\bibinfo
		{title} {Engineering vibrationally assisted energy transfer in a trapped-ion
			quantum simulator},\ }\href {https://doi.org/10.1103/PhysRevX.8.011038}
	{\bibfield  {journal} {\bibinfo  {journal} {Phys. Rev. X}\ }\textbf {\bibinfo
			{volume} {8}},\ \bibinfo {pages} {011038} (\bibinfo {year}
		{2018})}\BibitemShut {NoStop}%
	\bibitem [{\citenamefont {Zhang}\ \emph {et~al.}(2018)\citenamefont {Zhang},
		\citenamefont {Zhang}, \citenamefont {Shen}, \citenamefont {Zhang},
		\citenamefont {Zhang}, \citenamefont {Yung}, \citenamefont {Casanova},
		\citenamefont {Pedernales}, \citenamefont {Lamata}, \citenamefont {Solano},\
		and\ \citenamefont {Kim}}]{Zhang2018}%
	\BibitemOpen
	\bibfield  {author} {\bibinfo {author} {\bibfnamefont {X.}~\bibnamefont
			{Zhang}}, \bibinfo {author} {\bibfnamefont {K.}~\bibnamefont {Zhang}},
		\bibinfo {author} {\bibfnamefont {Y.}~\bibnamefont {Shen}}, \bibinfo {author}
		{\bibfnamefont {S.}~\bibnamefont {Zhang}}, \bibinfo {author} {\bibfnamefont
			{J.-N.}\ \bibnamefont {Zhang}}, \bibinfo {author} {\bibfnamefont {M.-H.}\
			\bibnamefont {Yung}}, \bibinfo {author} {\bibfnamefont {J.}~\bibnamefont
			{Casanova}}, \bibinfo {author} {\bibfnamefont {J.~S.}\ \bibnamefont
			{Pedernales}}, \bibinfo {author} {\bibfnamefont {L.}~\bibnamefont {Lamata}},
		\bibinfo {author} {\bibfnamefont {E.}~\bibnamefont {Solano}},\ and\ \bibinfo
		{author} {\bibfnamefont {K.}~\bibnamefont {Kim}},\ }\bibfield  {title}
	{\bibinfo {title} {Experimental quantum simulation of fermion-antifermion
			scattering via boson exchange in a trapped ion},\ }\href
	{https://doi.org/10.1038/s41467-017-02507-y} {\bibfield  {journal} {\bibinfo
			{journal} {Nat. Commun.}\ }\textbf {\bibinfo {volume} {9}},\ \bibinfo {pages}
		{195} (\bibinfo {year} {2018})}\BibitemShut {NoStop}%
	\bibitem [{\citenamefont {Kokail}\ \emph {et~al.}(2019)\citenamefont {Kokail},
		\citenamefont {Maier}, \citenamefont {van Bijnen}, \citenamefont {Brydges},
		\citenamefont {Joshi}, \citenamefont {Jurcevic}, \citenamefont {Muschik},
		\citenamefont {Silvi}, \citenamefont {Blatt}, \citenamefont {Roos},\ and\
		\citenamefont {Zoller}}]{Kokail2019}%
	\BibitemOpen
	\bibfield  {author} {\bibinfo {author} {\bibfnamefont {C.}~\bibnamefont
			{Kokail}}, \bibinfo {author} {\bibfnamefont {C.}~\bibnamefont {Maier}},
		\bibinfo {author} {\bibfnamefont {R.}~\bibnamefont {van Bijnen}}, \bibinfo
		{author} {\bibfnamefont {T.}~\bibnamefont {Brydges}}, \bibinfo {author}
		{\bibfnamefont {M.~K.}\ \bibnamefont {Joshi}}, \bibinfo {author}
		{\bibfnamefont {P.}~\bibnamefont {Jurcevic}}, \bibinfo {author}
		{\bibfnamefont {C.~A.}\ \bibnamefont {Muschik}}, \bibinfo {author}
		{\bibfnamefont {P.}~\bibnamefont {Silvi}}, \bibinfo {author} {\bibfnamefont
			{R.}~\bibnamefont {Blatt}}, \bibinfo {author} {\bibfnamefont {C.~F.}\
			\bibnamefont {Roos}},\ and\ \bibinfo {author} {\bibfnamefont
			{P.}~\bibnamefont {Zoller}},\ }\bibfield  {title} {\bibinfo {title}
		{Self-verifying variational quantum simulation of lattice models},\ }\href
	{https://doi.org/10.1038/s41586-019-1177-4} {\bibfield  {journal} {\bibinfo
			{journal} {Nature}\ }\textbf {\bibinfo {volume} {569}},\ \bibinfo {pages}
		{355} (\bibinfo {year} {2019})}\BibitemShut {NoStop}%
	\bibitem [{\citenamefont {Tamura}\ \emph {et~al.}(2020)\citenamefont {Tamura},
		\citenamefont {Mukaiyama},\ and\ \citenamefont {Toyoda}}]{Tamura2020}%
	\BibitemOpen
	\bibfield  {author} {\bibinfo {author} {\bibfnamefont {M.}~\bibnamefont
			{Tamura}}, \bibinfo {author} {\bibfnamefont {T.}~\bibnamefont {Mukaiyama}},\
		and\ \bibinfo {author} {\bibfnamefont {K.}~\bibnamefont {Toyoda}},\
	}\bibfield  {title} {\bibinfo {title} {Quantum walks of a phonon in trapped
			ions},\ }\href {https://doi.org/10.1103/PhysRevLett.124.200501} {\bibfield
		{journal} {\bibinfo  {journal} {Phys. Rev. Lett.}\ }\textbf {\bibinfo
			{volume} {124}},\ \bibinfo {pages} {200501} (\bibinfo {year}
		{2020})}\BibitemShut {NoStop}%
	\bibitem [{\citenamefont {Jurcevic}\ \emph {et~al.}(2017)\citenamefont
		{Jurcevic}, \citenamefont {Shen}, \citenamefont {Hauke}, \citenamefont
		{Maier}, \citenamefont {Brydges}, \citenamefont {Hempel}, \citenamefont
		{Lanyon}, \citenamefont {Heyl}, \citenamefont {Blatt},\ and\ \citenamefont
		{Roos}}]{Jurcevic2017}%
	\BibitemOpen
	\bibfield  {author} {\bibinfo {author} {\bibfnamefont {P.}~\bibnamefont
			{Jurcevic}}, \bibinfo {author} {\bibfnamefont {H.}~\bibnamefont {Shen}},
		\bibinfo {author} {\bibfnamefont {P.}~\bibnamefont {Hauke}}, \bibinfo
		{author} {\bibfnamefont {C.}~\bibnamefont {Maier}}, \bibinfo {author}
		{\bibfnamefont {T.}~\bibnamefont {Brydges}}, \bibinfo {author} {\bibfnamefont
			{C.}~\bibnamefont {Hempel}}, \bibinfo {author} {\bibfnamefont {B.~P.}\
			\bibnamefont {Lanyon}}, \bibinfo {author} {\bibfnamefont {M.}~\bibnamefont
			{Heyl}}, \bibinfo {author} {\bibfnamefont {R.}~\bibnamefont {Blatt}},\ and\
		\bibinfo {author} {\bibfnamefont {C.~F.}\ \bibnamefont {Roos}},\ }\bibfield
	{title} {\bibinfo {title} {Direct observation of dynamical quantum phase
			transitions in an interacting many-body system},\ }\href
	{https://doi.org/10.1103/PhysRevLett.119.080501} {\bibfield  {journal}
		{\bibinfo  {journal} {Phys. Rev. Lett.}\ }\textbf {\bibinfo {volume} {119}},\
		\bibinfo {pages} {080501} (\bibinfo {year} {2017})}\BibitemShut {NoStop}%
	\bibitem [{\citenamefont {Kibble}(1976)}]{Kibble1976}%
	\BibitemOpen
	\bibfield  {author} {\bibinfo {author} {\bibfnamefont {T.~W.}\ \bibnamefont
			{Kibble}},\ }\bibfield  {title} {\bibinfo {title} {Topology of cosmic domains
			and strings},\ }\href {https://doi.org/10.1088/0305-4470/9/8/029} {\bibfield
		{journal} {\bibinfo  {journal} {J. Phys. A: Math. Gen.}\ }\textbf {\bibinfo
			{volume} {9}},\ \bibinfo {pages} {1387} (\bibinfo {year} {1976})}\BibitemShut
	{NoStop}%
	\bibitem [{\citenamefont {Zurek}(1985)}]{Zurek1985}%
	\BibitemOpen
	\bibfield  {author} {\bibinfo {author} {\bibfnamefont {W.~H.}\ \bibnamefont
			{Zurek}},\ }\bibfield  {title} {\bibinfo {title} {Cosmological experiments in
			superfluid helium?},\ }\href {https://doi.org//10.1038/317505a0} {\bibfield
		{journal} {\bibinfo  {journal} {Nature}\ }\textbf {\bibinfo {volume} {317}},\
		\bibinfo {pages} {505} (\bibinfo {year} {1985})}\BibitemShut {NoStop}%
	\bibitem [{\citenamefont {del Campo}\ \emph {et~al.}(2010)\citenamefont {del
			Campo}, \citenamefont {De~Chiara}, \citenamefont {Morigi}, \citenamefont
		{Plenio},\ and\ \citenamefont {Retzker}}]{DelCampo2010}%
	\BibitemOpen
	\bibfield  {author} {\bibinfo {author} {\bibfnamefont {A.}~\bibnamefont {del
				Campo}}, \bibinfo {author} {\bibfnamefont {G.}~\bibnamefont {De~Chiara}},
		\bibinfo {author} {\bibfnamefont {G.}~\bibnamefont {Morigi}}, \bibinfo
		{author} {\bibfnamefont {M.~B.}~\bibnamefont {Plenio}},\ and\ \bibinfo {author}
		{\bibfnamefont {A.}~\bibnamefont {Retzker}},\ }\bibfield  {title} {\bibinfo
		{title} {Structural defects in ion chains by quenching the external
			potential: the inhomogeneous {Kibble-Zurek} mechanism},\ }\href
	{https://doi.org/10.1103/PhysRevLett.105.075701} {\bibfield  {journal}
		{\bibinfo  {journal} {Phys. Rev. Lett.}\ }\textbf {\bibinfo {volume} {105}},\
		\bibinfo {pages} {075701} (\bibinfo {year} {2010})}\BibitemShut {NoStop}%
	\bibitem [{\citenamefont {Ulm}\ \emph {et~al.}(2013)\citenamefont {Ulm},
		\citenamefont {Ro{\ss}nagel}, \citenamefont {Jacob}, \citenamefont
		{Deg\"{u}nther}, \citenamefont {Dawkins}, \citenamefont {Poschinger},
		\citenamefont {Nigmatullin}, \citenamefont {Retzker}, \citenamefont {Plenio},
		\citenamefont {Schmidt-Kaler},\ and\ \citenamefont {Singer}}]{Ulm2013}%
	\BibitemOpen
	\bibfield  {author} {\bibinfo {author} {\bibfnamefont {S.}~\bibnamefont
			{Ulm}}, \bibinfo {author} {\bibfnamefont {J.}~\bibnamefont {Ro{\ss}nagel}},
		\bibinfo {author} {\bibfnamefont {G.}~\bibnamefont {Jacob}}, \bibinfo
		{author} {\bibfnamefont {C.}~\bibnamefont {Deg\"{u}nther}}, \bibinfo {author}
		{\bibfnamefont {S.~T.}\ \bibnamefont {Dawkins}}, \bibinfo {author}
		{\bibfnamefont {U.~G.}\ \bibnamefont {Poschinger}}, \bibinfo {author}
		{\bibfnamefont {R.}~\bibnamefont {Nigmatullin}}, \bibinfo {author}
		{\bibfnamefont {A.}~\bibnamefont {Retzker}}, \bibinfo {author} {\bibfnamefont
			{M.~B.}\ \bibnamefont {Plenio}}, \bibinfo {author} {\bibfnamefont
			{F.}~\bibnamefont {Schmidt-Kaler}},\ and\ \bibinfo {author} {\bibfnamefont
			{K.}~\bibnamefont {Singer}},\ }\bibfield  {title} {\bibinfo {title}
		{Observation of the {Kibble-Zurek} scaling law for defect formation in ion
			crystals},\ }\href {https://doi.org/10.1038/ncomms3290} {\bibfield  {journal}
		{\bibinfo  {journal} {Nat. Commun.}\ }\textbf {\bibinfo {volume} {4}},\
		\bibinfo {pages} {2290} (\bibinfo {year} {2013})}\BibitemShut {NoStop}%
	\bibitem [{\citenamefont {Pyka}\ \emph {et~al.}(2013)\citenamefont {Pyka},
		\citenamefont {Keller}, \citenamefont {Partner}, \citenamefont {Nigmatullin},
		\citenamefont {Burgermeister}, \citenamefont {Meier}, \citenamefont
		{Kuhlmann}, \citenamefont {Retzker}, \citenamefont {Plenio}, \citenamefont
		{Zurek}, \citenamefont {del Campo},\ and\ \citenamefont
		{Mehlst\"{a}ubler}}]{Pyka2013}%
	\BibitemOpen
	\bibfield  {author} {\bibinfo {author} {\bibfnamefont {K.}~\bibnamefont
			{Pyka}}, \bibinfo {author} {\bibfnamefont {J.}~\bibnamefont {Keller}},
		\bibinfo {author} {\bibfnamefont {H.~L.}\ \bibnamefont {Partner}}, \bibinfo
		{author} {\bibfnamefont {R.}~\bibnamefont {Nigmatullin}}, \bibinfo {author}
		{\bibfnamefont {T.}~\bibnamefont {Burgermeister}}, \bibinfo {author}
		{\bibfnamefont {D.~M.}\ \bibnamefont {Meier}}, \bibinfo {author}
		{\bibfnamefont {K.}~\bibnamefont {Kuhlmann}}, \bibinfo {author}
		{\bibfnamefont {A.}~\bibnamefont {Retzker}}, \bibinfo {author} {\bibfnamefont
			{M.~B.}\ \bibnamefont {Plenio}}, \bibinfo {author} {\bibfnamefont {W.~H.}\
			\bibnamefont {Zurek}}, \bibinfo {author} {\bibfnamefont {A.}~\bibnamefont
			{del Campo}},\ and\ \bibinfo {author} {\bibfnamefont {T.~E.}\ \bibnamefont
			{Mehlst\"{a}ubler}},\ }\bibfield  {title} {\bibinfo {title} {Topological
			defect formation and spontaneous symmetry breaking in ion {Coulomb}
			crystals},\ }\href {https://doi.org/10.1038/ncomms3291} {\bibfield  {journal}
		{\bibinfo  {journal} {Nat. Commun.}\ }\textbf {\bibinfo {volume} {4}},\
		\bibinfo {pages} {2291} (\bibinfo {year} {2013})}\BibitemShut {NoStop}%
	\bibitem [{\citenamefont {Mielenz}\ \emph {et~al.}(2013)\citenamefont
		{Mielenz}, \citenamefont {Brox}, \citenamefont {Kahra}, \citenamefont
		{Leschhorn}, \citenamefont {Albert}, \citenamefont {Schaetz}, \citenamefont
		{Landa},\ and\ \citenamefont {Reznik}}]{Mielenz2013}%
	\BibitemOpen
	\bibfield  {author} {\bibinfo {author} {\bibfnamefont {M.}~\bibnamefont
			{Mielenz}}, \bibinfo {author} {\bibfnamefont {J.}~\bibnamefont {Brox}},
		\bibinfo {author} {\bibfnamefont {S.}~\bibnamefont {Kahra}}, \bibinfo
		{author} {\bibfnamefont {G.}~\bibnamefont {Leschhorn}}, \bibinfo {author}
		{\bibfnamefont {M.}~\bibnamefont {Albert}}, \bibinfo {author} {\bibfnamefont
			{T.}~\bibnamefont {Schaetz}}, \bibinfo {author} {\bibfnamefont
			{H.}~\bibnamefont {Landa}},\ and\ \bibinfo {author} {\bibfnamefont
			{B.}~\bibnamefont {Reznik}},\ }\bibfield  {title} {\bibinfo {title} {Trapping
			of topological-structural defects in {Coulomb} crystals},\ }\href
	{https://doi.org/10.1103/PhysRevLett.110.133004} {\bibfield  {journal}
		{\bibinfo  {journal} {Phys. Rev. Lett.}\ }\textbf {\bibinfo {volume} {110}},\
		\bibinfo {pages} {133004} (\bibinfo {year} {2013})}\BibitemShut {NoStop}%
	\bibitem [{\citenamefont {Ejtemaee}\ and\ \citenamefont
		{Haljan}(2013)}]{Ejtemaee2013}%
	\BibitemOpen
	\bibfield  {author} {\bibinfo {author} {\bibfnamefont {S.}~\bibnamefont
			{Ejtemaee}}\ and\ \bibinfo {author} {\bibfnamefont {P.~C.}~\bibnamefont
			{Haljan}},\ }\bibfield  {title} {\bibinfo {title} {Spontaneous nucleation and
			dynamics of kink defects in zigzag arrays of trapped ions},\ }\href
	{https://doi.org/10.1103/PhysRevA.87.051401} {\bibfield  {journal} {\bibinfo
			{journal} {Phys. Rev. A}\ }\textbf {\bibinfo {volume} {87}},\ \bibinfo
		{pages} {051401(R)} (\bibinfo {year} {2013})}\BibitemShut {NoStop}%
	\bibitem [{\citenamefont {Dubin}(1993)}]{Dubin1993}%
	\BibitemOpen
	\bibfield  {author} {\bibinfo {author} {\bibfnamefont {D.~H.~E.}\
			\bibnamefont {Dubin}},\ }\bibfield  {title} {\bibinfo {title} {Theory of
			structural phase transitions in a trapped {Coulomb} crystal},\ }\href
	{https://doi.org/10.1103/PhysRevLett.71.2753} {\bibfield  {journal} {\bibinfo
			{journal} {Phys. Rev. Lett.}\ }\textbf {\bibinfo {volume} {71}},\ \bibinfo
		{pages} {2753} (\bibinfo {year} {1993})}\BibitemShut {NoStop}%
	\bibitem [{\citenamefont {Schiffer}(1993)}]{Schiffer1993}%
	\BibitemOpen
	\bibfield  {author} {\bibinfo {author} {\bibfnamefont {J.~P.}\ \bibnamefont
			{Schiffer}},\ }\bibfield  {title} {\bibinfo {title} {Phase transitions in
			anisotropically confined ionic crystals},\ }\href
	{https://doi.org/10.1103/PhysRevLett.70.818} {\bibfield  {journal} {\bibinfo
			{journal} {Phys. Rev. Lett.}\ }\textbf {\bibinfo {volume} {70}},\ \bibinfo
		{pages} {818} (\bibinfo {year} {1993})}\BibitemShut {NoStop}%
	\bibitem [{\citenamefont {Fishman}\ \emph {et~al.}(2008)\citenamefont
		{Fishman}, \citenamefont {De~Chiara}, \citenamefont {Calarco},\ and\
		\citenamefont {Morigi}}]{Fishman2008}%
	\BibitemOpen
	\bibfield  {author} {\bibinfo {author} {\bibfnamefont {S.}~\bibnamefont
			{Fishman}}, \bibinfo {author} {\bibfnamefont {G.}~\bibnamefont {De~Chiara}},
		\bibinfo {author} {\bibfnamefont {T.}~\bibnamefont {Calarco}},\ and\ \bibinfo
		{author} {\bibfnamefont {G.}~\bibnamefont {Morigi}},\ }\bibfield  {title}
	{\bibinfo {title} {Structural phase transitions in low-dimensional ion
			crystals},\ }\href {https://doi.org/10.1103/PhysRevB.77.064111} {\bibfield
		{journal} {\bibinfo  {journal} {Phys. Rev. B}\ }\textbf {\bibinfo {volume}
			{77}},\ \bibinfo {pages} {064111} (\bibinfo {year} {2008})}\BibitemShut
	{NoStop}%
	\bibitem [{\citenamefont {Piacente}\ \emph {et~al.}(2010)\citenamefont
		{Piacente}, \citenamefont {Hai},\ and\ \citenamefont
		{Peeters}}]{Piacente2010}%
	\BibitemOpen
	\bibfield  {author} {\bibinfo {author} {\bibfnamefont {G.}~\bibnamefont
			{Piacente}}, \bibinfo {author} {\bibfnamefont {G.~Q.}\ \bibnamefont {Hai}},\
		and\ \bibinfo {author} {\bibfnamefont {F.~M.}\ \bibnamefont {Peeters}},\
	}\bibfield  {title} {\bibinfo {title} {Continuous structural transitions in
			quasi-one-dimensional classical {Wigner} crystals},\ }\href
	{https://doi.org/10.1103/PhysRevB.81.024108} {\bibfield  {journal} {\bibinfo
			{journal} {Phys. Rev. B}\ }\textbf {\bibinfo {volume} {81}},\ \bibinfo
		{pages} {024108} (\bibinfo {year} {2010})}\BibitemShut {NoStop}%
	\bibitem [{\citenamefont {Garc{\'i}a-Mata}\ \emph {et~al.}(2007)\citenamefont
		{Garc{\'i}a-Mata}, \citenamefont {Zhirov},\ and\ \citenamefont
		{Shepelyansky}}]{Garcia-Mata2007}%
	\BibitemOpen
	\bibfield  {author} {\bibinfo {author} {\bibfnamefont {I.}~\bibnamefont
			{Garc{\'i}a-Mata}}, \bibinfo {author} {\bibfnamefont {O.}~\bibnamefont
			{Zhirov}},\ and\ \bibinfo {author} {\bibfnamefont {D.}~\bibnamefont
			{Shepelyansky}},\ }\bibfield  {title} {\bibinfo {title} {Frenkel-kontorova
			model with cold trapped ions},\ }\href
	{https://doi.org/10.1140/epjd/e2006-00220-2} {\bibfield  {journal} {\bibinfo
			{journal} {Eur. Phys. J. D}\ }\textbf {\bibinfo {volume} {41}},\ \bibinfo
		{pages} {325} (\bibinfo {year} {2007})}\BibitemShut {NoStop}%
	\bibitem [{\citenamefont {Benassi}\ \emph {et~al.}(2011)\citenamefont
		{Benassi}, \citenamefont {Vanossi},\ and\ \citenamefont
		{Tosatti}}]{Benasi2011}%
	\BibitemOpen
	\bibfield  {author} {\bibinfo {author} {\bibfnamefont {A.}~\bibnamefont
			{Benassi}}, \bibinfo {author} {\bibfnamefont {A.}~\bibnamefont {Vanossi}},\
		and\ \bibinfo {author} {\bibfnamefont {E.}~\bibnamefont {Tosatti}},\
	}\bibfield  {title} {\bibinfo {title} {Nanofriction in cold ion traps},\
	}\href {https://doi.org/10.1038/ncomms1230} {\bibfield  {journal} {\bibinfo
			{journal} {Nat. Commun.}\ }\textbf {\bibinfo {volume} {2}},\ \bibinfo {pages}
		{236} (\bibinfo {year} {2011})}\BibitemShut {NoStop}%
	\bibitem [{\citenamefont {Mandelli}\ \emph {et~al.}(2013)\citenamefont
		{Mandelli}, \citenamefont {Vanossi},\ and\ \citenamefont
		{Tosatti}}]{Mandelli2013}%
	\BibitemOpen
	\bibfield  {author} {\bibinfo {author} {\bibfnamefont {D.}~\bibnamefont
			{Mandelli}}, \bibinfo {author} {\bibfnamefont {A.}~\bibnamefont {Vanossi}},\
		and\ \bibinfo {author} {\bibfnamefont {E.}~\bibnamefont {Tosatti}},\
	}\bibfield  {title} {\bibinfo {title} {Stick-slip nanofriction in trapped
			cold ion chains},\ }\href {https://doi.org/10.1103/PhysRevB.87.195418}
	{\bibfield  {journal} {\bibinfo  {journal} {Phys. Rev. B}\ }\textbf {\bibinfo
			{volume} {87}},\ \bibinfo {pages} {195418} (\bibinfo {year}
		{2013})}\BibitemShut {NoStop}%
	\bibitem [{\citenamefont {Aubry}(1983)}]{Aubry1983}%
	\BibitemOpen
	\bibfield  {author} {\bibinfo {author} {\bibfnamefont {S.}~\bibnamefont
			{Aubry}},\ }\bibfield  {title} {\bibinfo {title} {The twist map, the extended
			{Frenkel-Kontorova} model and the devil's staircase},\ }\href
	{https://doi.org/10.1016/0167-2789(83)90129-X} {\bibfield  {journal}
		{\bibinfo  {journal} {Physica D}\ }\textbf {\bibinfo {volume} {7}},\ \bibinfo
		{pages} {240} (\bibinfo {year} {1983})}\BibitemShut {NoStop}%
	\bibitem [{\citenamefont {Gong}\ \emph {et~al.}(2010)\citenamefont {Gong},
		\citenamefont {Lin},\ and\ \citenamefont {Duan}}]{Gong2010}%
	\BibitemOpen
	\bibfield  {author} {\bibinfo {author} {\bibfnamefont {Z.-X.}\ \bibnamefont
			{Gong}}, \bibinfo {author} {\bibfnamefont {G.-D.}\ \bibnamefont {Lin}},\ and\
		\bibinfo {author} {\bibfnamefont {L.-M.}\ \bibnamefont {Duan}},\ }\bibfield
	{title} {\bibinfo {title} {Temperature-driven structural phase transition for
			trapped ions and a proposal for its experimental detection},\ }\href
	{https://doi.org/10.1103/PhysRevLett.105.265703} {\bibfield  {journal}
		{\bibinfo  {journal} {Phys. Rev. Lett.}\ }\textbf {\bibinfo {volume} {105}},\
		\bibinfo {pages} {265703} (\bibinfo {year} {2010})}\BibitemShut {NoStop}%
	\bibitem [{\citenamefont {Li}\ \emph {et~al.}(2019)\citenamefont {Li},
		\citenamefont {Yan}, \citenamefont {Chen}, \citenamefont {Liu}, \citenamefont
		{Zhou}, \citenamefont {Zhang}, \citenamefont {Yang},\ and\ \citenamefont
		{Feng}}]{Li2019}%
	\BibitemOpen
	\bibfield  {author} {\bibinfo {author} {\bibfnamefont {J.}~\bibnamefont
			{Li}}, \bibinfo {author} {\bibfnamefont {L.~L.}\ \bibnamefont {Yan}},
		\bibinfo {author} {\bibfnamefont {L.}~\bibnamefont {Chen}}, \bibinfo {author}
		{\bibfnamefont {Z.~C.}\ \bibnamefont {Liu}}, \bibinfo {author} {\bibfnamefont
			{F.}~\bibnamefont {Zhou}}, \bibinfo {author} {\bibfnamefont {J.~Q.}\
			\bibnamefont {Zhang}}, \bibinfo {author} {\bibfnamefont {W.~L.}\ \bibnamefont
			{Yang}},\ and\ \bibinfo {author} {\bibfnamefont {M.}~\bibnamefont {Feng}},\
	}\bibfield  {title} {\bibinfo {title} {Ion-crystal demonstration of a
			temperature-driven structural phase transition},\ }\href
	{https://doi.org/10.1103/PhysRevA.99.063402} {\bibfield  {journal} {\bibinfo
			{journal} {Phys. Rev. A}\ }\textbf {\bibinfo {volume} {99}},\ \bibinfo
		{pages} {063402} (\bibinfo {year} {2019})}\BibitemShut {NoStop}%
	\bibitem [{\citenamefont {Birkl}\ \emph {et~al.}(1992)\citenamefont {Birkl},
		\citenamefont {Kassner},\ and\ \citenamefont {Walther}}]{Birkl1992}%
	\BibitemOpen
	\bibfield  {author} {\bibinfo {author} {\bibfnamefont {G.}~\bibnamefont
			{Birkl}}, \bibinfo {author} {\bibfnamefont {S.}~\bibnamefont {Kassner}},\
		and\ \bibinfo {author} {\bibfnamefont {H.}~\bibnamefont {Walther}},\
	}\bibfield  {title} {\bibinfo {title} {Multiple-shell structures of
			laser-cooled $^{24}${Mg}$^{+}$ ions in a quadrupole storage ring},\ }\href
	{https://doi.org/10.1038/357310a0} {\bibfield  {journal} {\bibinfo  {journal}
			{Nature}\ }\textbf {\bibinfo {volume} {357}},\ \bibinfo {pages} {310}
		(\bibinfo {year} {1992})}\BibitemShut {NoStop}%
	\bibitem [{\citenamefont {Piacente}\ \emph
		{et~al.}(2004{\natexlab{a}})\citenamefont {Piacente}, \citenamefont
		{Schweigert}, \citenamefont {Betouras},\ and\ \citenamefont
		{Peeters}}]{Piacente2004}%
	\BibitemOpen
	\bibfield  {author} {\bibinfo {author} {\bibfnamefont {G.}~\bibnamefont
			{Piacente}}, \bibinfo {author} {\bibfnamefont {I.~V.}\ \bibnamefont
			{Schweigert}}, \bibinfo {author} {\bibfnamefont {J.~J.}\ \bibnamefont
			{Betouras}},\ and\ \bibinfo {author} {\bibfnamefont {F.~M.}\ \bibnamefont
			{Peeters}},\ }\bibfield  {title} {\bibinfo {title} {Generic properties of a
			quasi-one-dimensional classical {Wigner} crystal},\ }\href
	{https://doi.org/10.1103/PhysRevB.69.045324} {\bibfield  {journal} {\bibinfo
			{journal} {Phys. Rev. B}\ }\textbf {\bibinfo {volume} {69}},\ \bibinfo
		{pages} {045324} (\bibinfo {year} {2004}{\natexlab{a}})}\BibitemShut
	{NoStop}%
	\bibitem [{\citenamefont {Steane}(1997)}]{Steane1997}%
	\BibitemOpen
	\bibfield  {author} {\bibinfo {author} {\bibfnamefont {A.}~\bibnamefont
			{Steane}},\ }\bibfield  {title} {\bibinfo {title} {The ion trap quantum
			information processor},\ }\href {https://doi.org/10.1007/s003400050225}
	{\bibfield  {journal} {\bibinfo  {journal} {Appl. Phys. B}\ }\textbf
		{\bibinfo {volume} {64}},\ \bibinfo {pages} {623} (\bibinfo {year}
		{1997})}\BibitemShut {NoStop}%
	\bibitem [{\citenamefont {Enzer}\ \emph {et~al.}(2000)\citenamefont {Enzer},
		\citenamefont {Schauer}, \citenamefont {Gomez}, \citenamefont {Gulley},
		\citenamefont {Holzscheiter}, \citenamefont {Kwiat}, \citenamefont
		{Lamoreaux}, \citenamefont {Peterson}, \citenamefont {Sandberg},
		\citenamefont {Tupa}, \citenamefont {White}, \citenamefont {Hughes},\ and\
		\citenamefont {James}}]{Enzer2000}%
	\BibitemOpen
	\bibfield  {author} {\bibinfo {author} {\bibfnamefont {D.~G.}\ \bibnamefont
			{Enzer}}, \bibinfo {author} {\bibfnamefont {M.~M.}\ \bibnamefont {Schauer}},
		\bibinfo {author} {\bibfnamefont {J.~J.}\ \bibnamefont {Gomez}}, \bibinfo
		{author} {\bibfnamefont {M.~S.}\ \bibnamefont {Gulley}}, \bibinfo {author}
		{\bibfnamefont {M.~H.}\ \bibnamefont {Holzscheiter}}, \bibinfo {author}
		{\bibfnamefont {P.~G.}\ \bibnamefont {Kwiat}}, \bibinfo {author}
		{\bibfnamefont {S.~K.}\ \bibnamefont {Lamoreaux}}, \bibinfo {author}
		{\bibfnamefont {C.~G.}\ \bibnamefont {Peterson}}, \bibinfo {author}
		{\bibfnamefont {V.~D.}\ \bibnamefont {Sandberg}}, \bibinfo {author}
		{\bibfnamefont {D.}~\bibnamefont {Tupa}}, \bibinfo {author} {\bibfnamefont
			{A.~G.}\ \bibnamefont {White}}, \bibinfo {author} {\bibfnamefont {R.~J.}\
			\bibnamefont {Hughes}},\ and\ \bibinfo {author} {\bibfnamefont {D.~F.~V.}\
			\bibnamefont {James}},\ }\bibfield  {title} {\bibinfo {title} {Observation of
			power-law scaling for phase transitions in linear trapped ion crystals},\
	}\href {https://doi.org/10.1103/PhysRevLett.85.2466} {\bibfield  {journal}
		{\bibinfo  {journal} {Phys. Rev. Lett.}\ }\textbf {\bibinfo {volume} {85}},\
		\bibinfo {pages} {2466} (\bibinfo {year} {2000})}\BibitemShut {NoStop}%
	\bibitem [{\citenamefont {Silvi}\ \emph {et~al.}(2014)\citenamefont {Silvi},
		\citenamefont {Calarco}, \citenamefont {Morigi},\ and\ \citenamefont
		{Montangero}}]{Silvi2014}%
	\BibitemOpen
	\bibfield  {author} {\bibinfo {author} {\bibfnamefont {P.}~\bibnamefont
			{Silvi}}, \bibinfo {author} {\bibfnamefont {T.}~\bibnamefont {Calarco}},
		\bibinfo {author} {\bibfnamefont {G.}~\bibnamefont {Morigi}},\ and\ \bibinfo
		{author} {\bibfnamefont {S.}~\bibnamefont {Montangero}},\ }\bibfield  {title}
	{\bibinfo {title} {Ab initio characterization of the quantum linear-zigzag
			transition using density matrix renormalization group calculations},\ }\href
	{https://doi.org/10.1103/PhysRevB.89.094103} {\bibfield  {journal} {\bibinfo
			{journal} {Phys. Rev. B}\ }\textbf {\bibinfo {volume} {89}},\ \bibinfo
		{pages} {094103} (\bibinfo {year} {2014})}\BibitemShut {NoStop}%
	\bibitem [{\citenamefont {Podolsky}\ \emph {et~al.}(2014)\citenamefont
		{Podolsky}, \citenamefont {Shimshoni}, \citenamefont {Silvi}, \citenamefont
		{Montangero}, \citenamefont {Calarco}, \citenamefont {Morigi},\ and\
		\citenamefont {Fishman}}]{Podolsky2014}%
	\BibitemOpen
	\bibfield  {author} {\bibinfo {author} {\bibfnamefont {D.}~\bibnamefont
			{Podolsky}}, \bibinfo {author} {\bibfnamefont {E.}~\bibnamefont {Shimshoni}},
		\bibinfo {author} {\bibfnamefont {P.}~\bibnamefont {Silvi}}, \bibinfo
		{author} {\bibfnamefont {S.}~\bibnamefont {Montangero}}, \bibinfo {author}
		{\bibfnamefont {T.}~\bibnamefont {Calarco}}, \bibinfo {author} {\bibfnamefont
			{G.}~\bibnamefont {Morigi}},\ and\ \bibinfo {author} {\bibfnamefont
			{S.}~\bibnamefont {Fishman}},\ }\bibfield  {title} {\bibinfo {title} {From
			classical to quantum criticality},\ }\href
	{https://doi.org/10.1103/PhysRevB.89.214408} {\bibfield  {journal} {\bibinfo
			{journal} {Phys. Rev. B}\ }\textbf {\bibinfo {volume} {89}},\ \bibinfo
		{pages} {214408} (\bibinfo {year} {2014})}\BibitemShut {NoStop}%
	\bibitem [{\citenamefont {Morigi}\ and\ \citenamefont
		{Fishman}(2004)}]{Morigi2004}%
	\BibitemOpen
	\bibfield  {author} {\bibinfo {author} {\bibfnamefont {G.}~\bibnamefont
			{Morigi}}\ and\ \bibinfo {author} {\bibfnamefont {S.}~\bibnamefont
			{Fishman}},\ }\bibfield  {title} {\bibinfo {title} {Dynamics of an ion chain
			in a harmonic potential},\ }\href
	{https://doi.org/10.1103/PhysRevE.70.066141} {\bibfield  {journal} {\bibinfo
			{journal} {Phys. Rev. E}\ }\textbf {\bibinfo {volume} {70}},\ \bibinfo
		{pages} {066141} (\bibinfo {year} {2004})}\BibitemShut {NoStop}%
	\bibitem [{\citenamefont {Stenholm}(1986)}]{Stenholm1986}%
	\BibitemOpen
	\bibfield  {author} {\bibinfo {author} {\bibfnamefont {S.}~\bibnamefont
			{Stenholm}},\ }\bibfield  {title} {\bibinfo {title} {The semiclassical theory
			of laser cooling},\ }\href {https://doi.org/10.1103/RevModPhys.58.699}
	{\bibfield  {journal} {\bibinfo  {journal} {Rev. Mod. Phys.}\ }\textbf
		{\bibinfo {volume} {58}},\ \bibinfo {pages} {699} (\bibinfo {year}
		{1986})}\BibitemShut {NoStop}%
	\bibitem [{\citenamefont {Shimshoni}\ \emph {et~al.}(2011)\citenamefont
		{Shimshoni}, \citenamefont {Morigi},\ and\ \citenamefont
		{Fishman}}]{Shimshoni2011}%
	\BibitemOpen
	\bibfield  {author} {\bibinfo {author} {\bibfnamefont {E.}~\bibnamefont
			{Shimshoni}}, \bibinfo {author} {\bibfnamefont {G.}~\bibnamefont {Morigi}},\
		and\ \bibinfo {author} {\bibfnamefont {S.}~\bibnamefont {Fishman}},\
	}\bibfield  {title} {\bibinfo {title} {Quantum zigzag transition in ion
			chains},\ }\href {https://doi.org/10.1103/PhysRevLett.106.010401} {\bibfield
		{journal} {\bibinfo  {journal} {Phys. Rev. Lett.}\ }\textbf {\bibinfo
			{volume} {106}},\ \bibinfo {pages} {010401} (\bibinfo {year}
		{2011})}\BibitemShut {NoStop}%
	\bibitem [{\citenamefont {Sachdev}(2000)}]{Sachdev2000}%
	\BibitemOpen
	\bibfield  {author} {\bibinfo {author} {\bibfnamefont {S.}~\bibnamefont
			{Sachdev}},\ }\href {https://doi.org/10.1017/cbo9780511622540} {\emph
		{\bibinfo {title} {Quantum Phase Transitions}}}\ (\bibinfo  {publisher}
	{Cambridge University Press},\ \bibinfo {year} {2000})\BibitemShut {NoStop}%
	\bibitem [{\citenamefont {Pyka}\ \emph {et~al.}(2014)\citenamefont {Pyka},
		\citenamefont {Herschbach}, \citenamefont {Keller},\ and\ \citenamefont
		{Mehlst\"{a}ubler}}]{Pyka2014}%
	\BibitemOpen
	\bibfield  {author} {\bibinfo {author} {\bibfnamefont {K.}~\bibnamefont
			{Pyka}}, \bibinfo {author} {\bibfnamefont {N.}~\bibnamefont {Herschbach}},
		\bibinfo {author} {\bibfnamefont {J.}~\bibnamefont {Keller}},\ and\ \bibinfo
		{author} {\bibfnamefont {T.~E.}\ \bibnamefont {Mehlst\"{a}ubler}},\
	}\bibfield  {title} {\bibinfo {title} {A high-precision segmented {Paul} trap
			with minimized micromotion for an optical multiple-ion clock},\ }\href
	{https://doi.org/10.1007/s00340-013-5580-5} {\bibfield  {journal} {\bibinfo
			{journal} {Appl. Phys. B}\ }\textbf {\bibinfo {volume} {114}},\ \bibinfo
		{pages} {231} (\bibinfo {year} {2014})}\BibitemShut {NoStop}%
	\bibitem [{\citenamefont {Keller}\ \emph {et~al.}(2019)\citenamefont {Keller},
		\citenamefont {Burgermeister}, \citenamefont {Kalincev}, \citenamefont
		{Didier}, \citenamefont {Kulosa}, \citenamefont {Nordmann}, \citenamefont
		{Kiethe},\ and\ \citenamefont {Mehlst{\"a}ubler}}]{Keller2019}%
	\BibitemOpen
	\bibfield  {author} {\bibinfo {author} {\bibfnamefont {J.}~\bibnamefont
			{Keller}}, \bibinfo {author} {\bibfnamefont {T.}~\bibnamefont
			{Burgermeister}}, \bibinfo {author} {\bibfnamefont {D.}~\bibnamefont
			{Kalincev}}, \bibinfo {author} {\bibfnamefont {A.}~\bibnamefont {Didier}},
		\bibinfo {author} {\bibfnamefont {A.~P.}\ \bibnamefont {Kulosa}}, \bibinfo
		{author} {\bibfnamefont {T.}~\bibnamefont {Nordmann}}, \bibinfo {author}
		{\bibfnamefont {J.}~\bibnamefont {Kiethe}},\ and\ \bibinfo {author}
		{\bibfnamefont {T.~E.}\ \bibnamefont {Mehlst{\"a}ubler}},\ }\bibfield
	{title} {\bibinfo {title} {Controlling systematic frequency uncertainties at
			the $10^{-19}$ level in linear {Coulomb} crystals},\ }\href
	{https://doi.org/10.1103/PhysRevA.99.013405} {\bibfield  {journal} {\bibinfo
			{journal} {Phys. Rev. A}\ }\textbf {\bibinfo {volume} {99}},\ \bibinfo
		{pages} {013405} (\bibinfo {year} {2019})}\BibitemShut {NoStop}%
	\bibitem [{\citenamefont {Liu}\ \emph {et~al.}(2003)\citenamefont {Liu},
		\citenamefont {Avinash},\ and\ \citenamefont {Goree}}]{Liu2003}%
	\BibitemOpen
	\bibfield  {author} {\bibinfo {author} {\bibfnamefont {B.}~\bibnamefont
			{Liu}}, \bibinfo {author} {\bibfnamefont {K.}~\bibnamefont {Avinash}},\ and\
		\bibinfo {author} {\bibfnamefont {J.}~\bibnamefont {Goree}},\ }\bibfield
	{title} {\bibinfo {title} {Transverse optical mode in a one-dimensional
			{Yukawa} chain},\ }\href {https://doi.org/10.1103/PhysRevLett.91.255003}
	{\bibfield  {journal} {\bibinfo  {journal} {Phys. Rev. Lett.}\ }\textbf
		{\bibinfo {volume} {91}},\ \bibinfo {pages} {255003} (\bibinfo {year}
		{2003})}\BibitemShut {NoStop}%
	\bibitem [{\citenamefont {Piacente}\ \emph
		{et~al.}(2004{\natexlab{b}})\citenamefont {Piacente}, \citenamefont
		{Peeters},\ and\ \citenamefont {Betouras}}]{Piacente2004a}%
	\BibitemOpen
	\bibfield  {author} {\bibinfo {author} {\bibfnamefont {G.}~\bibnamefont
			{Piacente}}, \bibinfo {author} {\bibfnamefont {F.~M.}\ \bibnamefont
			{Peeters}},\ and\ \bibinfo {author} {\bibfnamefont {J.~J.}\ \bibnamefont
			{Betouras}},\ }\bibfield  {title} {\bibinfo {title} {Normal modes of a
			quasi-one-dimensional multichain complex plasma},\ }\href
	{https://doi.org/10.1103/PhysRevE.70.036406} {\bibfield  {journal} {\bibinfo
			{journal} {Phys. Rev. E}\ }\textbf {\bibinfo {volume} {70}},\ \bibinfo
		{pages} {036406} (\bibinfo {year} {2004}{\natexlab{b}})}\BibitemShut
	{NoStop}%
	\bibitem [{\citenamefont {Morigi}\ and\ \citenamefont
		{Eschner}(2001)}]{Morigi2001}%
	\BibitemOpen
	\bibfield  {author} {\bibinfo {author} {\bibfnamefont {G.}~\bibnamefont
			{Morigi}}\ and\ \bibinfo {author} {\bibfnamefont {J.}~\bibnamefont
			{Eschner}},\ }\bibfield  {title} {\bibinfo {title} {Doppler cooling of a
			{Coulomb} crystal},\ }\href {https://doi.org/10.1103/PhysRevA.64.063407}
	{\bibfield  {journal} {\bibinfo  {journal} {Phys. Rev. A}\ }\textbf {\bibinfo
			{volume} {64}},\ \bibinfo {pages} {063407} (\bibinfo {year}
		{2001})}\BibitemShut {NoStop}%
	\bibitem [{\citenamefont {Kubo}\ \emph {et~al.}(2012)\citenamefont {Kubo},
		\citenamefont {Toda},\ and\ \citenamefont {Hashitsume}}]{Kubo2012}%
	\BibitemOpen
	\bibfield  {author} {\bibinfo {author} {\bibfnamefont {R.}~\bibnamefont
			{Kubo}}, \bibinfo {author} {\bibfnamefont {M.}~\bibnamefont {Toda}},\ and\
		\bibinfo {author} {\bibfnamefont {N.}~\bibnamefont {Hashitsume}},\ }\href
	{https://doi.org/10.1007/978-3-642-96701-6} {\emph {\bibinfo {title}
			{Statistical physics II: nonequilibrium statistical mechanics}}},\
	Vol.~\bibinfo {volume} {31}\ (\bibinfo  {publisher} {Springer Science \&
		Business Media},\ \bibinfo {year} {2012})\BibitemShut {NoStop}%
	\bibitem [{\citenamefont {H\"anggi}\ \emph {et~al.}(1990)\citenamefont
		{H\"anggi}, \citenamefont {Talkner},\ and\ \citenamefont
		{Borkovec}}]{Haenggi1990}%
	\BibitemOpen
	\bibfield  {author} {\bibinfo {author} {\bibfnamefont {P.}~\bibnamefont
			{H\"anggi}}, \bibinfo {author} {\bibfnamefont {P.}~\bibnamefont {Talkner}},\
		and\ \bibinfo {author} {\bibfnamefont {M.}~\bibnamefont {Borkovec}},\
	}\bibfield  {title} {\bibinfo {title} {Reaction-rate theory: fifty years
			after {Kramers}},\ }\href {https://doi.org/10.1103/RevModPhys.62.251}
	{\bibfield  {journal} {\bibinfo  {journal} {Rev. Mod. Phys.}\ }\textbf
		{\bibinfo {volume} {62}},\ \bibinfo {pages} {251} (\bibinfo {year}
		{1990})}\BibitemShut {NoStop}%
	\bibitem [{\citenamefont {Partner}\ \emph {et~al.}(2013)\citenamefont
		{Partner}, \citenamefont {Nigmatullin}, \citenamefont {Burgermeister},
		\citenamefont {Pyka}, \citenamefont {Keller}, \citenamefont {Retzker},
		\citenamefont {Plenio},\ and\ \citenamefont
		{Mehlst\"{a}ubler}}]{Partner2013}%
	\BibitemOpen
	\bibfield  {author} {\bibinfo {author} {\bibfnamefont {H.~L.}\ \bibnamefont
			{Partner}}, \bibinfo {author} {\bibfnamefont {R.}~\bibnamefont
			{Nigmatullin}}, \bibinfo {author} {\bibfnamefont {T.}~\bibnamefont
			{Burgermeister}}, \bibinfo {author} {\bibfnamefont {K.}~\bibnamefont {Pyka}},
		\bibinfo {author} {\bibfnamefont {J.}~\bibnamefont {Keller}}, \bibinfo
		{author} {\bibfnamefont {A.}~\bibnamefont {Retzker}}, \bibinfo {author}
		{\bibfnamefont {M.~B.}\ \bibnamefont {Plenio}},\ and\ \bibinfo {author}
		{\bibfnamefont {T.~E.}\ \bibnamefont {Mehlst\"{a}ubler}},\ }\bibfield
	{title} {\bibinfo {title} {Dynamics of topological defects in ion {Coulomb}
			crystals},\ }\href {https://doi.org/10.1088/1367-2630/15/10/103013}
	{\bibfield  {journal} {\bibinfo  {journal} {New J. Phys.}\ }\textbf {\bibinfo
			{volume} {15}},\ \bibinfo {pages} {103013} (\bibinfo {year}
		{2013})}\BibitemShut {NoStop}%
	\bibitem [{\citenamefont {Delfau}\ \emph {et~al.}(2013)\citenamefont {Delfau},
		\citenamefont {Coste},\ and\ \citenamefont {Saint~Jean}}]{Delfau2013}%
	\BibitemOpen
	\bibfield  {author} {\bibinfo {author} {\bibfnamefont {J.-B.}\ \bibnamefont
			{Delfau}}, \bibinfo {author} {\bibfnamefont {C.}~\bibnamefont {Coste}},\ and\
		\bibinfo {author} {\bibfnamefont {M.}~\bibnamefont {Saint~Jean}},\ }\bibfield
	{title} {\bibinfo {title} {Noisy zigzag transition, fluctuations, and
			thermal bifurcation threshold},\ }\href
	{https://doi.org/10.1103/PhysRevE.87.062135} {\bibfield  {journal} {\bibinfo
			{journal} {Phys. Rev. E}\ }\textbf {\bibinfo {volume} {87}},\ \bibinfo
		{pages} {062135} (\bibinfo {year} {2013})}\BibitemShut {NoStop}%
	\bibitem [{\citenamefont {Kramers}(1940)}]{Kramers1940}%
	\BibitemOpen
	\bibfield  {author} {\bibinfo {author} {\bibfnamefont {H.~A.}\ \bibnamefont
			{Kramers}},\ }\bibfield  {title} {\bibinfo {title} {Brownian motion in a
			field of force and the diffusion model of chemical reactions},\ }\href
	{https://doi.org/10.1016/S0031-8914(40)90098-2} {\bibfield  {journal}
		{\bibinfo  {journal} {Physica}\ }\textbf {\bibinfo {volume} {7}},\ \bibinfo
		{pages} {284} (\bibinfo {year} {1940})}\BibitemShut {NoStop}%
	\bibitem [{\citenamefont {Retzker}\ \emph {et~al.}(2008)\citenamefont
		{Retzker}, \citenamefont {Thompson}, \citenamefont {Segal},\ and\
		\citenamefont {Plenio}}]{Retzker2008}%
	\BibitemOpen
	\bibfield  {author} {\bibinfo {author} {\bibfnamefont {A.}~\bibnamefont
			{Retzker}}, \bibinfo {author} {\bibfnamefont {R.~C.}\ \bibnamefont
			{Thompson}}, \bibinfo {author} {\bibfnamefont {D.~M.}\ \bibnamefont
			{Segal}},\ and\ \bibinfo {author} {\bibfnamefont {M.~B.}\ \bibnamefont
			{Plenio}},\ }\bibfield  {title} {\bibinfo {title} {Double well potentials and
			quantum phase transitions in ion traps},\ }\href
	{https://doi.org/10.1103/PhysRevLett.101.260504} {\bibfield  {journal}
		{\bibinfo  {journal} {Phys. Rev. Lett.}\ }\textbf {\bibinfo {volume} {101}},\
		\bibinfo {pages} {260504} (\bibinfo {year} {2008})}\BibitemShut {NoStop}%
	\bibitem [{\citenamefont {De~Chiara}\ \emph {et~al.}(2010)\citenamefont
		{De~Chiara}, \citenamefont {del Campo}, \citenamefont {Morigi}, \citenamefont
		{Plenio},\ and\ \citenamefont {Retzker}}]{DeChiara2010}%
	\BibitemOpen
	\bibfield  {author} {\bibinfo {author} {\bibfnamefont {G.}~\bibnamefont
			{De~Chiara}}, \bibinfo {author} {\bibfnamefont {A.}~\bibnamefont {del
				Campo}}, \bibinfo {author} {\bibfnamefont {G.}~\bibnamefont {Morigi}},
		\bibinfo {author} {\bibfnamefont {M.~B.}\ \bibnamefont {Plenio}},\ and\
		\bibinfo {author} {\bibfnamefont {A.}~\bibnamefont {Retzker}},\ }\bibfield
	{title} {\bibinfo {title} {Spontaneous nucleation of structural defects in
			inhomogeneous ion chains},\ }\href
	{https://doi.org/10.1088/1367-2630/12/11/115003} {\bibfield  {journal}
		{\bibinfo  {journal} {New J. Phys.}\ }\textbf {\bibinfo {volume} {12}},\
		\bibinfo {pages} {115003} (\bibinfo {year} {2010})}\BibitemShut {NoStop}%
	\bibitem [{\citenamefont {Ibaraki}\ \emph {et~al.}(2011)\citenamefont
		{Ibaraki}, \citenamefont {Tanaka},\ and\ \citenamefont
		{Urabe}}]{Ibaraki2011}%
	\BibitemOpen
	\bibfield  {author} {\bibinfo {author} {\bibfnamefont {Y.}~\bibnamefont
			{Ibaraki}}, \bibinfo {author} {\bibfnamefont {U.}~\bibnamefont {Tanaka}},\
		and\ \bibinfo {author} {\bibfnamefont {S.}~\bibnamefont {Urabe}},\ }\bibfield
	{title} {\bibinfo {title} {Detection of parametric resonance of trapped ions
			for micromotion compensation},\ }\href
	{https://doi.org/10.1007/s00340-011-4463-x} {\bibfield  {journal} {\bibinfo
			{journal} {Appl. Phys. B}\ }\textbf {\bibinfo {volume} {105}},\ \bibinfo
		{pages} {219} (\bibinfo {year} {2011})}\BibitemShut {NoStop}%
	\bibitem [{\citenamefont {Bergamini}\ \emph {et~al.}(2004)\citenamefont
		{Bergamini}, \citenamefont {Darqui\'{e}}, \citenamefont {Jones},
		\citenamefont {Jacubowiez}, \citenamefont {Browaeys},\ and\ \citenamefont
		{Grangier}}]{Bergamini2004}%
	\BibitemOpen
	\bibfield  {author} {\bibinfo {author} {\bibfnamefont {S.}~\bibnamefont
			{Bergamini}}, \bibinfo {author} {\bibfnamefont {B.}~\bibnamefont
			{Darqui\'{e}}}, \bibinfo {author} {\bibfnamefont {M.}~\bibnamefont {Jones}},
		\bibinfo {author} {\bibfnamefont {L.}~\bibnamefont {Jacubowiez}}, \bibinfo
		{author} {\bibfnamefont {A.}~\bibnamefont {Browaeys}},\ and\ \bibinfo
		{author} {\bibfnamefont {P.}~\bibnamefont {Grangier}},\ }\bibfield  {title}
	{\bibinfo {title} {Holographic generation of microtrap arrays for single
			atoms by use of a programmable phase modulator},\ }\href
	{https://doi.org/10.1364/JOSAB.21.001889} {\bibfield  {journal} {\bibinfo
			{journal} {J. Opt. Soc. Am. B}\ }\textbf {\bibinfo {volume} {21}},\ \bibinfo
		{pages} {1889} (\bibinfo {year} {2004})}\BibitemShut {NoStop}%
	\bibitem [{\citenamefont {Zupancic}\ \emph {et~al.}(2016)\citenamefont
		{Zupancic}, \citenamefont {Preiss}, \citenamefont {Ma}, \citenamefont
		{Lukin}, \citenamefont {Tai}, \citenamefont {Rispoli}, \citenamefont
		{Islam},\ and\ \citenamefont {Greiner}}]{Zupancic2016}%
	\BibitemOpen
	\bibfield  {author} {\bibinfo {author} {\bibfnamefont {P.}~\bibnamefont
			{Zupancic}}, \bibinfo {author} {\bibfnamefont {P.~M.}\ \bibnamefont
			{Preiss}}, \bibinfo {author} {\bibfnamefont {R.}~\bibnamefont {Ma}}, \bibinfo
		{author} {\bibfnamefont {A.}~\bibnamefont {Lukin}}, \bibinfo {author}
		{\bibfnamefont {M.~E.}\ \bibnamefont {Tai}}, \bibinfo {author} {\bibfnamefont
			{M.}~\bibnamefont {Rispoli}}, \bibinfo {author} {\bibfnamefont
			{R.}~\bibnamefont {Islam}},\ and\ \bibinfo {author} {\bibfnamefont
			{M.}~\bibnamefont {Greiner}},\ }\bibfield  {title} {\bibinfo {title}
		{Ultra-precise holographic beam shaping for microscopic quantum control},\
	}\href {https://doi.org/10.1364/OE.24.013881} {\bibfield  {journal} {\bibinfo
			{journal} {Opt. Express}\ }\textbf {\bibinfo {volume} {24}},\ \bibinfo
		{pages} {13881} (\bibinfo {year} {2016})}\BibitemShut {NoStop}%
	\bibitem [{\citenamefont {Marquet}\ \emph {et~al.}(2003)\citenamefont
		{Marquet}, \citenamefont {Schmidt-Kaler},\ and\ \citenamefont
		{James}}]{Marquet2003}%
	\BibitemOpen
	\bibfield  {author} {\bibinfo {author} {\bibfnamefont {C.}~\bibnamefont
			{Marquet}}, \bibinfo {author} {\bibfnamefont {F.}~\bibnamefont
			{Schmidt-Kaler}},\ and\ \bibinfo {author} {\bibfnamefont {D.~F.~V.}\
			\bibnamefont {James}},\ }\bibfield  {title} {\bibinfo {title} {Phonon--phonon
			interactions due to non-linear effects in a linear ion trap},\ }\href
	{https://doi.org/10.1007/s00340-003-1097-7} {\bibfield  {journal} {\bibinfo
			{journal} {Appl. Phys. B}\ }\textbf {\bibinfo {volume} {76}},\ \bibinfo
		{pages} {199} (\bibinfo {year} {2003})}\BibitemShut {NoStop}%
	\bibitem [{\citenamefont {Kaufmann}\ \emph {et~al.}(2012)\citenamefont
		{Kaufmann}, \citenamefont {Ulm}, \citenamefont {Jacob}, \citenamefont
		{Poschinger}, \citenamefont {Landa}, \citenamefont {Retzker}, \citenamefont
		{Plenio},\ and\ \citenamefont {Schmidt-Kaler}}]{Kaufmann2012}%
	\BibitemOpen
	\bibfield  {author} {\bibinfo {author} {\bibfnamefont {H.}~\bibnamefont
			{Kaufmann}}, \bibinfo {author} {\bibfnamefont {S.}~\bibnamefont {Ulm}},
		\bibinfo {author} {\bibfnamefont {G.}~\bibnamefont {Jacob}}, \bibinfo
		{author} {\bibfnamefont {U.}~\bibnamefont {Poschinger}}, \bibinfo {author}
		{\bibfnamefont {H.}~\bibnamefont {Landa}}, \bibinfo {author} {\bibfnamefont
			{A.}~\bibnamefont {Retzker}}, \bibinfo {author} {\bibfnamefont
			{M.~B.}~\bibnamefont {Plenio}},\ and\ \bibinfo {author} {\bibfnamefont
			{F.}~\bibnamefont {Schmidt-Kaler}},\ }\bibfield  {title} {\bibinfo {title}
		{Precise experimental investigation of eigenmodes in a planar ion crystal},\
	}\href {https://doi.org/10.1103/PhysRevLett.109.263003} {\bibfield  {journal}
		{\bibinfo  {journal} {Phys. Rev. Lett.}\ }\textbf {\bibinfo {volume} {109}},\
		\bibinfo {pages} {263003} (\bibinfo {year} {2012})}\BibitemShut {NoStop}%
	\bibitem [{\citenamefont {Wineland}\ \emph {et~al.}(1992)\citenamefont
		{Wineland}, \citenamefont {Dalibard},\ and\ \citenamefont
		{Cohen-Tannoudji}}]{Wineland1992}%
	\BibitemOpen
	\bibfield  {author} {\bibinfo {author} {\bibfnamefont {D.~J.}\ \bibnamefont
			{Wineland}}, \bibinfo {author} {\bibfnamefont {J.}~\bibnamefont {Dalibard}},\
		and\ \bibinfo {author} {\bibfnamefont {C.}~\bibnamefont {Cohen-Tannoudji}},\
	}\bibfield  {title} {\bibinfo {title} {Sisyphus cooling of a bound atom},\
	}\href {https://doi.org/10.1364/josab.9.000032} {\bibfield  {journal}
		{\bibinfo  {journal} {J Opt Soc Am B}\ }\textbf {\bibinfo {volume} {9}},\
		\bibinfo {pages} {32} (\bibinfo {year} {1992})}\BibitemShut {NoStop}%
	\bibitem [{\citenamefont {Ejtemaee}\ and\ \citenamefont
		{Haljan}(2017)}]{Ejtemaee2017}%
	\BibitemOpen
	\bibfield  {author} {\bibinfo {author} {\bibfnamefont {S.}~\bibnamefont
			{Ejtemaee}}\ and\ \bibinfo {author} {\bibfnamefont {P.~C.}~\bibnamefont
			{Haljan}},\ }\bibfield  {title} {\bibinfo {title} {3d sisyphus cooling of
			trapped ions},\ }\href {https://doi.org/10.1103/PhysRevLett.119.043001}
	{\bibfield  {journal} {\bibinfo  {journal} {Phys. Rev. Lett.}\ }\textbf
		{\bibinfo {volume} {119}},\ \bibinfo {pages} {043001} (\bibinfo {year}
		{2017})}\BibitemShut {NoStop}%
	\bibitem [{\citenamefont {Joshi}\ \emph {et~al.}(2020)\citenamefont {Joshi},
		\citenamefont {Fabre}, \citenamefont {Maier}, \citenamefont {Brydges},
		\citenamefont {Kiesenhofer}, \citenamefont {Hainzer}, \citenamefont {Blatt},\
		and\ \citenamefont {Roos}}]{Joshi2020}%
	\BibitemOpen
	\bibfield  {author} {\bibinfo {author} {\bibfnamefont {M.~K.}\ \bibnamefont
			{Joshi}}, \bibinfo {author} {\bibfnamefont {A.}~\bibnamefont {Fabre}},
		\bibinfo {author} {\bibfnamefont {C.}~\bibnamefont {Maier}}, \bibinfo
		{author} {\bibfnamefont {T.}~\bibnamefont {Brydges}}, \bibinfo {author}
		{\bibfnamefont {D.}~\bibnamefont {Kiesenhofer}}, \bibinfo {author}
		{\bibfnamefont {H.}~\bibnamefont {Hainzer}}, \bibinfo {author} {\bibfnamefont
			{R.}~\bibnamefont {Blatt}},\ and\ \bibinfo {author} {\bibfnamefont {C.~F.}\
			\bibnamefont {Roos}},\ }\bibfield  {title} {\bibinfo {title}
		{Polarization-gradient cooling of {1D} and {2D} ion {Coulomb} crystals},\
	}\href {https://doi.org/10.1088/1367-2630/abb912} {\bibfield  {journal}
		{\bibinfo  {journal} {New J. Phys.}\ }\textbf {\bibinfo {volume} {22}},\
		\bibinfo {pages} {103013} (\bibinfo {year} {2020})},\ \bibinfo {note}
	{publisher: IOP Publishing}\BibitemShut {NoStop}%
	\bibitem [{\citenamefont {Morigi}\ \emph {et~al.}(2000)\citenamefont {Morigi},
		\citenamefont {Eschner},\ and\ \citenamefont {Keitel}}]{Morigi2000}%
	\BibitemOpen
	\bibfield  {author} {\bibinfo {author} {\bibfnamefont {G.}~\bibnamefont
			{Morigi}}, \bibinfo {author} {\bibfnamefont {J.}~\bibnamefont {Eschner}},\
		and\ \bibinfo {author} {\bibfnamefont {C.~H.}\ \bibnamefont {Keitel}},\
	}\bibfield  {title} {\bibinfo {title} {Ground state laser cooling using
			electromagnetically induced transparency},\ }\href
	{https://doi.org/10.1103/PhysRevLett.85.4458} {\bibfield  {journal} {\bibinfo
			{journal} {Phys. Rev. Lett.}\ }\textbf {\bibinfo {volume} {85}},\ \bibinfo
		{pages} {4458} (\bibinfo {year} {2000})}\BibitemShut {NoStop}%
	\bibitem [{\citenamefont {Morigi}(2003)}]{Morigi2003}%
	\BibitemOpen
	\bibfield  {author} {\bibinfo {author} {\bibfnamefont {G.}~\bibnamefont
			{Morigi}},\ }\bibfield  {title} {\bibinfo {title} {Cooling atomic motion with
			quantum interference},\ }\href {https://doi.org/10.1103/PhysRevA.67.033402}
	{\bibfield  {journal} {\bibinfo  {journal} {Phys. Rev. A}\ }\textbf {\bibinfo
			{volume} {67}},\ \bibinfo {pages} {033402} (\bibinfo {year}
		{2003})}\BibitemShut {NoStop}%
	\bibitem [{\citenamefont {Lechner}\ \emph {et~al.}(2016)\citenamefont
		{Lechner}, \citenamefont {Maier}, \citenamefont {Hempel}, \citenamefont
		{Jurcevic}, \citenamefont {Lanyon}, \citenamefont {Monz}, \citenamefont
		{Brownnutt}, \citenamefont {Blatt},\ and\ \citenamefont
		{Roos}}]{Lechner2016}%
	\BibitemOpen
	\bibfield  {author} {\bibinfo {author} {\bibfnamefont {R.}~\bibnamefont
			{Lechner}}, \bibinfo {author} {\bibfnamefont {C.}~\bibnamefont {Maier}},
		\bibinfo {author} {\bibfnamefont {C.}~\bibnamefont {Hempel}}, \bibinfo
		{author} {\bibfnamefont {P.}~\bibnamefont {Jurcevic}}, \bibinfo {author}
		{\bibfnamefont {B.~P.}\ \bibnamefont {Lanyon}}, \bibinfo {author}
		{\bibfnamefont {T.}~\bibnamefont {Monz}}, \bibinfo {author} {\bibfnamefont
			{M.}~\bibnamefont {Brownnutt}}, \bibinfo {author} {\bibfnamefont
			{R.}~\bibnamefont {Blatt}},\ and\ \bibinfo {author} {\bibfnamefont {C.~F.}\
			\bibnamefont {Roos}},\ }\bibfield  {title} {\bibinfo {title}
		{Electromagnetically-induced-transparency ground-state cooling of long ion
			strings},\ }\href {https://doi.org/10.1103/PhysRevA.93.053401} {\bibfield
		{journal} {\bibinfo  {journal} {Phys. Rev. A}\ }\textbf {\bibinfo {volume}
			{93}},\ \bibinfo {pages} {053401} (\bibinfo {year} {2016})}\BibitemShut
	{NoStop}%
	\bibitem [{\citenamefont {Scharnhorst}\ \emph {et~al.}(2018)\citenamefont
		{Scharnhorst}, \citenamefont {Cerrillo}, \citenamefont {Kramer},
		\citenamefont {Leroux}, \citenamefont {W\"ubbena}, \citenamefont {Retzker},\
		and\ \citenamefont {Schmidt}}]{Scharnhorst2018}%
	\BibitemOpen
	\bibfield  {author} {\bibinfo {author} {\bibfnamefont {N.}~\bibnamefont
			{Scharnhorst}}, \bibinfo {author} {\bibfnamefont {J.}~\bibnamefont
			{Cerrillo}}, \bibinfo {author} {\bibfnamefont {J.}~\bibnamefont {Kramer}},
		\bibinfo {author} {\bibfnamefont {I.~D.}\ \bibnamefont {Leroux}}, \bibinfo
		{author} {\bibfnamefont {J.~B.}\ \bibnamefont {W\"ubbena}}, \bibinfo {author}
		{\bibfnamefont {A.}~\bibnamefont {Retzker}},\ and\ \bibinfo {author}
		{\bibfnamefont {P.~O.}\ \bibnamefont {Schmidt}},\ }\bibfield  {title}
	{\bibinfo {title} {Experimental and theoretical investigation of a multimode
			cooling scheme using multiple electromagnetically-induced-transparency
			resonances},\ }\href {https://doi.org/10.1103/physreva.98.023424} {\bibfield
		{journal} {\bibinfo  {journal} {Phys. Rev. A}\ }\textbf {\bibinfo {volume}
			{98}},\ \bibinfo {pages} {023424} (\bibinfo {year} {2018})}\BibitemShut
	{NoStop}%
	\bibitem [{\citenamefont {Jordan}\ \emph {et~al.}(2019)\citenamefont {Jordan},
		\citenamefont {Gilmore}, \citenamefont {Shankar}, \citenamefont
		{Safavi-Naini}, \citenamefont {Bohnet}, \citenamefont {Holland},\ and\
		\citenamefont {Bollinger}}]{Jordan2019}%
	\BibitemOpen
	\bibfield  {author} {\bibinfo {author} {\bibfnamefont {E.}~\bibnamefont
			{Jordan}}, \bibinfo {author} {\bibfnamefont {K.~A.}\ \bibnamefont {Gilmore}},
		\bibinfo {author} {\bibfnamefont {A.}~\bibnamefont {Shankar}}, \bibinfo
		{author} {\bibfnamefont {A.}~\bibnamefont {Safavi-Naini}}, \bibinfo {author}
		{\bibfnamefont {J.~G.}\ \bibnamefont {Bohnet}}, \bibinfo {author}
		{\bibfnamefont {M.~J.}\ \bibnamefont {Holland}},\ and\ \bibinfo {author}
		{\bibfnamefont {J.~J.}\ \bibnamefont {Bollinger}},\ }\bibfield  {title}
	{\bibinfo {title} {Near {Ground}-{State} {Cooling} of {Two}-{Dimensional}
			{Trapped}-{Ion} {Crystals} with {More} than 100 {Ions}},\ }\href
	{https://doi.org/10.1103/PhysRevLett.122.053603} {\bibfield  {journal}
		{\bibinfo  {journal} {Phys. Rev. Lett.}\ }\textbf {\bibinfo {volume} {122}},\
		\bibinfo {pages} {053603} (\bibinfo {year} {2019})}\BibitemShut {NoStop}%
	\bibitem [{\citenamefont {Kiethe}\ \emph {et~al.}(2018)\citenamefont {Kiethe},
		\citenamefont {Nigmatullin}, \citenamefont {Schmirander}, \citenamefont
		{Kalincev},\ and\ \citenamefont {Mehlst{\"a}ubler}}]{Kiethe2018}%
	\BibitemOpen
	\bibfield  {author} {\bibinfo {author} {\bibfnamefont {J.}~\bibnamefont
			{Kiethe}}, \bibinfo {author} {\bibfnamefont {R.}~\bibnamefont {Nigmatullin}},
		\bibinfo {author} {\bibfnamefont {T.}~\bibnamefont {Schmirander}}, \bibinfo
		{author} {\bibfnamefont {D.}~\bibnamefont {Kalincev}},\ and\ \bibinfo
		{author} {\bibfnamefont {T.~E.}\ \bibnamefont {Mehlst{\"a}ubler}},\
	}\bibfield  {title} {\bibinfo {title} {Nanofriction and motion of topological
			defects in self-organized ion {Coulomb} crystals},\ }\href
	{https://doi.org/10.1088/1367-2630/aaf3d5} {\bibfield  {journal} {\bibinfo
			{journal} {New J. Phys.}\ }\textbf {\bibinfo {volume} {20}},\ \bibinfo
		{pages} {123017} (\bibinfo {year} {2018})}\BibitemShut {NoStop}%
	\bibitem [{\citenamefont {Timm}\ \emph {et~al.}(2020)\citenamefont {Timm},
		\citenamefont {Weimer}, \citenamefont {Santos},\ and\ \citenamefont
		{Mehlst\"aubler}}]{Timm2020}%
	\BibitemOpen
	\bibfield  {author} {\bibinfo {author} {\bibfnamefont {L.}~\bibnamefont
			{Timm}}, \bibinfo {author} {\bibfnamefont {H.}~\bibnamefont {Weimer}},
		\bibinfo {author} {\bibfnamefont {L.}~\bibnamefont {Santos}},\ and\ \bibinfo
		{author} {\bibfnamefont {T.~E.}\ \bibnamefont {Mehlst\"aubler}},\ }\bibfield
	{title} {\bibinfo {title} {Energy localization in an atomic chain with a
			topological soliton},\ }\href
	{https://doi.org/10.1103/PhysRevResearch.2.033198} {\bibfield  {journal}
		{\bibinfo  {journal} {Phys. Rev. Research}\ }\textbf {\bibinfo {volume}
			{2}},\ \bibinfo {pages} {033198} (\bibinfo {year} {2020})}\BibitemShut
	{NoStop}%
	\bibitem [{\citenamefont {Landa}\ \emph {et~al.}(2010)\citenamefont {Landa},
		\citenamefont {Marcovitch}, \citenamefont {Retzker}, \citenamefont {Plenio},\
		and\ \citenamefont {Reznik}}]{Landa2010}%
	\BibitemOpen
	\bibfield  {author} {\bibinfo {author} {\bibfnamefont {H.}~\bibnamefont
			{Landa}}, \bibinfo {author} {\bibfnamefont {S.}~\bibnamefont {Marcovitch}},
		\bibinfo {author} {\bibfnamefont {A.}~\bibnamefont {Retzker}}, \bibinfo
		{author} {\bibfnamefont {M.~B.}\ \bibnamefont {Plenio}},\ and\ \bibinfo
		{author} {\bibfnamefont {B.}~\bibnamefont {Reznik}},\ }\bibfield  {title}
	{\bibinfo {title} {Quantum coherence of discrete kink solitons in ion
			traps},\ }\href {https://doi.org/10.1103/PhysRevLett.104.043004} {\bibfield
		{journal} {\bibinfo  {journal} {Phys. Rev. Lett.}\ }\textbf {\bibinfo
			{volume} {104}},\ \bibinfo {pages} {043004} (\bibinfo {year}
		{2010})}\BibitemShut {NoStop}%
	\bibitem [{\citenamefont {Landa}\ \emph {et~al.}(2014)\citenamefont {Landa},
		\citenamefont {Retzker}, \citenamefont {Schaetz},\ and\ \citenamefont
		{Reznik}}]{Landa2014}%
	\BibitemOpen
	\bibfield  {author} {\bibinfo {author} {\bibfnamefont {H.}~\bibnamefont
			{Landa}}, \bibinfo {author} {\bibfnamefont {A.}~\bibnamefont {Retzker}},
		\bibinfo {author} {\bibfnamefont {T.}~\bibnamefont {Schaetz}},\ and\ \bibinfo
		{author} {\bibfnamefont {B.}~\bibnamefont {Reznik}},\ }\bibfield  {title}
	{\bibinfo {title} {Entanglement generation using discrete solitons in
			{Coulomb} crystals},\ }\href {https://doi.org/10.1103/PhysRevLett.113.053001}
	{\bibfield  {journal} {\bibinfo  {journal} {Phys. Rev. Lett.}\ }\textbf
		{\bibinfo {volume} {113}},\ \bibinfo {pages} {053001} (\bibinfo {year}
		{2014})}\BibitemShut {NoStop}%
	\bibitem [{\citenamefont {Shankar}\ \emph {et~al.}(2020)\citenamefont
		{Shankar}, \citenamefont {Tang}, \citenamefont {Affolter}, \citenamefont
		{Gilmore}, \citenamefont {Dubin}, \citenamefont {Parker}, \citenamefont
		{Holland},\ and\ \citenamefont {Bollinger}}]{Shankar2020}%
	\BibitemOpen
	\bibfield  {author} {\bibinfo {author} {\bibfnamefont {A.}~\bibnamefont
			{Shankar}}, \bibinfo {author} {\bibfnamefont {C.}~\bibnamefont {Tang}},
		\bibinfo {author} {\bibfnamefont {M.}~\bibnamefont {Affolter}}, \bibinfo
		{author} {\bibfnamefont {K.}~\bibnamefont {Gilmore}}, \bibinfo {author}
		{\bibfnamefont {D.~H.~E.}\ \bibnamefont {Dubin}}, \bibinfo {author}
		{\bibfnamefont {S.}~\bibnamefont {Parker}}, \bibinfo {author} {\bibfnamefont
			{M.~J.}\ \bibnamefont {Holland}},\ and\ \bibinfo {author} {\bibfnamefont
			{J.~J.}\ \bibnamefont {Bollinger}},\ }\bibfield  {title} {\bibinfo {title}
		{Broadening of the drumhead-mode spectrum due to in-plane thermal
			fluctuations of two-dimensional trapped ion crystals in a {Penning} trap},\
	}\href {https://doi.org/10.1103/PhysRevA.102.053106} {\bibfield  {journal}
		{\bibinfo  {journal} {Phys. Rev. A}\ }\textbf {\bibinfo {volume} {102}},\
		\bibinfo {pages} {053106} (\bibinfo {year} {2020})}\BibitemShut {NoStop}%
\end{thebibliography}
%

\end{document}